%% file: main.tex
\documentclass[acmsmall]{acmart}
\AtBeginDocument{%
  \providecommand\BibTeX{{%
    \normalfont B\kern-0.5em{\scshape i\kern-0.25em b}\kern-0.8em\TeX}}}

\setcopyright{acmcopyright}
\copyrightyear{2018}
\acmYear{2018}
\acmDOI{XXXXXXX.XXXXXXX}

\newcommand{\asplossubmissionnumber}{1296}
\usepackage[normalem]{ulem}

\usepackage{mdframed}

\DeclareUrlCommand\ULurl{%
  }%

\usepackage{pdfpages}

\usepackage{caption}
\usepackage{subcaption}
\usepackage{multirow}
\usepackage{algorithm}
\usepackage[noend]{algpseudocode}
\usepackage{array}
\usepackage{fixltx2e}
\usepackage{stfloats}
\usepackage{url}

\usepackage{times}
\usepackage{color}
\usepackage{verbatim}
\usepackage{subfiles}
\usepackage{url}
\usepackage{lipsum}
\usepackage{xr}
\usepackage{hyperref}
\usepackage{zref-xr}
\usepackage{blindtext}
\usepackage{tabularx}
\usepackage{etoolbox}
\usepackage{setspace}
\usepackage{export}
\usepackage{nameref}
\usepackage{inputenc}
\usepackage{mathtools}
\usepackage{esvect}
\usepackage{xspace}	
\usepackage{soul}
\usepackage[export]{adjustbox}
\usepackage{amsmath, amsthm}
\usepackage[toc,page]{appendix}
\graphicspath{{../pdf/}{../jpeg/}}
\DeclareGraphicsExtensions{.pdf,.jpeg,.png}
\setcopyright{none}
\renewcommand\footnotetextcopyrightpermission[1]{} 
\renewcommand\footnotetextauthorsaddresses[1]{}
\makeatletter
\patchcmd{\@classz}
  {\CT@row@color}
  {\oldCT@column@color}
  {}
  {}
\patchcmd{\@classz}
  {\CT@column@color}
  {\CT@row@color}
  {}
  {}
\patchcmd{\@classz}
  {\oldCT@column@color}
  {\CT@column@color}
  {}
  {}
\makeatother

\definecolor{fxnote}{rgb}{0.8000,0.0000,0.0000}
\hypersetup{
    colorlinks=true,
    linkcolor=red,
    filecolor=magenta,      
    urlcolor=cyan,
}
\renewcommand{\paragraph}[1]{\textbf{Brief: #1}}

\newcommand{\red}[1]{\textcolor{red}{#1}}

\usepackage{hyperref}

\begin{document}

\title{FARSI: An Agile, Early-Stage Design Space Exploration Framework to Tame the Complexity of DSSoCs}
\title{FARSI: Facebook AR System Investigator for Agile Domain-Specific System-on-Chip Exploration}

\author{Behzad Boroujerdian}
\email{behzadboro@gmail.com}
\affiliation{%
  \institution{University of Texas at Austin and Facebook Inc.}
  \country{USA}
}

\author{Ying Jing}
\email{yingj4@illinois.edu}
\affiliation{%
  \institution{University of Illinois Urbana-Champaign
}
  \country{USA}
}

\author{Amit Kumar}
\email{amit34@fb.com}
\affiliation{%
  \institution{Facebook Inc.}
    \country{USA}
}

\author{Lavanya Subramanian}
\email{lavanyas@fb.com}
\affiliation{%
  \institution{Facebook Inc.}
    \country{USA}
}

\author{Luke Yen}
\email{lyen@tenstorrent.com}
\affiliation{%
  \institution{Tenstorrent Inc.}
    \country{USA}
}

\author{Vincent Lee}
\email{vtlee@fb.com}
\affiliation{%
  \institution{Facebook Inc.}
    \country{USA}
}

\author{Vivek Venkatesan}
\email{vvenkatesan@fb.com}
\affiliation{%
  \institution{Facebook Inc.}
    \country{USA}
}

\author{Amit Jindal}
\email{amit.jindal@onepeloton.com}
\affiliation{%
  \institution{Peloton Interactive Inc.}
    \country{USA}
}

\author{Robert Shearer}
\email{robshearer@fb.com}
\affiliation{%
  \institution{Facebook Inc.}
    \country{USA}
}

\author{Vijay Janapa Reddi}
\email{vj@eecs.harvard.edu}
\affiliation{%
\institution{Harvard University and The University of Texas at Austin
  \country{USA}\\
\ULurl{https://github.com/facebookresearch/Project_FARSI}}
}


\date{}
\begin{abstract}
\input{tex/abstract}
\end{abstract}
\maketitle

\thispagestyle{empty}

\input{tex/introduction.tex}

\input{tex/background.tex}

\input{tex/FARSI}

\input{tex/evaluation.tex}

\input{tex/case_study}

\input{tex/related.tex}
\input{tex/conclusion.tex}
\bibliographystyle{plain}
\bibliography{tex/main.bib}


\end{document}

%% file: tex/abstract.tex
\label{sec:abstract} 

Domain-specific SoCs (DSSoCs) are attractive solutions for domains with stringent power/performance/area constraints; however, they suffer from two fundamental complexities. On the one hand, their many specialized hardware blocks result in complex systems and thus high development effort. On the other, their many system knobs expand the complexity of design space, making the search for the optimal design difficult. Thus to reach prevalence, taming such complexities is necessary. 
This work identifies necessary features of an early-stage design space exploration (DSE) framework that targets the complex design space of DSSoCs and further provides an instance of one called FARSI, (F)acebook (AR) (S)ystem (I)nvestigator. Concretely, FARSI provides an agile system-level simulator with speed up and accuracy of 8,400X and 98.5\% comparing to Synopsys Platform Architect.
FARSI also provides an efficient exploration heuristic and achieves up to 16X improvement in convergence time comparing to naive simulated annealing (SA). This is done by augmenting SA with architectural reasoning such as locality exploitation and bottleneck relaxation. Furthermore, we embed various co-design capabilities and show that on average, they have a 32\% impact on the convergence rate. 
Finally, we demonstrate that using simple development-cost-aware policies can lower the system complexity, both in terms of the component count and variation by as much as 150\% and 118\% (e,g., for Network-on-a-Chip subsystem).

%% file: tex/introduction.tex
\section{Introduction}
\label{sec:introduction} 

With the end of Moore's law, simple systems with a few general-purpose processors do not provide a path forward for domains with stringent power/performance/area constraints. This is due to the performance and energy inefficiencies of these systems/processors, such as the high instruction fetch and decode overhead. Domain-specific SoCs (DSSoCs) are an attractive and adaptive alternative as they provide the required performance and energy scalability. This is due to their domain awareness and specialized hardware blocks to keep the computation/communication fast and efficient.

Although efficient, DSSoCs are complex systems and demand high development efforts to design. This complexity is driven by number and variation (heterogeneity) of the processing elements, and the intricate topological structure connecting them. 
Concretely, from the compute perspective, increasingly customized accelerators accompany general-purpose processors; from the communication perspective, both the Network-on-a-Chip (NoC) and memory subsystems see a complexity rise to keep the data local and the movement low energy~\cite{diephoto}. 
In addition to the system complexity, the sheer number of design knobs per component dramatically expands the design space making the search for the optimal design difficult. For example, a simple system with five simple workloads and six total knobs (e.g., frequency, bus width, accelerator hardening knobs) totals over a million design points. Naive brute force sweeps are infeasible for the design space exploration of DSSoCs.

\begin{figure}[t]
    \centering
        \includegraphics[trim=0 0 0 0, clip, width=.6\linewidth]{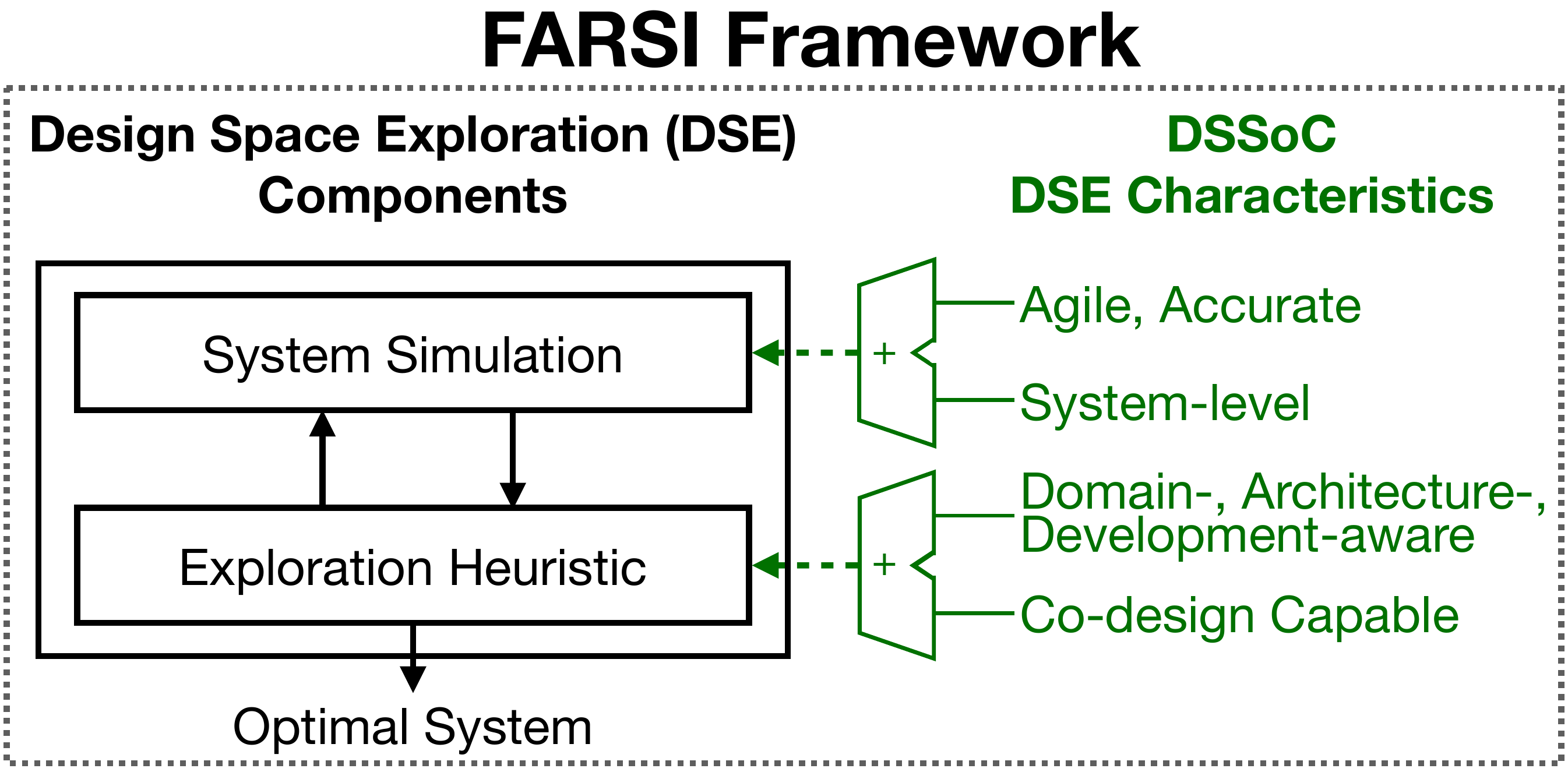}
        \caption{DSE components to the left need the characteristics on the right to tame the DSSoC design space complexities. We introduce FARSI, an early-stage DSE framework that has those characteristics. Hence, it enables an agile DSE for DSSoCs.}
         \label{fig:dse}
\end{figure}


To manage all said complexities, we need a design space exploration (DSE) technique that is agile/efficient to navigate the space effectively, and further is development cost-aware to dampen the design effort.
DSEs typically consist of a simulator (Figure~\ref{fig:dse}, top left) to capture the design behavior and an exploration heuristic (Figure~\ref{fig:dse}, bottom left) to navigate the design space. Both of these components need to be efficient. To that end, in this work, we provide an efficient early-stage (pre-synthesis) DSE, called FARSI, that concretely targets DSSoCs and their complexity. Our work is open-sourced and free for public use: \url{https://github.com/facebookresearch/Project\_FARSI}. 


FARSI meets the requirement that DSSoC simulators need to be agile, accurate, and capable of system-level estimation (Figure~\ref{fig:dse}, top right); DSSoC simulators must be capable of estimating system-level dynamics as profiling components in isolation cannot accurately measure their system-level impact.   Furthermore, they must be agile to enable sufficient coverage of the vast design space. 
FARSI provides a system-level simulator to capture the complexity of accelerator-rich systems (e.g., with more than 50 hardware blocks). It also achieves high agility/accuracy by combining analytical models' speed and the event-driven simulation's accuracy.
Our analytical models (for the first time) build on top and further expand Gables~\cite{gables}, a  roofline-based bottleneck analysis for SoCs. We augment these analytical models with a lightweight phase-driven simulator to estimate system dynamics. FARSI splits the program progression into phases (the longest unit of time within which the system bottleneck stays constant), calculates their bottlenecks, and advances the simulation accordingly. Such an approach enables fast simulation as the phases are not fixed and adaptively expand or shrink based on the program characteristics. 

We quantify our simulator fidelity by comparing it against Synopsys' Platform Architect~\cite{PA}, which is an industry-strength early-stage analysis and optimization tool for SoC architectures. In all of our studies, we target the Augmented reality (AR) domain without the loss of generality as its rigid constraints demand accelerator-rich and complex DSSoCs. Specifically, we compare against over 200 designs of different complexities and show an average speedup of 8,400X (.032 seconds on average for FARSI and 235 seconds for PA) and an accuracy of 98.5\%. We further quantify the agility scalability of our simulator with respect to the system complexity, showing a slight slowdown of 3X when the number of blocks is increased by 20X. Lastly, we show that using our simulator in the DSE can improve the exploration convergence time\footnote{Convergence time is the time that it takes for the DSE to meet the design's budgets.} from (estimated) 3 years to 3 hours. 

In addition, DSSoC exploration heuristics need to be domain-, architecture-, and development-aware and also co-design capable (Figure~\ref{fig:dse}, bottom right). Embedding these characteristics into a search heuristic (e.g., simulated annealing, although we can expand to other heuristics) can improve the convergence time and system complexity. 
DSSoC exploration heuristics need to be ``domain-aware'', meaning that they must extract and further exploit the opportunities provided by their workload set (e.g., loop/task-level parallelism). We equip FARSI with this feature and show how parallelism and computation/communication boundedness can effectively guide the search. DSSoC exploration heuristics also need to be ``architecture-aware'', i.e., deploy architectural reasoning during the search and efficiently prune the space. For example, they should exploit spatial locality to bring the data closer for the accelerator or use bottleneck analysis to relax the congestion for an overloaded NoC. We equip FARSI with such reasoning and show that when embedded into naive search algorithms (e.g., simulated annealing), architectural insights can improve the convergence time by 16X.
We further quantify how different levels of reasoning can impact this convergence time and hypothesize that the said convergence gap will only grow with more complex workload sets.

DSSoC exploration heuristics also need to exploit co-design, e.g., simultaneous communication and computation optimization, as this approach provides additional optimization opportunities to accelerate the convergence.
FARSI exploits co-design heavily across four main vectors. It co-designs (1) across workloads, looking for holistic improvements across workloads boundaries.
This approach is also applied (2) across metrics, (3) across computation-communication, and (4) across optimizations (mapping, allocation, customization). 
We show that this feature can impact the convergence rate on average by 32\% and further detail the impact of each co-design vector.  
We also show that embedding the same co-design capabilities in regular simulated annealing (without architectural considerations) does not necessarily translate to design improvements.

Finally, DSSoC exploration heuristics need to be development cost-aware to keep system complexity and development effort low. We equip FARSI with this feature by prioritizing low development-cost optimizations (e.g., prioritizing software mapping over hardware customization) and provide a case study quantifying the impact.
We show that when system specifications (e.g., latency budget of SoC) are relaxed, FARSI decreases the output design complexity by lowering the number (e.g., 150\% reduction on the number of NoCs) and variation of components (118\% reduction in NoC frequency). 

We compare our methodology with the divide and conquer methodology, commonly used to manage complexity. Divide and conquer simply splits the design into sub-components and targets each locally.
We shed light on significant problems associated with this methodology and show that they can lead to systems with up to 56\% and 52\% system degradation in performance and power comparing to FARSI's designs. In short, FARSI reveals a methodology for efficiently managing the system and the search space complexity of DSSoCs. Our contributions include the following:
\begin{itemize}

    \item We provide the first Gables-based~\cite{gables} characterization of a host of AR workloads from three different AR application primitives---audio decoding~\cite{huzaifa2021illixr}, CAVA~\cite{cava}, and edge detection~\cite{hpvm}. 

    \item  We show how a hybrid estimation methodology that combines analytical models and phase-driven simulation can provide both the agility (8,400X speed up) and the accuracy (98.5\%)  much needed for DSSoCs. Our simulator lowers the convergence time from (estimated) 3 years to 3 hours comparing to Synopsys Platform Architect.

   \item We show how architectural reasoning (e.g., locality exploitation or bottleneck relaxation) and co-design 
   can improve the convergence time of simple search heuristics such as simulated annealing by more than an order of magnitude.
   %

   
   \item We show how development-aware policies (e.g., prioritizing incremental improvements) can exploit relax budgets and lower the final system complexity, both in terms of component counts and variations by more than 100\%.

   

\end{itemize}


%% file: tex/background.tex
\section{Background}
\label{sec:background}

This section presents an overview of the AR workloads that we test the FARSI DSE with. First, we motivate the suitability of the Augmented Reality (AR) domain for DSSoCs and then detail our workloads. Though we focus on the AR domain, the FARSI simulator is applicable in other domains. 

\subsection{Augmented Reality, a Great Match for DSSoC}
Augmented reality is emerging as one of the main drivers of technology. With use cases in medical training~\cite{AR_med}, repair and maintenance~\cite{AR_repair}, education~\cite{AR_education}, tourism~\cite{AR_tourism} and others, this domain is on the cusp of showing its full potential. Furthermore, the current COVID\_19 pandemic has accelerated this emergence as various use cases such as training and education continue migrating to the online domain~\cite{online_learning}. AR's promise to pervade various aspects of our lives has caught the attention of technology giants such as Facebook~\cite{Facebook_AR}, Google~\cite{Google_AR},  Apple~\cite{Apple_AR} and others with the effort to embed this technology in various form factors ranging from phones to glasses~\cite{Smart_glasses}.

Despite said promises, this domain faces various challenges from the compute perspective.
First of all, due to the domain's tight constraints, currently, there are several orders of magnitude gaps between the needed and achievable performance, power, and usability~\cite{huzaifa2021illixr}. 
Closing this gap is imperative as the performance constraints ensure timely interaction with the world while the power constraints ensure operational battery life.
Secondly, this domain demands SOCs with a diverse set of sub-domains, including video, graphics, computer vision, optics, and audio~\cite{huzaifa2021illixr}. The combination of these sub-systems forms a rich set of pipelines that continuously interact to deliver the real-time functionality needed. This diversity increases the development effort to design AR systems.

Said two challenges make AR an ideal candidate for domain-specific SOCs. 
On the one hand, the tight budgets of the domain require a host of specialized hardware units with high power and performance efficiency. On the other, the diversity of the sub-domains ask for holistic designs that go beyond each sub-domain and instead exploit cross sub-domain co-design opportunities. 

\begin{figure}[!t]
    \centering
    \includegraphics[trim=0 0 0 0, clip, width=.8\linewidth]{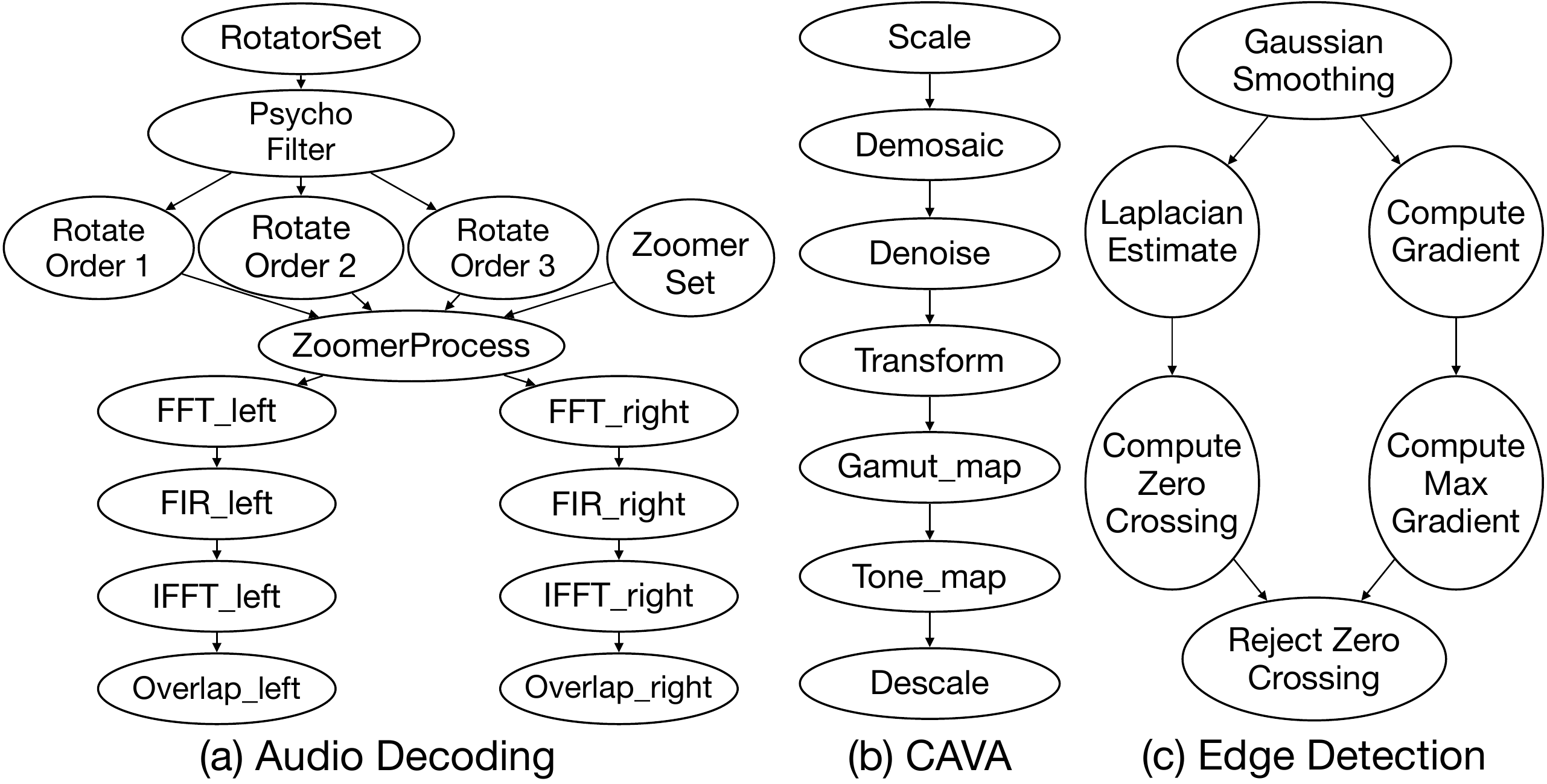}
    \caption{Our workloads' task dependency graphs (TDGs).}
    \label{fig:TDG}
\end{figure}

\subsection{Augmented Reality Workloads}
Augmented reality spans across many sub-domains and workloads. In this work, we target the primitive workloads, i.e., audio and image processing,
as they are present in almost any AR applications \cite{sponza, material, platformer}. Note that this work leaves out Facebook's proprietary internal workloads and only resorts to open-sourced libraries. 
Here we detail our workload's functionality and characteristics.

\textbf{Audio Decoder (Audio)} is used to playback the audio based on the user pose. Concretely, said pose, obtained from the IMU integrator, is used to rotate and zoom the soundfield which is then mapped to available speakers. This workload is from a new AR/VR testbed \cite{huzaifa2021illixr}.

\textbf{CAVA} simulates a configurable camera vision pipeline and is used to processes raw images (pixels) from sensors to feed the vision backend (e.g., a CNN). The default ISP kernel is modeled after Nikon-D7000 camera~\cite{Nikon} and is developed by Harvard~\cite{cava}.

\textbf{Edge Detection (ED)} processes the image and finds sharp changes in brightness to capture significant events and changes in properties of the world. It is one of the key steps used in image analysis and computer vision. Our workload appears in \cite{kotsifakou2018hpvm}. 

\subsection{Workload Characteristics}
To characterize each workload's execution flow,
we split them into a set of tasks (e.g., Tone Map task in CAVA workload) where
a task is the smallest unit of simulation and is selected from workload's functions. The execution dependency between the tasks is captured in the task dependency graph (TDG), where each node specifies tasks and edges specify their dependencies. Figure~\ref{fig:TDG} provides each workload's TDG. Audio has the most tasks (15), while Edge Detection has the least (6). In general, this distribution of tasks is in line with several internal proprietary workloads. 

To characterize the tasks, we use Gables~\cite{gables}. For the first time, we provide a Gables profile of a set of AR workloads. Gables is a set of abstract analytical models that captures high-level software computational and communicational characteristics. Concretely, it models each task's computation with its work ($f$ = instruction count) and its communication with operational intensity ($I$ =operation count per memory access which is split to $I_{read}$ and $I_{write}$ corresponding to read and write respectively).  Table~\ref{table:workCharacter} quantifies each workload's Gable relevant variables with each value averaged over all the tasks of a workload. TaLP and LLP values quantify (Ta)sk (L)evel and (L)oop (L)evel parallelism, respectively. The former quantifies the number of task combinations that can run in parallel, and the latter quantifies the average number of independent loop iterations.
 As seen, Audio uses both loop and task-level parallelism (highest TaLP among all), CAVA only uses loop-level parallelism, and Edge Detection has task level and highest LLP.  Edge detection is also the most communication-intensive benchmark. Note that task by task details for all the workloads are presented in the appendix.

\begin{table}[t!]
\footnotesize
\centering
\resizebox{.65\columnwidth}{!}{
\begin{tabular}{|c|c|c|c|}
\hline
Workload                        & Audio          & CAVA                       & ED           \\ \hline
$f$ (M.ops)                               & 13             & 24,252                & 1,098                 \\ \hline
($I_{read}$, $I_{write}$) (ops/byte)              & (8, 12) & (67K, 74K) & (126, 1.23M) \\ \hline
\begin{tabular}[c]{@{}c@{}}Average Data\\Movement (M.bytes)\end{tabular}          & 0.19             & 0.33                & 7.01                \\ \hline
Average LLP    & 2,392             & 151                        & 1,365,376                    \\ \hline
Average TaLP   & 48                      & 1                          & 4                          \\ \hline
\end{tabular}
}
\caption{Gables-based characteristics. K=kilo, M=million.}
\label{table:workCharacter}
\end{table}

%% file: tex/FARSI.tex
\begin{figure}[t!]
    \centering
   
        \includegraphics[trim=0 00 00 0, clip, width=.5\linewidth]{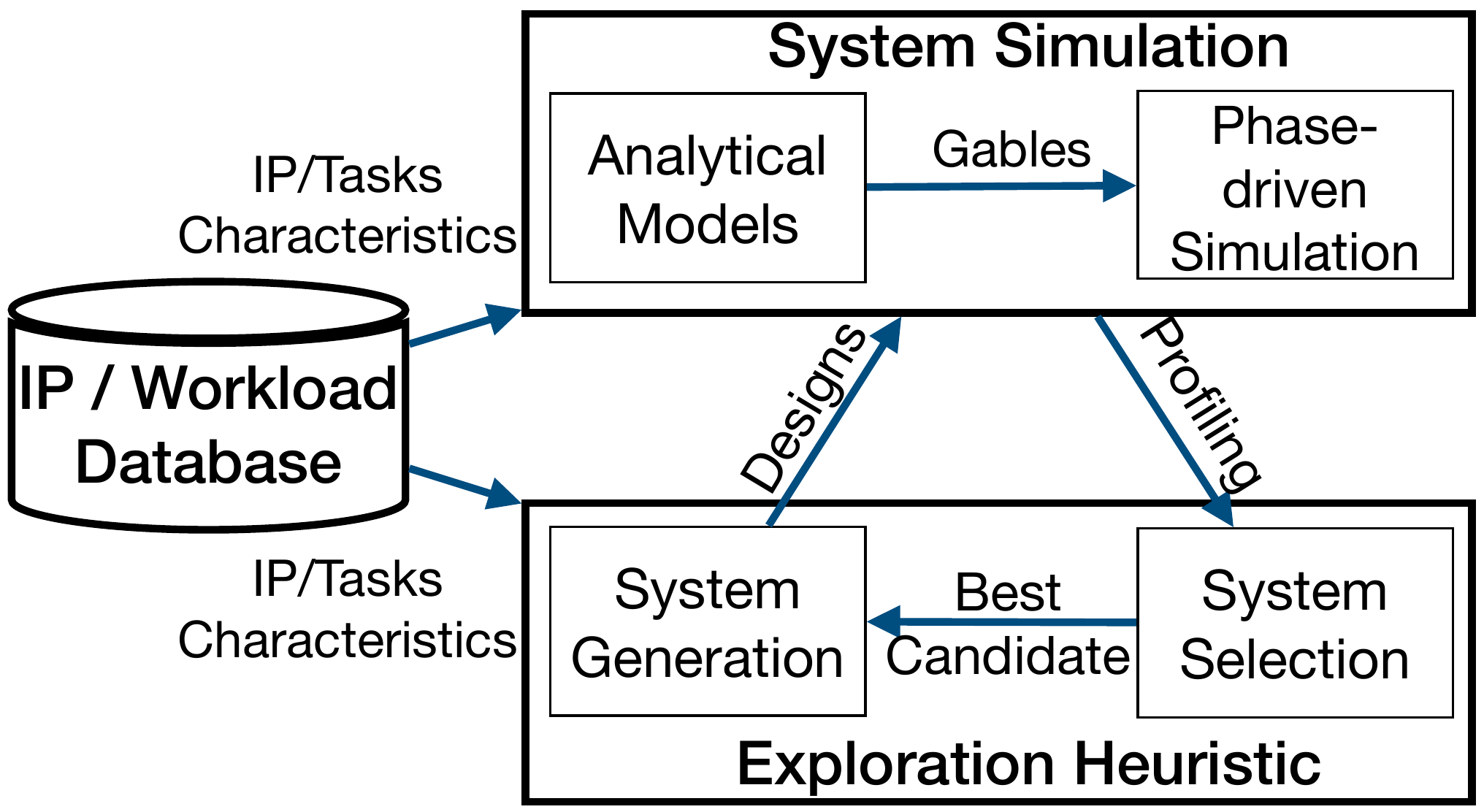}
        \caption{FARSI's end-to-end methodology.}
        \label{fig:system_analysis_methodology}
    
\end{figure}

\section{FARSI Methodology}
In this section, we detail FARSI DSE's methodology. FARSI is comprised of four stages (Figure~\ref{fig:system_analysis_methodology}), (1) database generation, (2) system simulation, (3) system generation, and (4) system selection. The first stage characterizes our workloads, collects (in isolation) IP performance/power/area (PPA) estimates, and populates our software/hardware database. This database is then continuously queried by our system simulator, system generator, and system selector as they work together to generate new systems and improve them. This process iteratively continues until the design's budgets (i.e., performance, power and area) are met.
FARSI's main contribution is in the last three stages as prior work such as ~\cite{accelSeeker, hpvm, aladdin}
sufficiently addresses the first. 


\subsection{Database Generation}
This stage populates a database with the characteristics of the domain's workloads and the PPA estimates of individual IPs. Both of these components are then used to estimate the system behavior while running various workloads. 

\textbf{Workload Analysis (Software Database):} This involves generating a task dependency graph (TDG) shown in  Figure~\ref{fig:TDG}. 
TDG's nodes are further augmented with computational characteristics (e.g., loop iteration and instruction counts), while edges augmented with communication data movement. Tasks are typically chosen from the functions with the highest computational/communication intensity as their optimization can significantly impact system behavior. Optimal task selection is a difficult problem and is outside of this work's scope; in our work, selection relies on application developers' insights. For TDG generation and profiling tasks' computational and communication characteristics, we use existing tools such as perf~\cite{perf}, AccelSeeker~\cite{accelSeeker} and HPVM~\cite{hpvm}.

\textbf{IP Analysis (Hardware Database):}
This involves PPA estimation of each task for different hardware mappings (e.g., to general-purpose processors or specialized accelerators). 
We build on top of an existing toolest called AccelSeeker~\cite{accelSeeker} and collect PPA estimates for each mapping. 
We augment our database with many memory and NoC widths and frequencies.

\subsection{System Simulation}

This stage raises the view to the system level and enables holistic analysis. System view is necessary since profiling components in isolation cannot accurately measure their overall impact within complex and accelerator rich DSSoCs.  This stage considers computational (general-purpose processors and accelerators) and communication IPs (NoCs and Memory) as the lowest abstraction unit.

We deploy a hybrid estimation methodology combining analytical models and a lightweight phase-driven simulator. The former enables agile traversal and thus an extensive design space coverage. The latter improves the simulation fidelity by capturing the system dynamics. Here we discuss each.


\begin{figure}[t]
\centering
 \captionsetup{type=table}
\includegraphics[trim=0 90 0 85, clip, width=.75\linewidth]{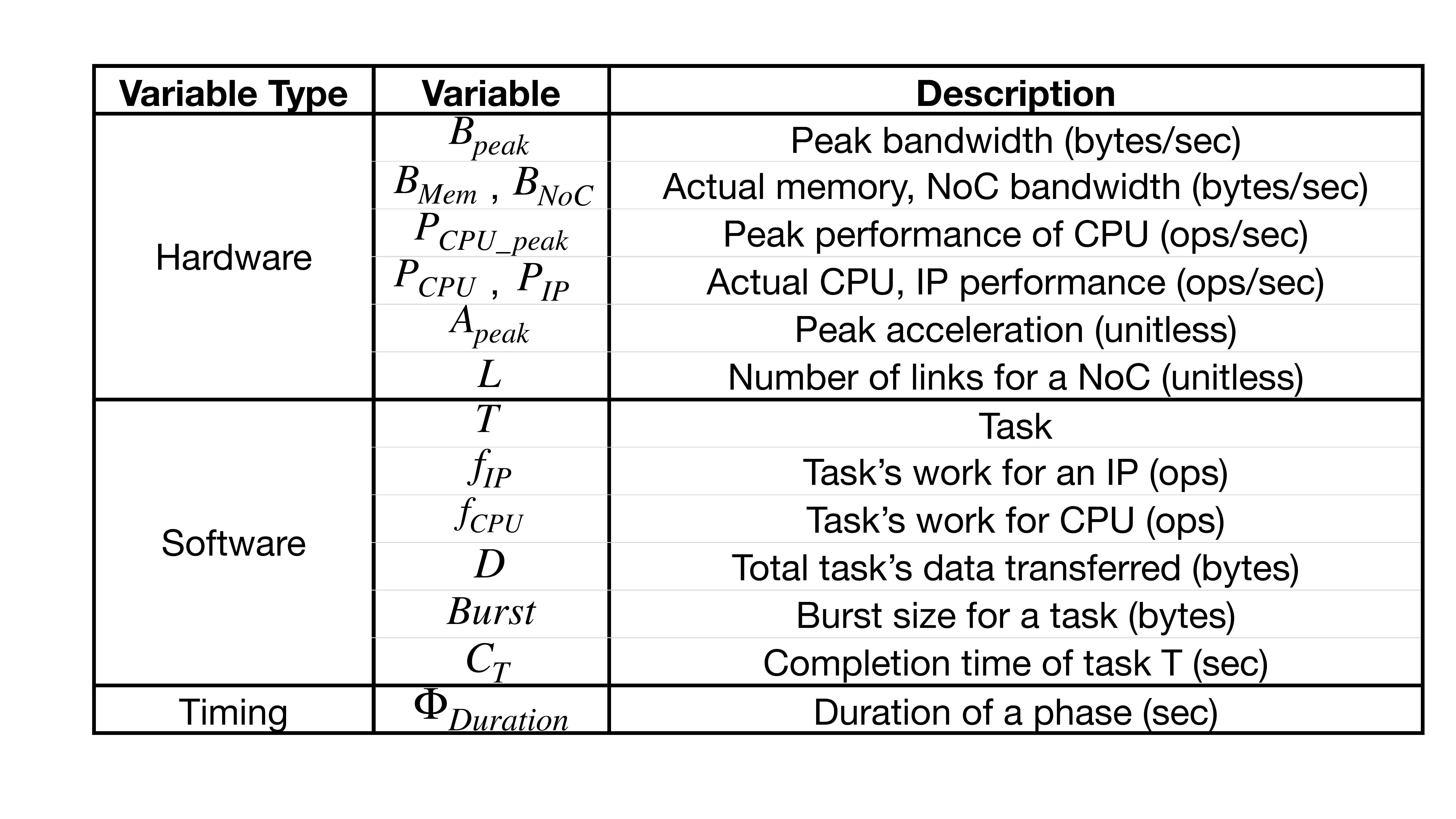}
\caption{Analytical model parameters.}
\label{table:anlytical_models_vars}
\end{figure}

\textbf{Software Analytical Models:}
We augment the Gables SoC-level roofline model as it provides a simple yet holistic/system-level set of analytical models~\cite{gables}. Gables combines bottleneck analysis and high-level computational/communication models. 
Our improvements include: 
\begin{itemize}
\item \textit{Finer Computation/Communication Granularity:} We lower the smallest execution unit from workload  to task. This captures the workload's lower-level compute, memory, and communication characteristics and further provides extra within-workload optimization opportunities. Also, we introduce a communication burst size parameter that captures NoC congestion behavior more accurately. 


\item \textit{Task Dependencies:} The Gables model assumes parallel execution of all workloads/tasks for simplicity. We use TDGs to model task-level parallelism (TaLP) accurately.


\item \textit{Computation/Communication Breakdown:} We augment Gables with loop iteration count and thus loop level parallelism (LLP). We also break down $I$ into read and write operational intensity ($I_{read}$, $I_{write}$) as modern routers/memories support separate channels for each.
 
\end{itemize}


\textbf{Hardware Analytical Models:}
Gabels models hardware with peak performance ($P_{peak}$) and bandwidth ($B$). It assumes a single NoC with a single channel. Our improvements include:
\begin{itemize}
\item \textit{Computation/Communication Resource Sharing:} We integrate CPU's multi-tasking models (pre-emptive scheduling),
and multi-channel routers for master-slave combinations.

\item \textit{Topology Improvement:}
We introduce ``many NoC'' topology systems to improve congestion/locality. 

\end{itemize}



\begin{figure*}[t!]
    \centering
    \subfloat[Phase driven simulation.]{
        \includegraphics[trim=0 0 0 0, clip, width=.6\linewidth]{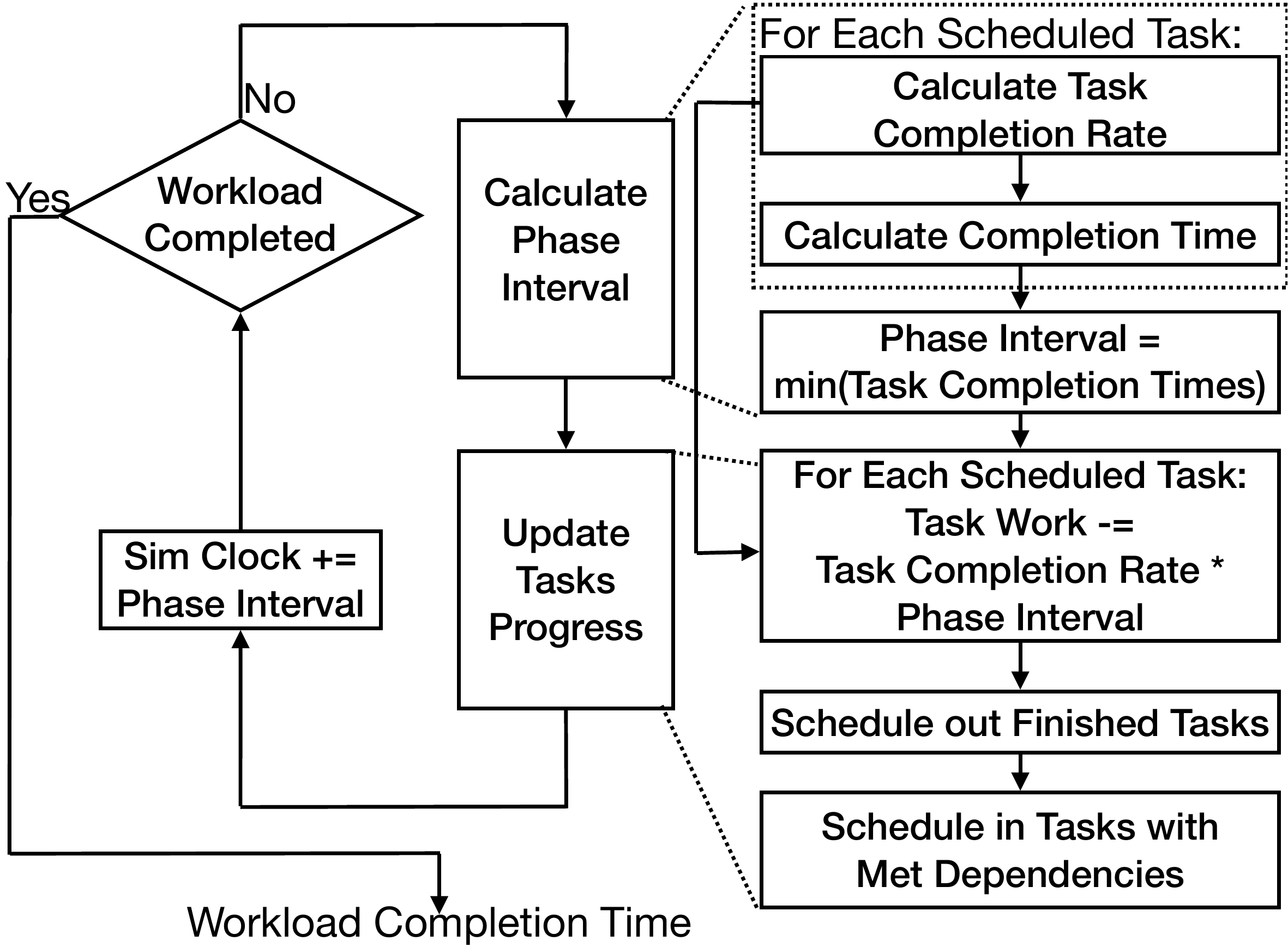}
        \label{fig:sys_sim_fc}
    }
    \subfloat[Phases for an example workload. $\phi$ and $T$ denote a phase and task respectively.]{
        \includegraphics[trim=0 -60 0 .5, clip, width=.4\linewidth]{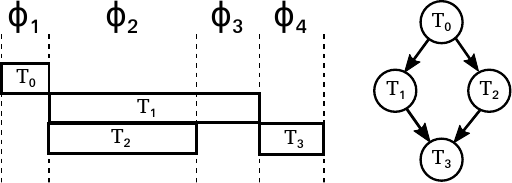}
        \label{fig:phase_example}
    }
    \caption{FARSI's phase-driven simulation. It is based on the concept of phase (right) where each tick moves the simulation forward one phase (left).}
\end{figure*}

\textbf{Phase-driven Simulation:}
Gables is a static model that cannot accurately capture the dynamic flow of complicated task graphs. So we wrap our analytical models with a lightweight \textit{phase-driven} simulation (Figure~\ref{fig:sys_sim_fc}). 
A \textit{phase} is a flexible time quantum with which we advance the simulation. It specifies the longest interval that the system behavior stays constant; Since our models use bottleneck analysis to estimate the system behavior, a new phase emerges when a task is completed and schedulued out (shifting the bottleneck by relaxing NoC/Memory/PEs pressure) or a new task is scheduled in (Figure~\ref{fig:phase_example}). 
At the highlevel, our simulator estimates a phase duration based on the fastest task's completion time, advances all tasks according to this interval, schedules out the completed task, and then schedules in
tasks whose dependencies are satisfied. This process continues iteratively until all the tasks/workloads(s) are completed (Figure~\ref{fig:sys_sim_fc}).

\textbf{Hybrid Estimation:}
Our hybrid approach combines the said analytical models and phase-driven simulation to estimate the SOC performance.
At the beginning of each phase, our phase-driven simulator schedules all the tasks whose dependencies are satisfied. Then, our analytical models determine each task's completion time by estimating their hardware blocks' processing rate (Equations~\ref{eq:Execurtion_rate_CPU} through~\ref{eq:Execurtion_rate_Mem}). $T$ and $Burst$ denote a task and its burst size, $P_{peak\_CPU}$ and $P_{CPU}$  denote a general-purpose processor's peak and actual performance, $A_{peak}$ and $P_{IP}$ denote peak accelerator speedup and its actual performance, $B_{peak\_Mem}$, 
$B_{peak\_NoC}$, $B_{Mem}$, $B_{NoC}$ denote  memory and NoC peak and actual bandwidth, $L$ denotes number of links in a NoC, and |\;\;| denotes a variable cardinality (e.g., |$T$| denote number of running tasks). To calculate the processing rate of CPU (Eq.~\ref{eq:Execurtion_rate_CPU}), IP (Eq.~\ref{eq:Execurtion_rate_IP}) and NoC (Eq.~\ref{eq:Execurtion_rate_NoC}) their peak rate is divided by the number of tasks running on them. This is because preemptive scheduling of CPU/IP and equal arbitration of a NoC will allocate the rate equally among the tasks.\footnote{This model can be easily extended for cooperative and non-equal arbitration; The details are left out for brevity.}
However, for memory (Eq.~\ref{eq:Execurtion_rate_Mem}) and NoC's links (Eq.~\ref{eq:Execurtion_rate_NoC}), this division is determined by the burst size ratio of the task over the total bursts of all running tasks. 

Once each block's processing rate for a task is determined, the completion time of the task is determined by its slowest block (Equation~\ref{eq:completion_time}). $T$ and $C$, and $f_{IP}$ denote a task, its completion time, and its work for an IP. Each component in the tuple calculates the execution time of a different block where each block's work (e.g., data or $D$ for memory) is divided by its execution rate (e.g., bandwidth for memory). The maximum function finds the slowest of blocks, and the number and type of its inputs are determined by the blocks hosting the task.

The end of a phase (phase duration) is determined by the fastest task, and thus estimated by the minimum of all tasks' completion times (Equation~\ref{eq:phase_duration}). $\Phi_{Duration}$, and $C_{T_i}$ denote the phase duration and the task i's completion time. Once said duration is calculated, the phase-driven simulator moves the simulation tick to the end of the phase, updates each task's progress according to the duration, schedules out the completed task, and schedules in ready tasks.\footnote{This paper currently only models the ``first ready first served'' scheduling.} This process continues until no more tasks are left to be scheduled.

\begin{align} 
 \label{eq:Execurtion_rate_CPU}
 P_{CPU} &= \frac{P_{peak\_CPU}}{|T|}  \;\;(\frac{ops}{sec}) \\
 \nonumber \\
 \label{eq:Execurtion_rate_IP}
 P_{IP} &= \frac{A_{peak}*P_{peak\_CPU}}{|T|}\;\;(\frac{ops}{sec}) \\
 \nonumber \\
   \label{eq:Execurtion_rate_NoC}
    B_{NoC} &= \frac{B_{peak\_NoC}} {(\frac {Burst_i} {\sum_i Burst}) *|L|} \;\;(\frac{byte}{sec}) \\
 \nonumber \\
  \label{eq:Execurtion_rate_Mem}
 B_{Mem} &= \frac{B_{peak\_Mem}} {(\frac {Burst_i} {\sum_i Burst})} \;\;(\frac{byte}{sec})
\end{align}

\begin{equation}
\label{eq:completion_time}
C_{T} = \max ( \frac{f_{IP}}{P_{IP}}, \frac{D} {B_{Mem}}, \frac{D}{B_{NoC}}, ...)  \; \;(sec)
\end{equation}

\begin{equation}
  \label{eq:phase_duration}
  \Phi_{Duration} = \min_{i}(C_{T_i}) 
  \; \;(sec)
\end{equation}

\textbf{Power/Area Modeling:}
For power and area of IPs, we use the verified AccelSeeker~\cite{accelSeeker} estimations, and for NoCs and memory, we have fully integrated CACTI~\cite{cacti} into FARSI. 

\subsection{Exploration Heuristic: System Generation}
\begin{figure}[!t]
\centering
\includegraphics[trim=0 0 0 0, clip, width=.55\linewidth]{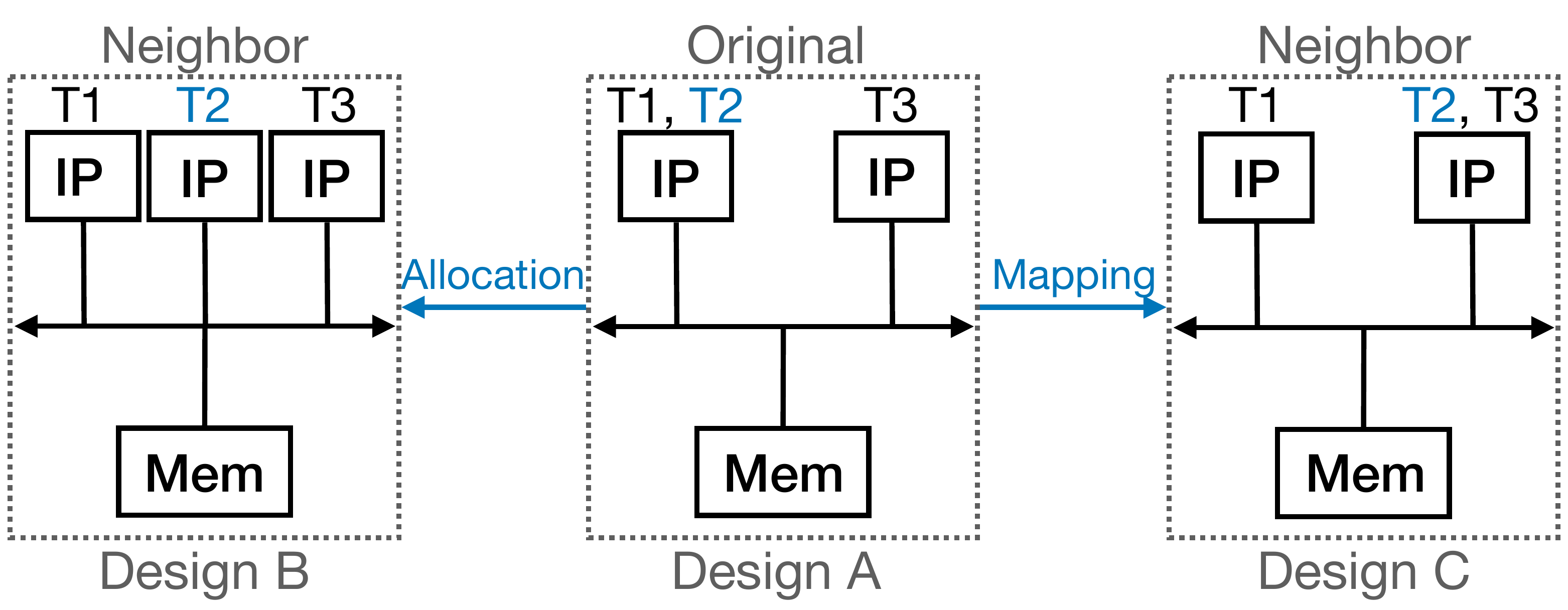}
\caption{Neighbor generation examples. Mem=Memory.}
\label{fig:FARSI_op:neighborGen}
\end{figure}

This stage exploits architectural insights to explore the design space. Without the loss of generality, we use simulated annealing (SA) as the search heuristic base and then augment it with various architectural insights. 
To explore the design space, we greedily generate a number of designs' neighbors (candidate designs in SA), simulate them and select the best one until the constraints are met (Figure~\ref{fig:system_analysis_methodology}). A design's \textit{neighbor}, is a design whose mapping, allocation or topology has been incrementally modified (Figure~\ref{fig:FARSI_op:neighborGen}). For example, designs B and C are design A's neighbors as the former has an allocation change (an extra IP) and the latter a mapping change (Task 2). 

We generate neighbors by selecting a five-tuple, i.e., a Metric, a Direction, a Task, a Block, and an Optimization Move to improve the said components.


\textbf{Metric/Direction:}
To generate a neighbor,  we target one metric per iteration (e.g., performance). We typically pick the metric farthest from its budget as it contributes to the distance (to goal) the most. Note that this metric can vary from iteration to iteration as we improve said metrics. Overall,  FARSI currently targets performance, power, and area. In addition to the metric, we also target an direction to improve the metric with (e.g., decrease or increase if we have overshot). 

\begin{figure*}[t!]
    \centering
    \subfloat[Swap]{
        \includegraphics[trim=0 0 0 0, clip, width=.4\linewidth]{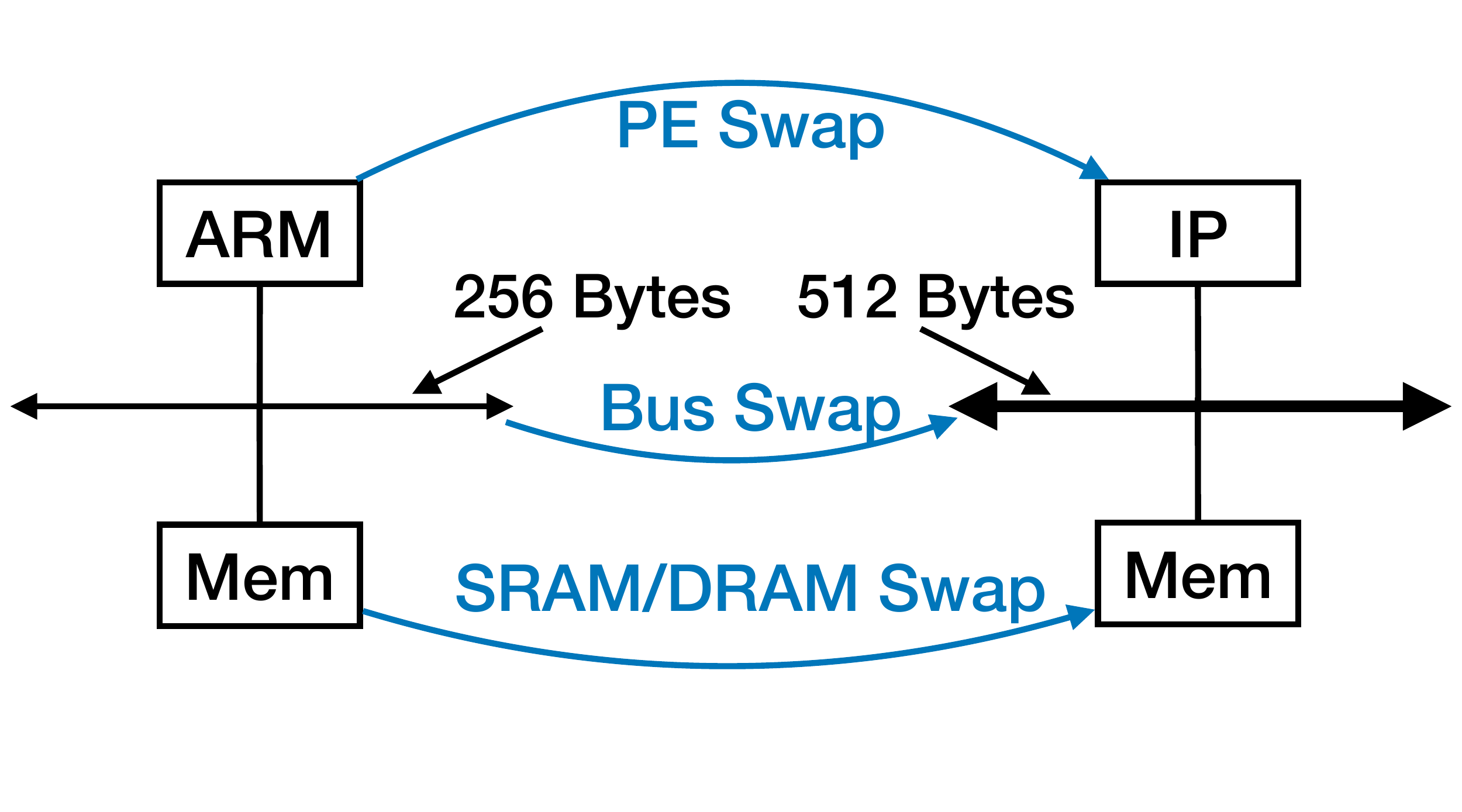}
        \label{fig:farsi_move:swap}
    }
   \hspace{-5pt}
    \subfloat[Fork/Join]{
        \includegraphics[trim=0 0 0 0, clip, width=.4\linewidth]{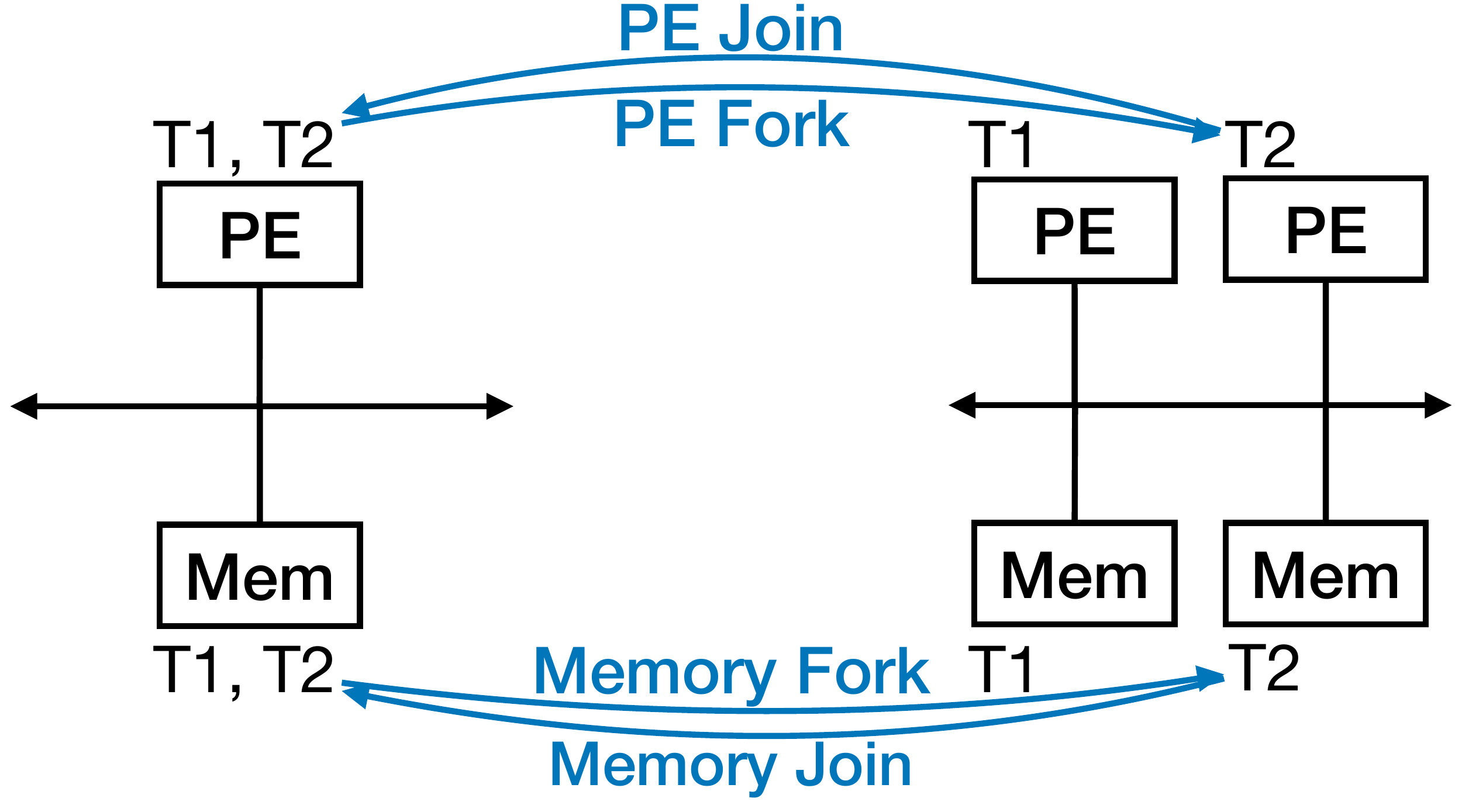}
        \label{fig:farsi_move:fork}
    }
      \vspace{15pt}
    \subfloat[Migrate]{
        \includegraphics[trim=0 0 0 0, clip, width=.4\linewidth]{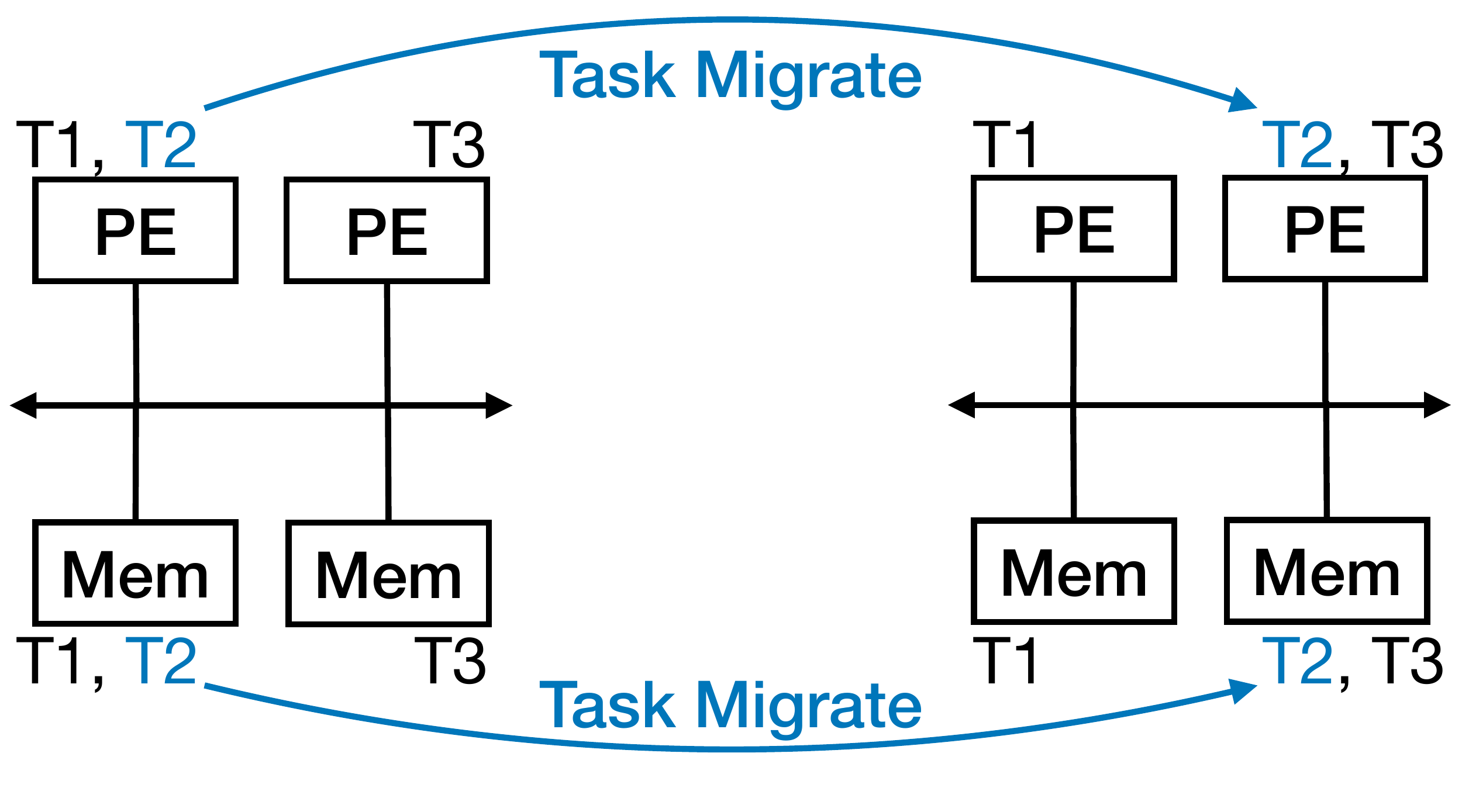}
        \label{fig:farsi_move:migrate}
    }
    \caption{``Optimization moves'' supported in FARSI. Each move incrementally modifies the design. We have provided a move for customization (Swap), allocation (Fork/Join) and software to hardware mapping (Migrate). PE = processing elements like CPU or accelerator (IP). Mem = Memory.}
    \label{fig:farsi_moves}
\end{figure*}

\textbf{Task/Hardware Block:}
To generate a neighbor, we target a task/block to improve, typically the one causing the highest distance to budget.
For example, if targeting latency, the highest latency task and its block bottleneck are selected.


\textbf{Optimization Move:}
\textit{Move} is an optimization on a task/block to improve a metric.
FARSI supports high-level optimizations such as hardware customization, hardware allocation, and software to hardware mapping and provides a move primitive for each, concretely swap, fork/join, and migrate.

\textit{Swap:} Enables customization by replacing a hardware block with a more specialized one (Figure~\ref{fig:farsi_move:swap}). For example, we replace an ARM core with an accelerator or a narrow bus with a wider one. Table~\ref{fig:swap_options} details our swap options.
Note that the swap only incrementally modifies the original block (e.g., 100MHz$\rightarrow$200MHz instead of 100MHz$\rightarrow$800MHz) to minimize the impact on competing metrics (e.g., performance vs power).



\textit{Fork/Join:} Impacts allocation by duplicating an existing block and migrating some of its tasks over (Figure~\ref{fig:farsi_move:fork}).
Duplication relaxes the block pressure, e.g., cutting down a NoC traffic and relaxing congestion.  
Join is the opposite of fork.

\textit{Migrate:} Modifies software to hardware mapping by migrating a task from one block to another (Figure~\ref{fig:farsi_move:migrate}).
Migration improves data locality (moving data closer to where it is used),
load balancing (mapping to an idling processor), and relaxes buses/memory pressure (mapping to bus/memory with the lower congestion).
Currently, mapping is done statically.



\textbf{Using Architectural Reasoning for Move Selection:} 
\begin{table}[b!]
\resizebox{.65\columnwidth}{!}{%
\begin{tabular}{|c|c|c|c|c|}
\hline
\begin{tabular}[c]{@{}c@{}}Block\\ Type\end{tabular} & Subtype                                              & \begin{tabular}[c]{@{}c@{}}Freq\\ (MHz)\end{tabular} & \begin{tabular}[c]{@{}c@{}}Bus Width\\ (Bytes)\end{tabular} & \begin{tabular}[c]{@{}c@{}}Loop\\ Unrolling\end{tabular}        \\ \hline
PE                                                   & \begin{tabular}[c]{@{}c@{}}GPP$\leftrightarrows$Acc\end{tabular}   & {[}100,...,800{]}                                       & N/A                                                         & \begin{tabular}[c]{@{}c@{}}According\\ to the task\end{tabular} \\ \hline
NoC                                                  & N/A                                                  & {[}100, ,...,800{]}                                       & {[}4, ..., 256{]}                                           & -                                                               \\ \hline
Mem                                                  & \begin{tabular}[c]{@{}c@{}}DRAM$\leftrightarrows$SRAM\end{tabular} & {[}100,...,800{]}                                       & {[}4, ..., 256{]}                                           & -                                                               \\ \hline
\end{tabular}
}
\caption{Current Swap types and ranges. PE=processing element, GPP=general purpose processors, Acc = accelerator.}
\label{fig:swap_options}
\end{table}

FARSI's exploration heuristic is equipped with architectural reasoning, concretely the capability to apply parallelism, locality, and customization according to the system needs to improve the convergence rate. For example, to improve power/performance, it automatically detects if a task's data is multiple hops away and chooses ``migrate'' to bring the data closer (spatial locality reasoning). Furthermore, to improve performance, it automatically recognizes if a block runs concurrent, parallel tasks and uses ``fork'' or ``migrate'' to relax the pressure/congestion (parallelism reasoning). Alternatively, if the metric to optimize is power, it uses ``join'' to ``serialize'' to improve power or reduce area. This automatic capability to reason about system/targeted metric and applying appropriate optimizations improves the convergence rate and is necessary for complex design spaces of DSSoCs. 

At a high level, to select optimization moves (Algorithm~\ref{alg:opt_move}), FARSI first applies architectural reasoning to find moves that can improve the design (step I), prioritizes them according to their development cost (detailed in next paragraph) (step II), and probabilisticly samples and applies the move (step III). This approach improves the random neighbour generation (or move selection) of SA and increases its convergence rate by more than an order of magnitude (section~\ref{subsec:architectural_awareness}).
Note that said reasonings is not tied to a specific search heuristic, and although this work applies them to simulated annealing, they can improve other heuristics' convergence rate.
Also note that FARSI users, i.e., system designers and architects, can easily extend said reasonings with their architectural insights to improve FARSI's intelligence.

\makeatletter
\renewcommand{\ALG@beginalgorithmic}{}
\makeatother

\begin{algorithm}[t]
    \caption{Optimization move selection.}\label{alg:opt_move}
    \begin{algorithmic}
    \State // Step I: Apply reasoning based on parallelism, locality and customization to find optimization moves that can improve the design.
    \If{$metric$ = "latency"}
        \If{$block$ has parallel tasks}
            \State $optimization\_moves$ $\gets$ ["migrate", "fork"]
        \Else
            \State $optimization\_moves$ $\gets$ ["swap", "fork\_swap"]
        \EndIf
    \EndIf
    \If{$metric$ = "power"}
        \If{task can run in parallel with other blocks' tasks}
            \If{$block$ has no parallel tasks}
                \State $optimization\_moves$ $\gets$ ["migrate"]
            \Else
                \State $optimization\_moves$ $\gets$ ["join"]
            \EndIf
        \Else
            \State $optimization\_moves$ $\gets$ ["swap", "fork\_swap"]
        \EndIf
    \EndIf
    \If{$metric$ = "area"}
        \If{$block$ = "PE"}
            \State $optimization\_moves$ $\gets$ ["join", "swap"]
        \Else
            \State $optimization\_moves$ $\gets$ ["migrate", "join", "swap"]
        \EndIf
    \EndIf
    \State 
    \State // Step II: Prioritize moves based on their development effort.
    \State  move\_precedence: join $>$ migrate $>$ fork $>$ swap $>$ fork\_swap
    \State 
    \State // Step III: Select moves.
    \State $move\_chosen$ $\gets$ probabilistically choose from $optimization\_moves$, based on $move\_precedence$ order
     \State 
    \State PS: fork\_swap is simply a fork followed by swap, and is introduced to accelerate navigation.
    \end{algorithmic}
\end{algorithm}

\textbf{Move Symmetry:}
Our optimization moves are symmetrical to enable backtracking and prevent the navigation from getting stuck. 
For example, we can swap up followed by a swap down (e.g., first widen and then narrow the bus), migrate back and forth, and join-in after forking-out.

\textbf{Development-cost Awareness:}
Since highly customized SoCs are complex and thus financially costly to develop, we embed various policies into our heuristic to keep the development-cost low:
\begin{itemize}
\item We introduce move-precedence to select low-effort moves over the expensive ones. For example, we prioritize join over other moves as it reduces the hardware complexity by eliminating a hardware block. Furthermore, we prioritize software moves (migrate) over the hardware ones (fork, swap) as software manipulation is cheaper. Finally, within hardware moves, we prioritize fork over swap for processing elements as the former only requires duplication whereas the latter involves porting (or hardening), which is more expensive. 
For memory and NoCs, the swap is prioritized if it does not increase the system heterogeneity, for example, creating a system with NoCs of different Frequency. A snipped of our move precedence is shown in  {\tt move\_precedence} of Algorithm~\ref{alg:opt_move}. 

\item FARSI starts with a very simple base design (one general-purpose processor, one NoC, and one DRAM bank) and advances its complexity incrementally only if needed. Since every new allocation or customization is costly (the former introduces new hardware and thus increases the development effort while the latter contributes to the system heterogeneity), we only apply one move at a time and thus incrementally modify the design. Furthermore, our moves are designed to only modify one knob at a time, e.g., migrate only one task. Overall, this approach allows us to increase the complexity in small steps and only if the design has not met the budget, thus keeping the development effort low.

\end{itemize}
We detail the impact of our development-aware policies in a case study (Section~\ref{sec:dev_effort_impact}). 

%


\subsection{Exploration Heuristic: System Selection}
Once we generate and simulate the neighbors (candidate generation in SA), we select the best. Neighbors' fitness is quantified
using their design's (metric) normalized distance to budget
with a dampening factor to the metrics already meeting budget (Equation~\ref{eq:dist}). $m$, $Des_m$, $Bud_m$ and $\alpha$ denote, a metric, design value for the metric, budget value for that metric, and dampening factor when the budget is met. 
\begin{align} \label{eq:dist}
    & distance\ to\ budget = \sum_m \alpha*\frac{(Des_m - Bud_m)}{Bud_m}\\
    & m \in \{ Performance, Power, Area\} \nonumber
\end{align}

If no improved designs are found, we inform the system generation stage for the next iteration to target the task/block with the next highest distance (comparing to the last). Note that similar to SA, we also use a temperature value to choose sub-optimal designs infrequently.






%% file: tex/evaluation.tex
\section{FARSI Validation}
This section sheds light on the performance breakdown of the FARSI components and further quantifies our simulation fidelity in terms of modeling accuracy.
In this study, we exploit our augmented simulated annealer to generate over 250 SoCs for our three workloads, namely, Audio Decoder, CAVA, and Edge Detection. These SoCs have different topology complexities ranging from 1 to 13 processing elements, 1 to 8 memory blocks, and 1 to 3 NoCs. All data are collected on a Xeon Skylake Machine with 2.00 GHz frequency, and PA's time interval parameter is set to 10 µs (corresponding to 1000-10000 cycles of our hardware blocks) to enable high accuracy. 


\begin{figure}[!t]
\vspace{-10pt}
    \centering
    \captionsetup{type=table}
    \subfloat[Workload's budget (5nm target technology).]{
        \includegraphics[trim=2 0 2 1, clip, width=.32\linewidth]{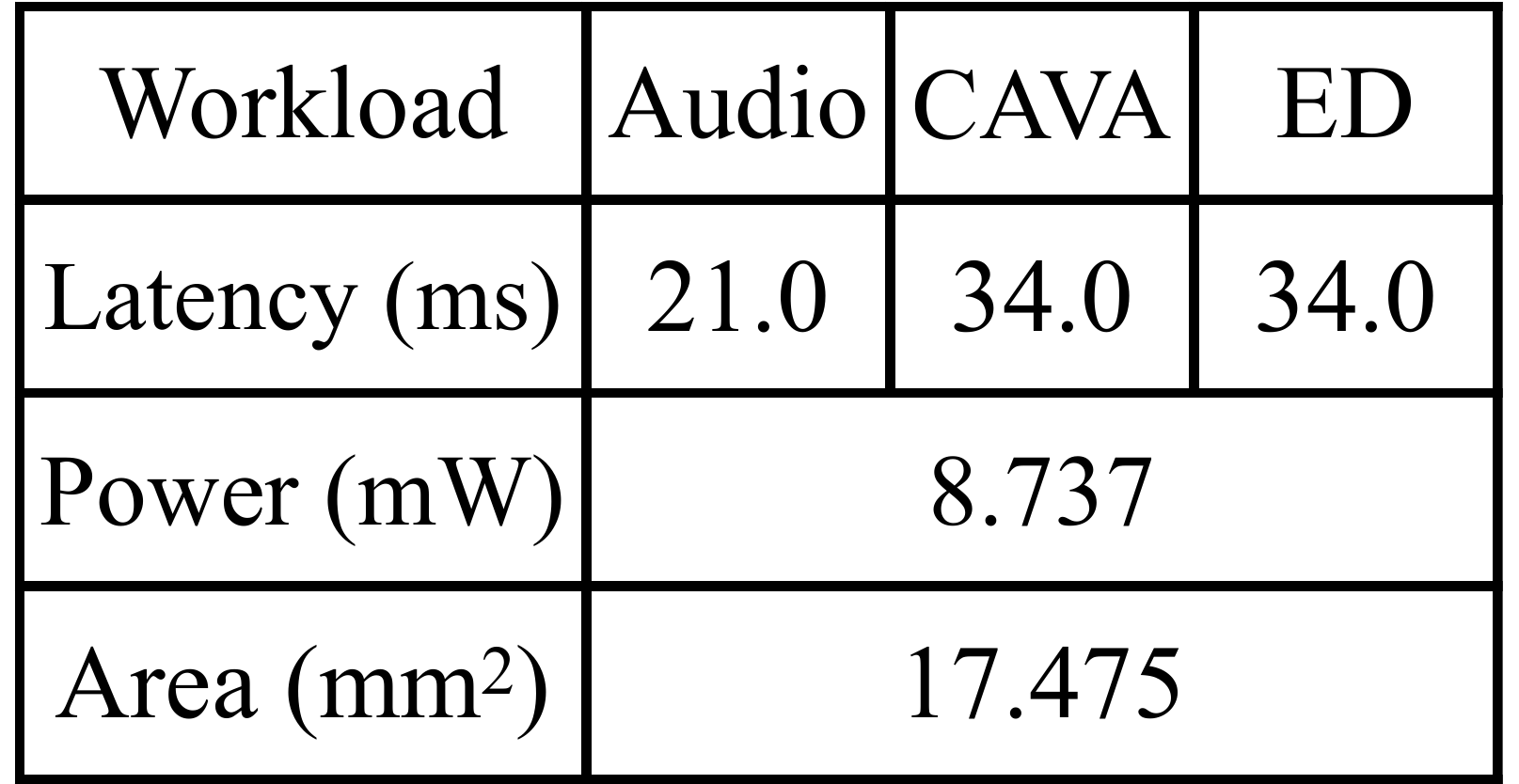}
        \label{table:budget}
    }
    \hspace{40pt}
    \subfloat[Validation results.]{
        \includegraphics[trim=2 0 2 1, clip, width=.32\linewidth]{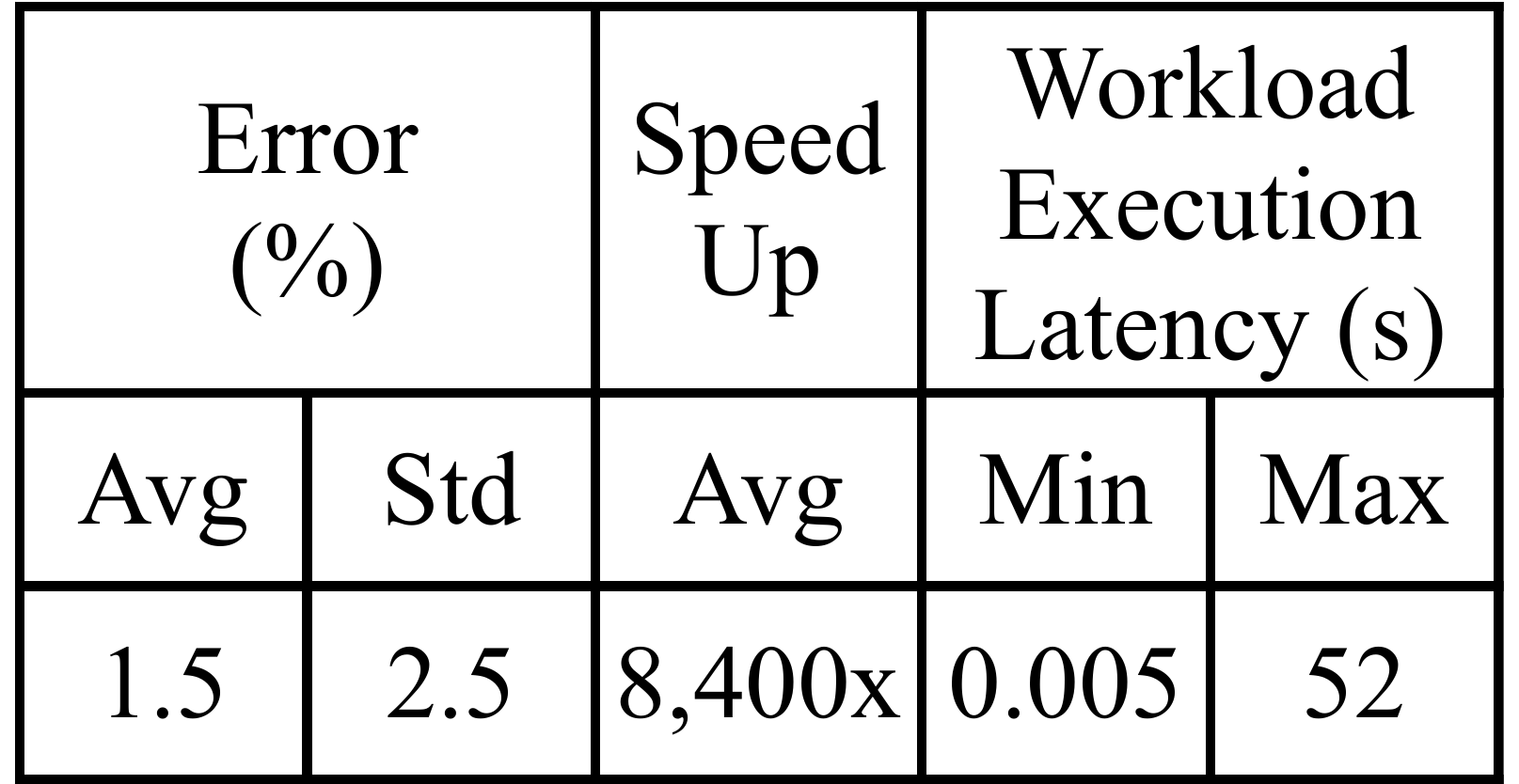}
        \label{table:overall_validation}
    }
    \caption{Workload's budget and validation results.}
    \label{table:budget_validation}
\end{figure}

\begin{figure}[!t]
\vspace{-10pt}
\centering
      \subfloat[Block type impact on error.]{
        \includegraphics[trim=0 0 0 0, clip, width=.32\linewidth]{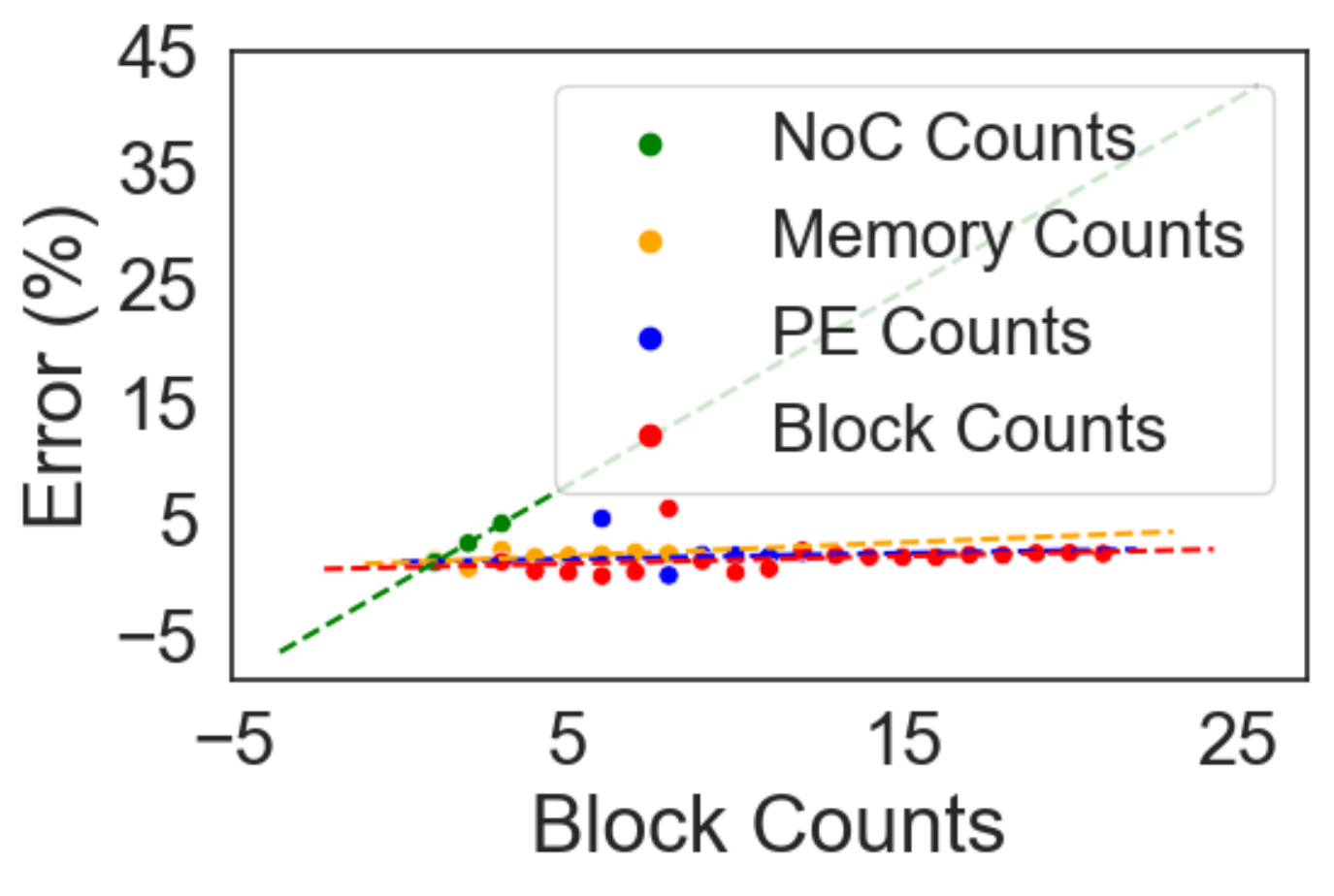}
        \label{fig:blck_cnt_vs_error}
    }
    \subfloat[Block count vs scalability.]{
        \includegraphics[trim=0 0 0 0, clip, width=.34\linewidth]{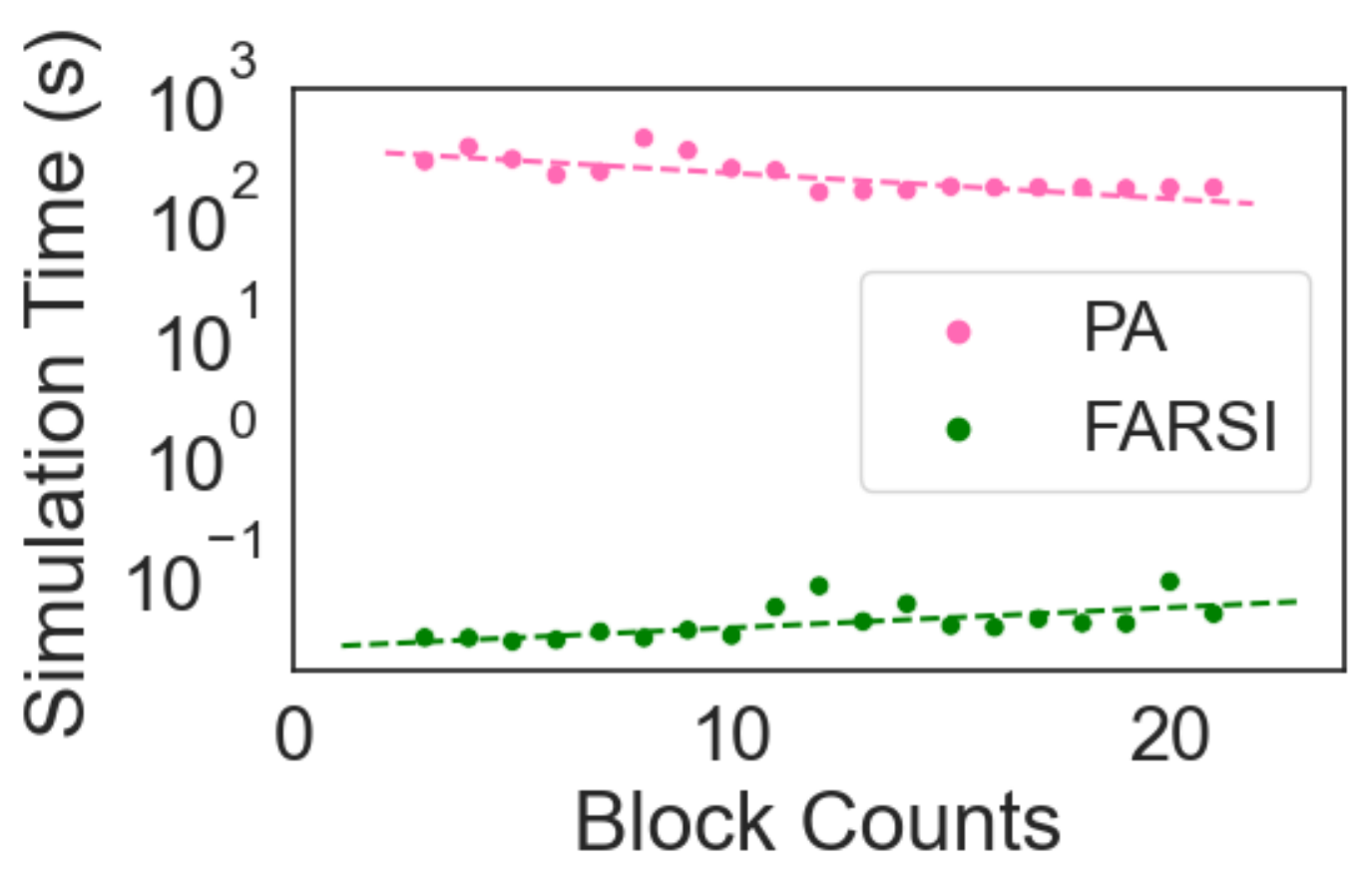}
        \label{fig:blck_cnt_vs_sim_time}
    }
    \subfloat[Wrokload's latency vs  scalability.]{
        \includegraphics[trim=0 0 0 0, clip, width=.33\linewidth]{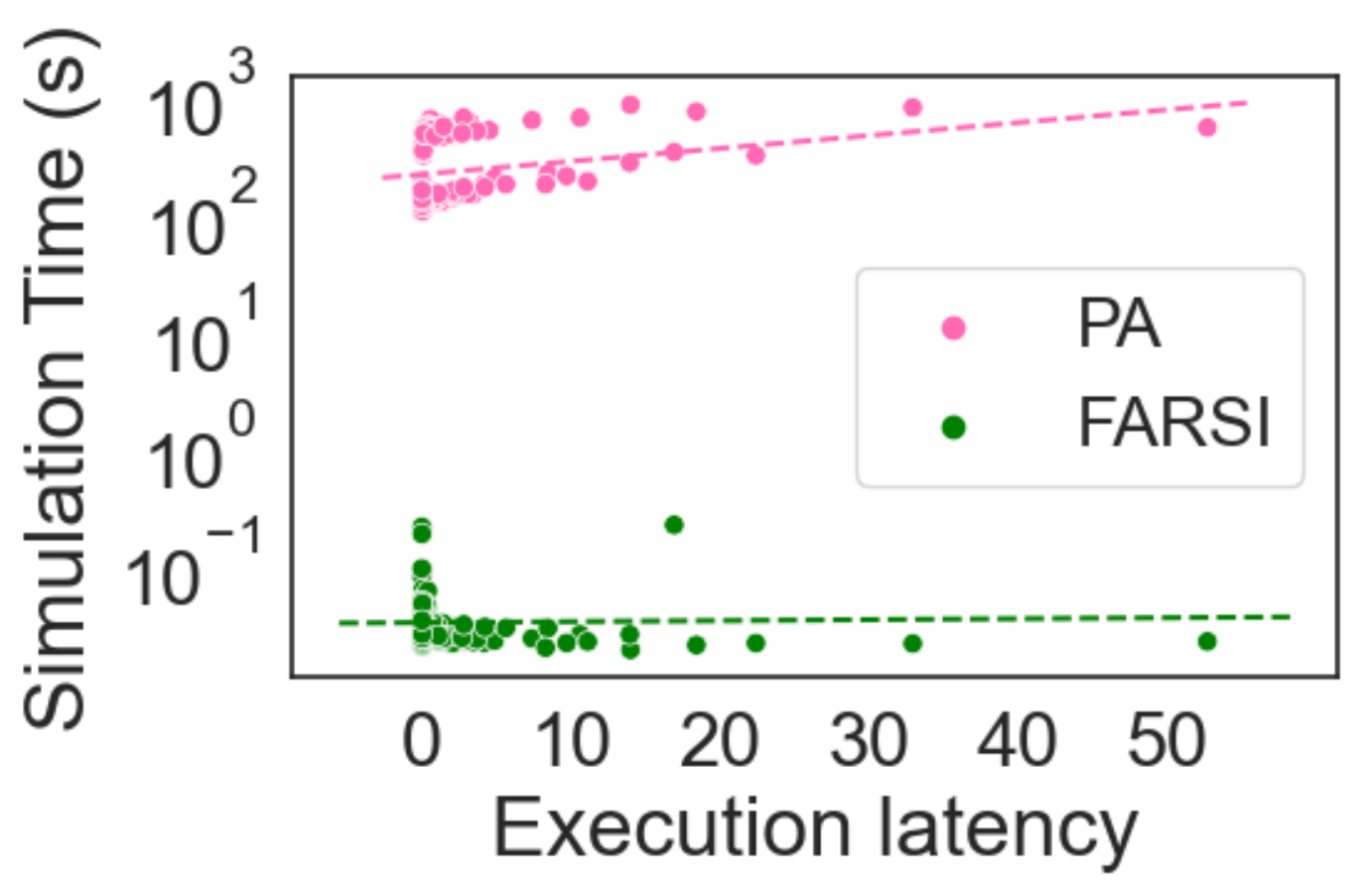}
        \label{fig:latency_vs_scalibilty}
    }
\caption{Simulation Scalibility. Fidelity scalability (left), and time scalibility (right two).}
\label{fig:farsiVSpa}
\end{figure}

\textbf{Simulation Validation:}
We validate our simulation framework against Synopsys Platform Architect (PA)~\cite{PA}, a widely used industry-grade performance simulator. Similar to FARSI, PA's approximately timed mode targets agile/early-stage system-level estimation. PA deploys event triggers (i.e., transactions and fixed time intervals) to advance the simulation. 
We achieve an accuracy of 98.5\%  (Table \ref{table:overall_validation}). 
We measure our simulator's fidelity sensitivity with respect to the types and quantity of hardware blocks within said designs (Figure~\ref{fig:blck_cnt_vs_error}). Processing elements (PEs) show the lowest sensitivity while buses show the highest. This is because we do not model intermittent congestion that can occur within a sequence of (NoCs') pipes and rather assume constant congestion for a phase.

\textbf{Simulation Speedup:}
We achieve an average speedup of
8,400X (compared to Synopsys PA), owing to our hybrid estimation methodology that combines analytical models and flexible phase-based simulation. We also measure our simulator's slowdown with respect to the number of system blocks (Figure~\ref{fig:blck_cnt_vs_sim_time}). We incur a small 3X slowdown for a 20X increase in the number of blocks. For Synopsys PA, we do not see a meaningful relationship as the simulation time sometimes lowers and then plateaus with a higher number of blocks.
However, PA shows a high sensitivity to the workload's execution latency (Figure~\ref{fig:latency_vs_scalibilty}) where an increase of .005(s) to 50(s) for this latency raises the simulation time from 101(s) to 814(s). FARSI also experiences a slow down of .018 (s) to .21(s) for the same increase, still maintaining high agility.

\begin{figure}[!t]
\centering
    \subfloat[FARSI system breakdown.]{
        \includegraphics[trim=0 0 0 0, clip, width=.33\linewidth]{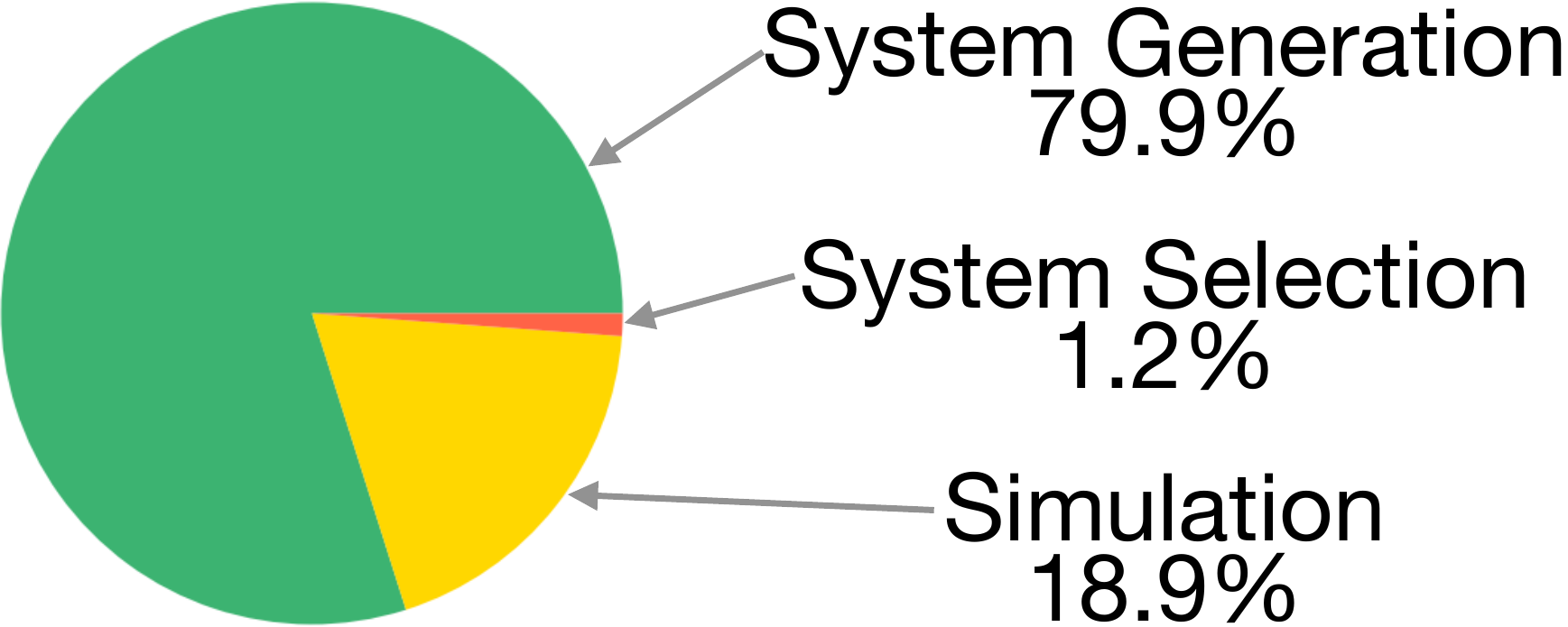}
        \label{fig:time_break_down:FA_components_time_breakdown}
    }
    \subfloat[System generation details.]{
        \includegraphics[trim=0 0 0 0, clip, width=.33\linewidth]{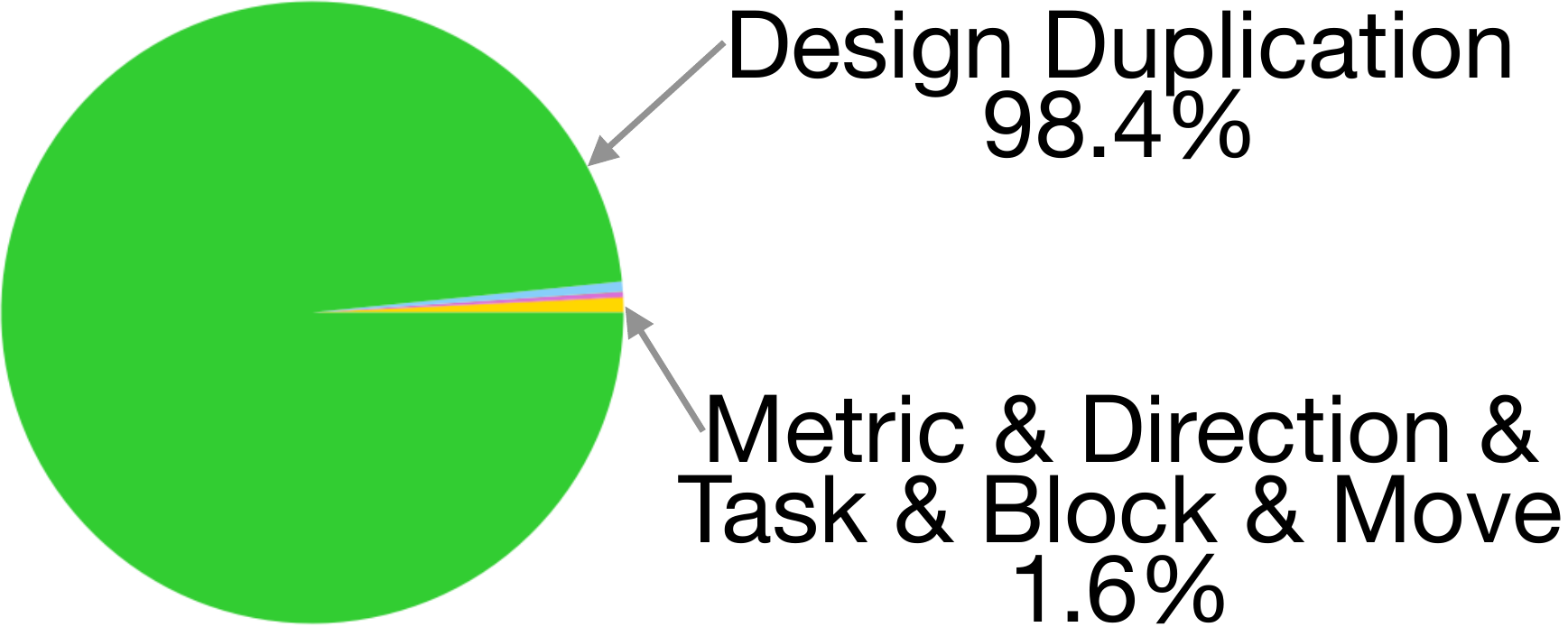}
        \label{fig:time_break_down:trans_gen_time_breakdown}
    }
\caption{FARSI time breakdown.}
\label{fig:time_break_down_all}
\end{figure}

\textbf{Performance Breakdown:}
We profile FARSI's last three iterative stages to show what needs to be sped up. Note that the first stage (Database Generation) occurs only once and further does not change across different runs. 
Figure~\ref{fig:time_break_down:FA_components_time_breakdown} shows that most of the time (79.9\%) is spent on the system generation stage. Further breakdown of this stage (Figure~\ref{fig:time_break_down:trans_gen_time_breakdown}) reveals that sub-stages such as task, block, or metric selection that do actual system modifications only consume a small percentage (2\%<). Instead, design duplication, which copies the original design object to be modified/improved, consumes the most time. This step can be sped up by lowering the memory footprint for a faster \texttt{memcpy}. 


\section{FARSI's Exploration Heuristic Analysis}
\label{sec:sys_gen_analysis}
 
As FARSI manages the design space complexity by using (1) agile system-level simulation, (2) architecture awareness, (3) heavy use of co-design, and (4) domain awareness, we quantify the impact of each and quantify their efficacy. 


For all four studies,  we lower the target SoC power and area budgets of 0.1 W, and 100 $mm^2$ ~\cite{takahashi2018} according to~\cite{huzaifa2021illixr}. Audio power budget was estimated based on the system power breakdown of~\cite{huzaifa2021illixr}, and CAVA and Edge Detection power budgets were estimated according to their relative power consumption to Audio. System power is then calculated as the sum of all workload components. 
For area budget, we use power as a proxy and similarly use the breakdown in~\cite{huzaifa2021illixr} to guide the budgeting.
These configurations are shown in Table~\ref{table:budget}.\footnote{Note that these values are specified for the target technology of 5nm TSMC.} For the first three studies, we set the individual workload's latencies to the values provided in~\cite{huzaifa2021illixr} and \cite{zedmini}. For the last study, these latencies are intentionally lowered to the workload's limits. This allows us to stress test FARSI toward exploring all possible optimization opportunities (e.g., TaLP and LLP).





\subsection{Simulation Agility's Impact on Convergence}
Here we quantify the impact of FARSI's agile simulator on the convergence time and fast coverage of the design space. Note that convergence time is the time that it takes for the DSE to meet the design budgets. 
We envision a SoC that runs all three workloads together (as they almost always are in the AR use case). We compare two DSEs that use the same exact heuristic, but one uses FARSI's simulator and the other uses PA. Note that for the latter, we estimate the convergence time by increasing the time of each iteration by the average slow-down mentioned in  Table~\ref{table:overall_validation}. This is because, as we will show, actually running PA takes too long and thus is infeasible.

\begin{figure}[!t]
    \centering
        \subfloat[FARSI vs PA.]{
        \vspace{-0.1cm}
        \includegraphics[trim=6 6 6 6, clip, width=0.28063\linewidth]{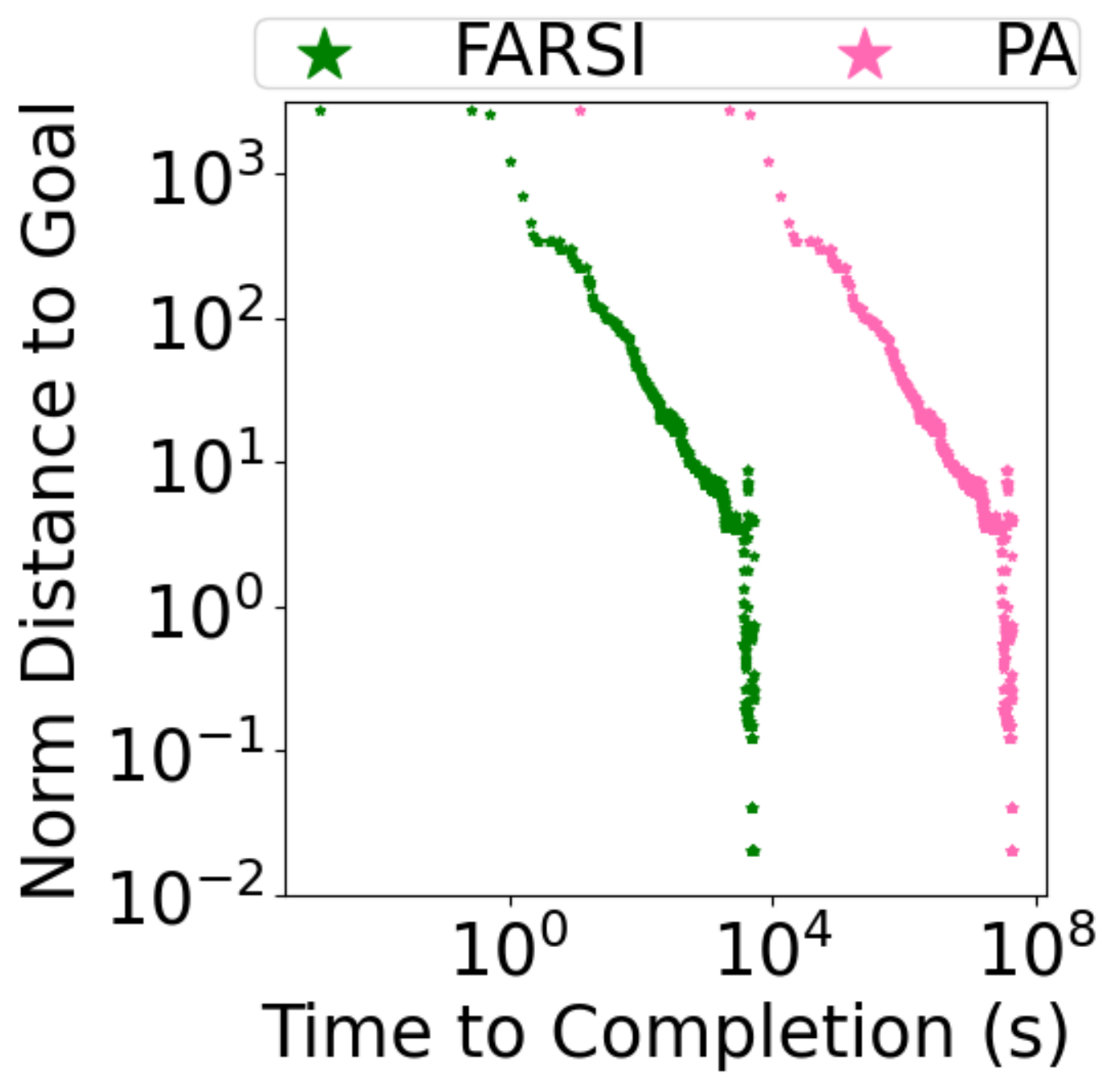}
        \label{fig:sim_agile:FARSI_vs_PA}
    }
    \hspace{0.35cm}
    \subfloat[FARSI vs simulated annealing.]{
        \includegraphics[trim=2 0 0 4, clip, width=0.3839\linewidth]{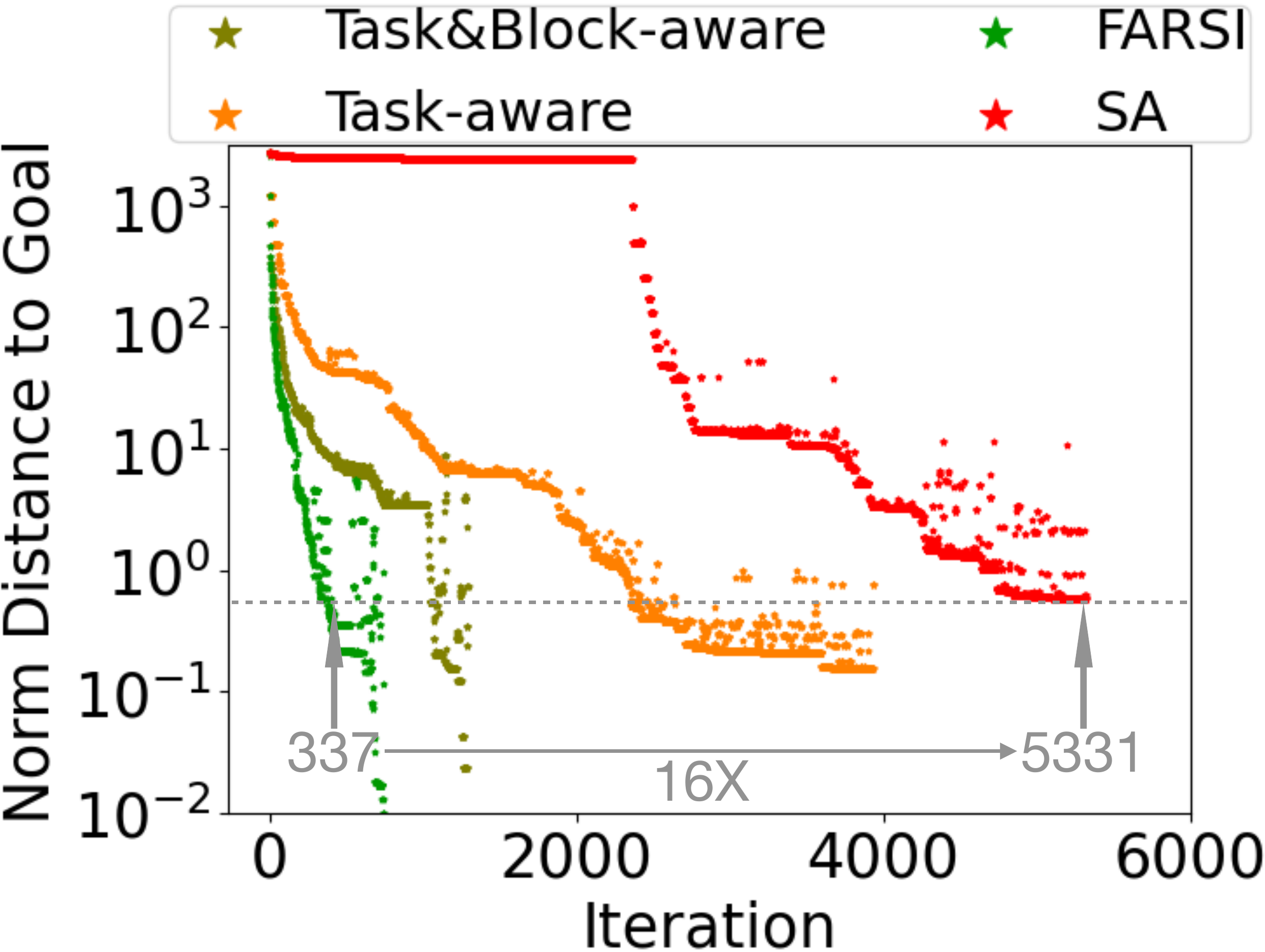}
        \label{fig:arch_aware:convergence}
    }
\vspace{-10pt}
\caption{FARSI's agility. Convergence comparison to PA simulator (left, log-log scale), and to naive simulated annealing DSEs (right, linear-log scale).}
\label{fig:arch_aware:FARSI_agility}
\end{figure}

Figure~\ref{fig:sim_agile:FARSI_vs_PA} shows the difference in the convergence time with y-axis and x-axis, showing the distance \footnote{Normalized city block distance to budget across all metrics.} to the goal or budget (Equation~\ref{eq:dist}) and time to convergence, respectively. As shown, FARSI takes 3 hours to converge by exploring more than 750 designs, and PA takes more than 3 years.
Note that by the time FARSI has converged, PA has only looked at 4 designs. FARSI's simulation agility makes it an ideal candidate for a DSSoC DSE as it enables high coverage of the complex/big design spaces.

\subsection{Architecture Awareness}
\label{subsec:architectural_awareness}
Given the number of knobs across allocation, mapping, and tuning, the design space is too great to be navigated blindly. For example, a system with 6 tasks and 2 knobs per processor/memory/NoC results in more than a million design points, and our AR complex with more than 28 tasks results in a space greater than the number of stars in the universe. Thus, optimal navigation demands architectural insights for guidance.



To highlight the importance of architecture-awareness, we compare FARSI against a simple simulated annealer (SA) that generates neighbors at random. We further vary the awareness level and show its impact on convergence. From least to most aware, SA uses random neighbor generation, Task-aware uses bottleneck analysis to select the task for such generation, Task\&Block-aware uses architectural reasoning to select the task and hardware block, and finally, FARSI uses architectural insights to select a task, a block, and an optimization move.

Figure~\ref{fig:arch_aware:convergence} shows the convergence rate of each DSE with the x-axis and y-axis denoting the iteration count and the city block distance to budget or goal (Equation~\ref{eq:dist}). As seen, our two least aware DSEs, i.e., SA and Task-aware, do not fully converge, as they get stuck without finding an improvement. These DSEs are, on average, about 16X and 6.5X slower than FARSI to reach the same distance. 
Task\&Block-aware adds block selection awareness and thus lowers the convergence difference to 2X. We believe these differences only worsen as workload set/complexity grow and thus, the architecture community insights must be embedded into the design automation community  heuristics.

\begin{figure}[!t]
\centering
 \subfloat[Co-design over time across metrics and workloads.]{
       \vspace{-6pt}
        \includegraphics[trim=0 0 0 0, clip, width=0.65\linewidth]{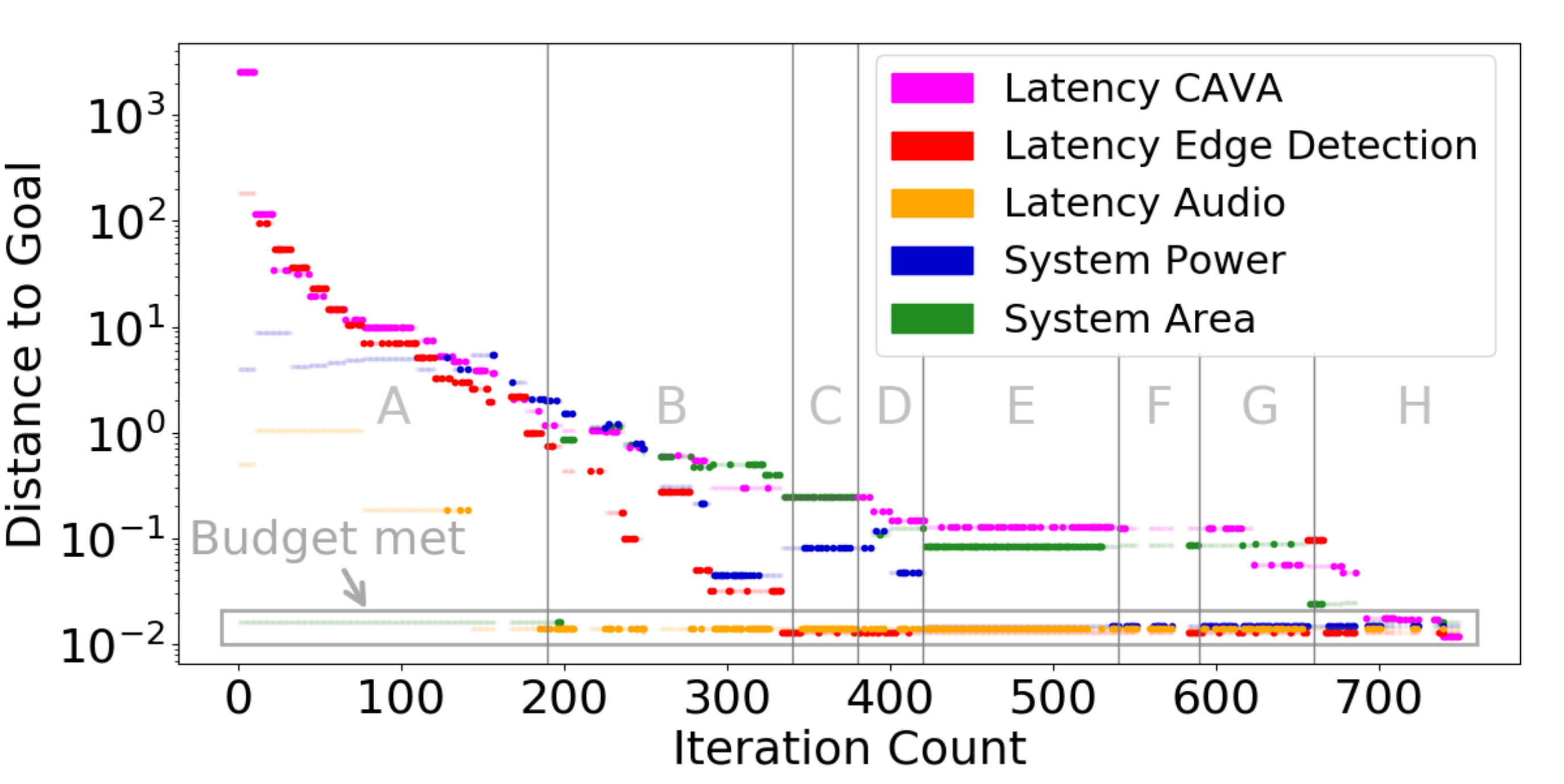}
       \label{fig:co-design-prog}
    }
  
  \vspace{6pt}
    \subfloat[Co-design deployment rate.]{
        \includegraphics[trim=0 0 0 0, clip, width=.3\linewidth]{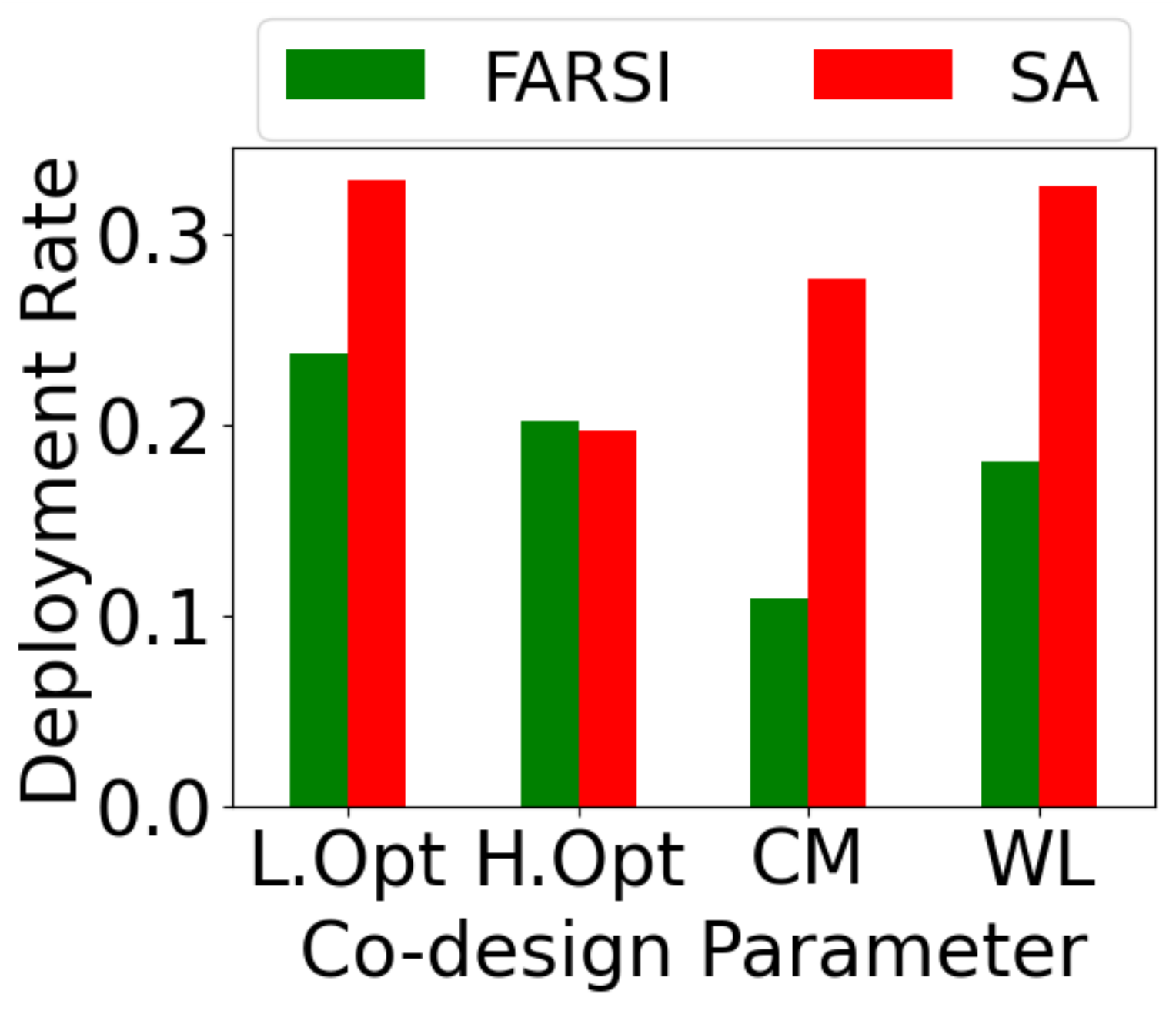}
        \label{fig:co_des:rate:cnt}
    }
   \hspace{5pt}
    \subfloat[Co-design resulted convergence rate.]{
        \includegraphics[trim=0 0 0 0, clip, width=.3\linewidth]{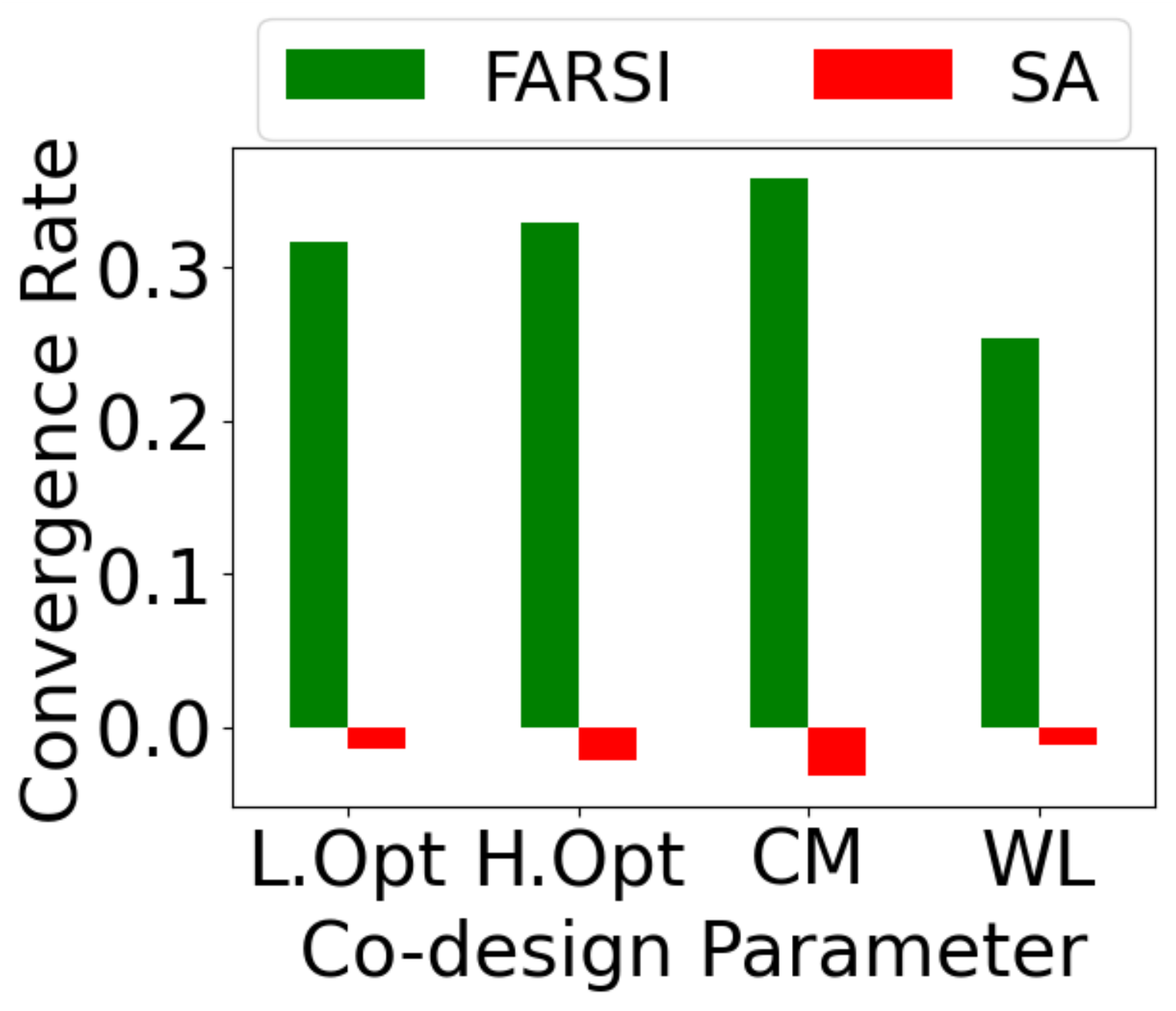}
        \label{fig:co_des:rate:efficacy}
    }

\caption{Co-design impact and rate. H.Opt=high-level optimization, L.Opt=low-level optimization, CM=communication\_computation, WL=workload.}
\label{fig:co_des_rate}
\end{figure}

\subsection{Co-design}
\label{subsec:co_design}
A DSE without a co-design cannot exploit cross-boundary optimization opportunities necessary for convergence. For example, without a cross-workload co-design, a DSE misses on area reduction enabled by inter workload memory sharing. FARSI uses co-design by not being fixated on one optimization for too long and rather continuously moving from one optimization to another. Our co-design occurs within each of the following vectors:
(1) across metrics (i.e., among power, performance, area), (2) across high-level optimization (i.e., among mapping/allocation/customization) and across low-level optimizations (frequency modulation, bandwidth modulation, hardening, etc.), (3) across computation/communication, and (4) across workloads. Here we  quantify their impact.

\begin{figure}[!t]
\centering
\includegraphics[trim=0 0 0 0, clip, width=.6\linewidth]{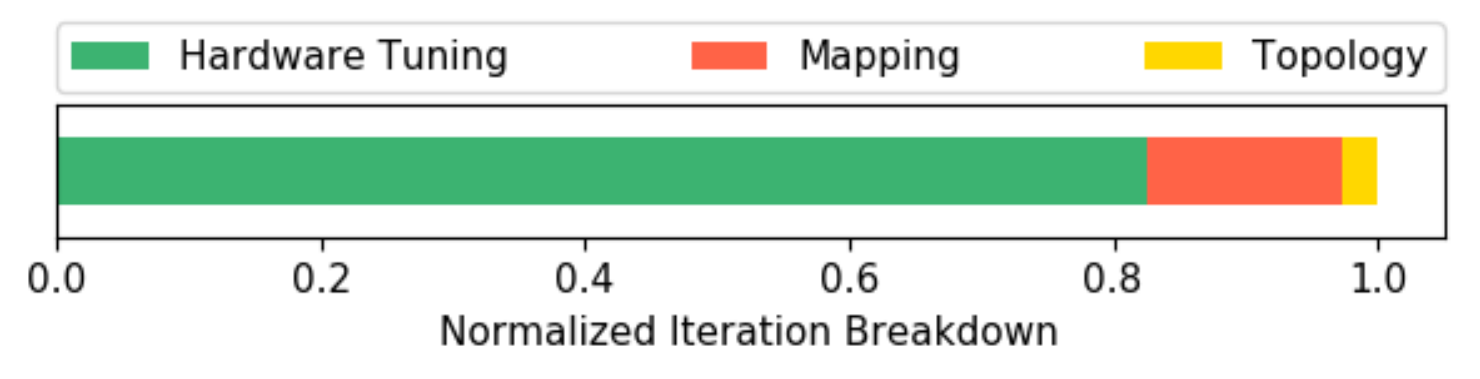}
\includegraphics[trim=0 0 0 0, clip, width=.6\linewidth]{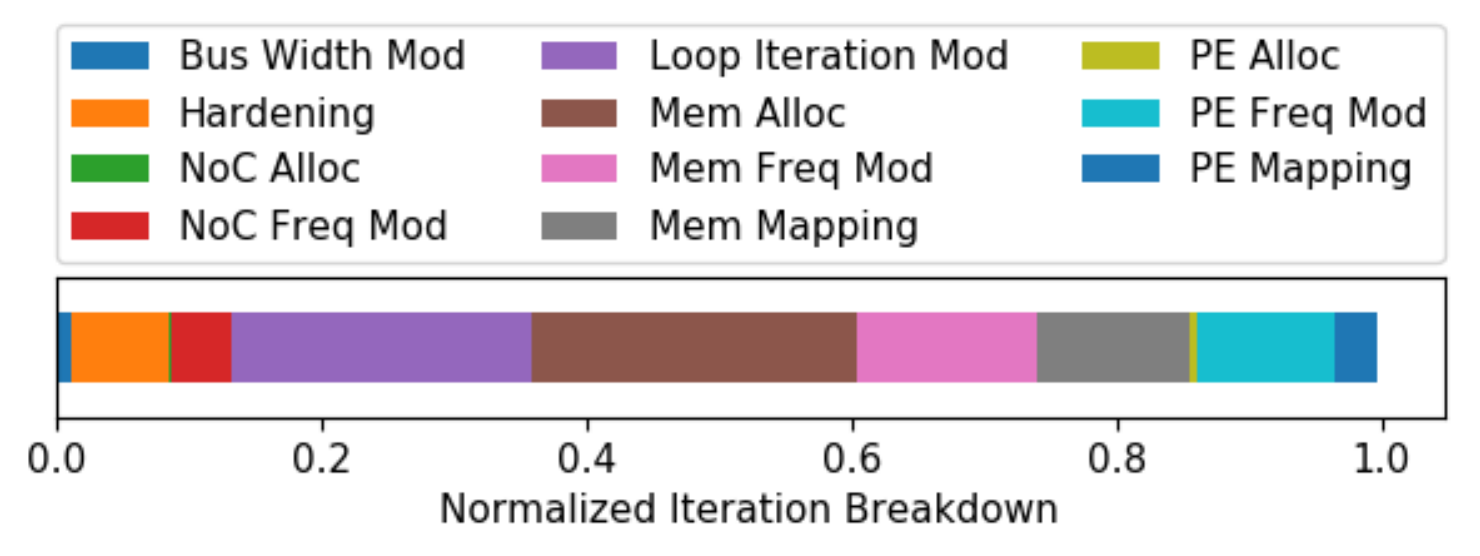}
\vspace{-10pt}
\caption{FARSI's use of high- (top) and low-level (bottom) optimizations. Alloc = Allocation, Mod = Modulation, \mbox{Freq = Frequency}, PE = Processing Element, Mem = Memory.}
\label{fig:exactOptNameDivNeighbour}
\end{figure}

Figure~\ref{fig:co-design-prog} as an example shows FARSI's co-design progression across two of these vectors, i.e., metrics and workloads.
X-axis and y-axis show iteration count and each metric's distance to the goal (log-scaled), respectively.
To emphasize when FARSI targets a metric/workload, we make the color of its curve bold; otherwise, make it see-through.
When the budget is met (distance=0), in the plot, we set it to $10^{-2}$ (as 0 cannot be shown with log).  As shown in the figure, the exploration goes through various zones (A through H), each with a different exploration focus. For example, zone A mainly targets latency and exploits cross-workload opportunities between Edge Detection and CAVA. Zone C shifts its focus toward system power and area co-design. Note that Audio latency is also targeted (although already met) as it can be increased to help with power/area. Finally, FARSI splits its attention between latency (of Audio and CAVA) and system area in zone E. Note that different metrics meet their budget (reach $10^{-2}$) throughout the exploration at different times, but a continuous trade-off between them is exploited until all distances are zero.
Such co-design exploitation is a significant contributor to the fast convergence previously shown in Figure~\ref{fig:arch_aware:convergence}.
Note that Figure~\ref{fig:exactOptNameDivNeighbour} shows the normalized iteration breakdown of our other co-design vector, namely high- and low-level optimizations. However, their co-design progression plots are left out due to space limitations.

Figure~\ref{fig:co_des:rate:cnt} shows how often (deployment rate) co-design occurs within each of our vectors. For example, a co-design rate of .2 for workloads means that FARSI changed its focus from one workload to another 2 times every 10 iterations on average.
Note that these rates are not tuned by the tool user but determined dynamically by FARSI as it explores the space. For example, if FARSI sees that one workload demands more attention (i.e., its distance to the PPA budget is high), it will select and focus on it long before switching to others.
FARSI applies co-design across both high-level (mapping, allocation, software to hardware mapping) and low-level (frequency modulation, memory allocation, etc.) optimization options the most. Concretely, it changes its focus between these options around every 4-5 iterations. Workloads co-design rate is lower, indicating a difference in each workload's convergence difficulty. Finally, FARSI applies even a longer attention length (>10 iterations) for computation and communication, indicating their higher imbalance. 
Note that SA deploys a higher co-design rate for all vectors as it randomly selects neighbors; however, next, we show that this strategy is not optimal.

To quantify the impact of the said co-design approaches, we measure the average convergence rate (i.e., average of improved distance per iteration)
resulted from each.
Figure~\ref{fig:co_des:rate:efficacy} reveals that computation/communication co-design leads to the highest improvement, i.e., an average convergence rate of 35\%.
Overall on average, co-design improves the distance by 32\%.
Note that the SA convergence rate due to co-design is very small, in fact, negative as random switching between, for example workloads can hurt the distance. This shows that a principled co-design is necessary.

\subsection{Domain Awareness}
\label{subsec:domain_awareness}
DSEs that target domain-specific SoCs need to be domain aware, i.e., 1) extract workloads' characteristics and then 2) direct their optimizations to exploit workload's inherent opportunities (e.g., TaLP). Here, we first detail our workload's characteristics and then showcase FARSI's awareness of them. 

\begin{figure}[!t]
    \centering
    \includegraphics[trim=2 2 2 2, clip, width=.68\linewidth]{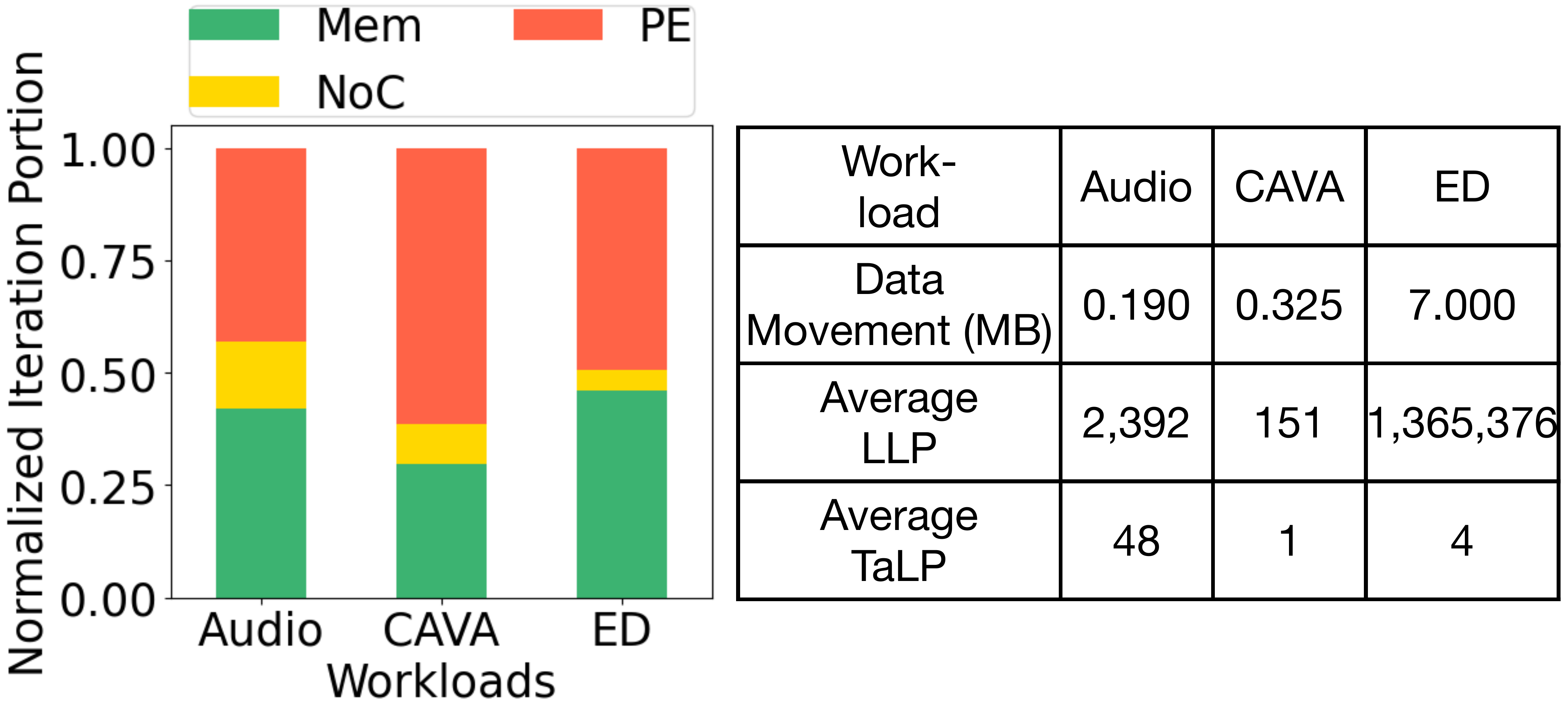}
    \caption{Communication(comm)/computation(comp) boundness (left) and Task (TaLP) and loop-level (LLP) parallelism (right), of each workload. ED shorten for Edge Detection.
    }
    \label{fig:workload_awareness}
\end{figure}

\begin{figure}[!t]
\centering
    \subfloat[Parallelism sensitivity.]{
        \includegraphics[trim=6 6 6 6, clip, width=.36\linewidth]{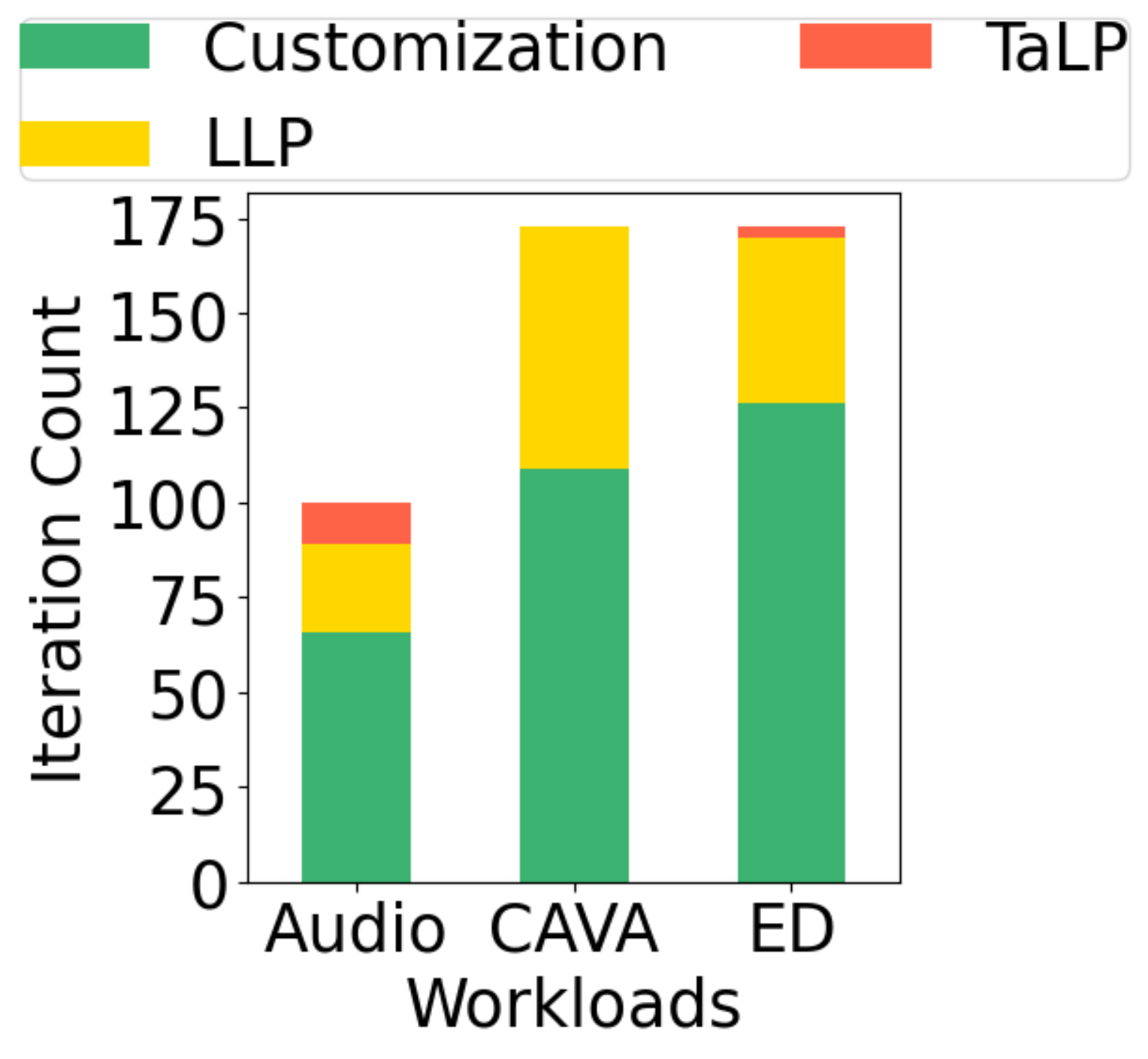}
        \label{fig:workload_aware:parallelism}
    }
    \hspace{20pt}
    \subfloat[Communication(comm) / computation(comp) boundness sensitivity.]{
        \includegraphics[trim=6 6 6 6, clip, width=.33\linewidth]{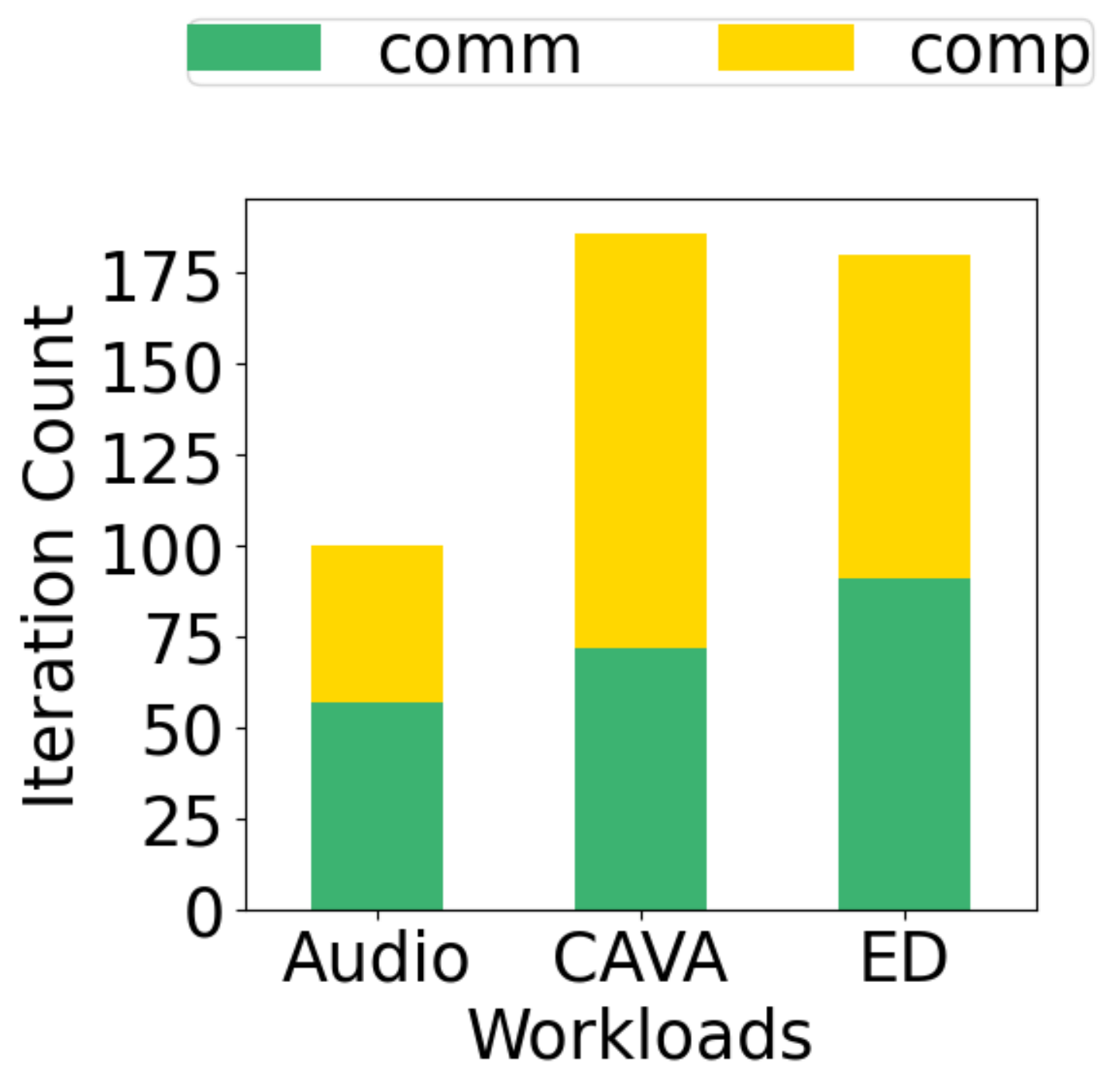}
        \label{fig:workload_aware:boundedness}
    }
    \vspace{-5pt}
    \caption{
    FARSI's domain awareness. FARSI exploits workload characteristics such as TaLP, LLP and computation/communication boundness for convergence.}
    \label{fig:workload_awareness_2}
\end{figure}

We run each workload individually and provide FARSI's response to them. Figure~\ref{fig:workload_awareness} (table to the right) sheds light on each workload's computation and communication characteristics, concretely, parallelism (loop- and task-level), and data movement. Edge Detection (ED) has the highest LLP and data movement, and Audio has the highest TaLP.
Figure~\ref{fig:workload_awareness} (left) further details the computation/communication boundedness of each workload. The y-axis shows the (normalized) breakdown of the hardware bottlenecks FARSI encounters during exploration. This information should guide the DSE in selecting their target stage (i.e., communication or computation). CAVA has the most computation boundness tendency
while the other two show more balanced boundedness.

Figure~\ref{fig:workload_awareness_2}
 illustrates FARSI's response to the above characteristics.
 Concretely, Figure~\ref{fig:workload_aware:parallelism} shows FARSI's uses of parallelism (in the number of iterations). FARSI applies task-level parallelism on Audio the most and CAVA the least, as Audio's TDG provides the highest TaLP opportunities and CAVA provides none. FARSI targets loop-level parallelism for Edge Detection more than Audio correlated with their LLP. Note that FARSI's excessive use of LLP for CAVA is inevitable as it has no other parallelism option. 
Moreover, Figure~\ref{fig:workload_aware:boundedness} shows FARSI's relative focus on computation and communication as a response to these bottlenecks.
CAVA has a high computation boundness tendency, and thus, FARSI targets its computation the most while a more balanced approach is used for the other two according to their needs. FARSI's focus on communication is correlated with each workload's data movement, with Edge Detection having the highest (7 MB on average) and Audio having the least (0.18 MB on average) data movement.

%% file: tex/case_study.tex
\section{Case Studies}
To show FARSI's capability with the design challenges of complex SoCs, we provide two case studies.
First, we show how FARSI's cost-aware methodology balances ``system complexity'' (and thus the development effort) with the product quality.  Then, we shed light on the inefficiencies of the ``divide and conquer'' approach often used to tame the complexity and show that FARSI can provide more optimal designs. For both case studies, we assume a system that runs all three workloads simultaneously.


\subsection{Budget Implications on Development Effort}
\label{sec:dev_effort_impact} 
Since a DSSoC requires a high-development effort, an optimal DSE framework must use various trade-offs to lower this design effort. An example of such trade-off is the product quality and development effort. Concretely, an optimal DSE must lower the development effort whenever a product quality is relaxed. This study showcases FARSI's capability in this balancing act. To this end, we proxy the product quality with performance/power/area (PPA) budgets. This is because for example in AR, performance budget, e.g., frame-rate, directly impacts user experience as it determines how smooth user-world interactions are; power budget impacts battery life and thus determines whether AR glasses can be reasonably operational; and finally area impacts the die cost and thus the overall product cost. We also proxy development effort by the number of hardware blocks in the system and their variations (heterogeneity) as an increase in either magnifies the development effort.
We relax Section~\ref{sec:sys_gen_analysis} PPA budgets (shown in Table~\ref{table:budget}) with 1X,2X, and 4X, run FARSI for each budget and 
illustrate how FARSI lowers hardware block counts and variations when faced with larger budgets (e.g., 4X budget).

\begin{figure}[t!]
\centering
    \subfloat[IP count.]{
        \includegraphics[trim=6.5 6.5 6.5 6.5, clip, width=.35\linewidth]{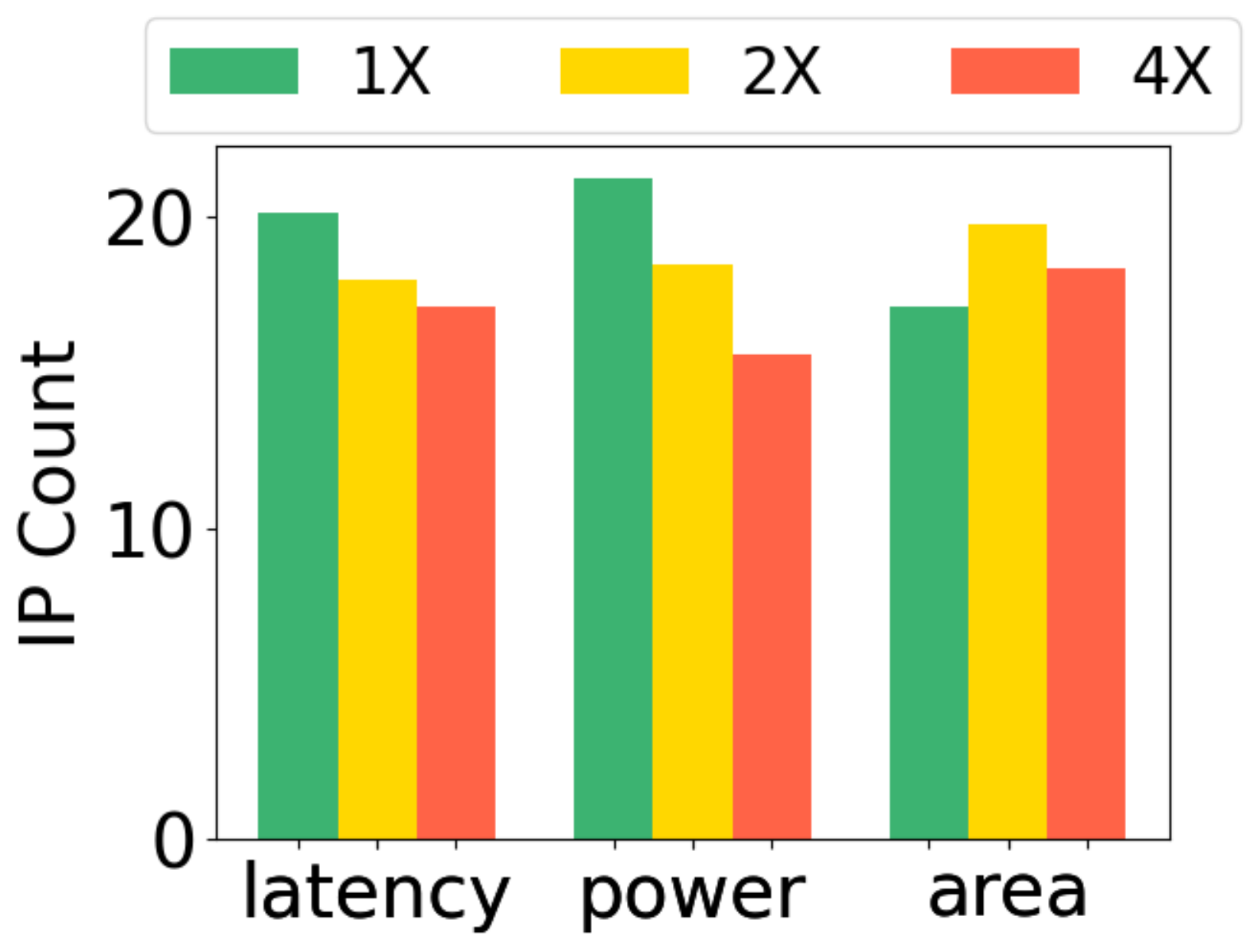}
        \label{fig:case_study:ip_cnt}
    }
    \subfloat[NoC count.]{
        \includegraphics[trim=6.5 6.5 6.5 6.5, clip, width=.345\linewidth]{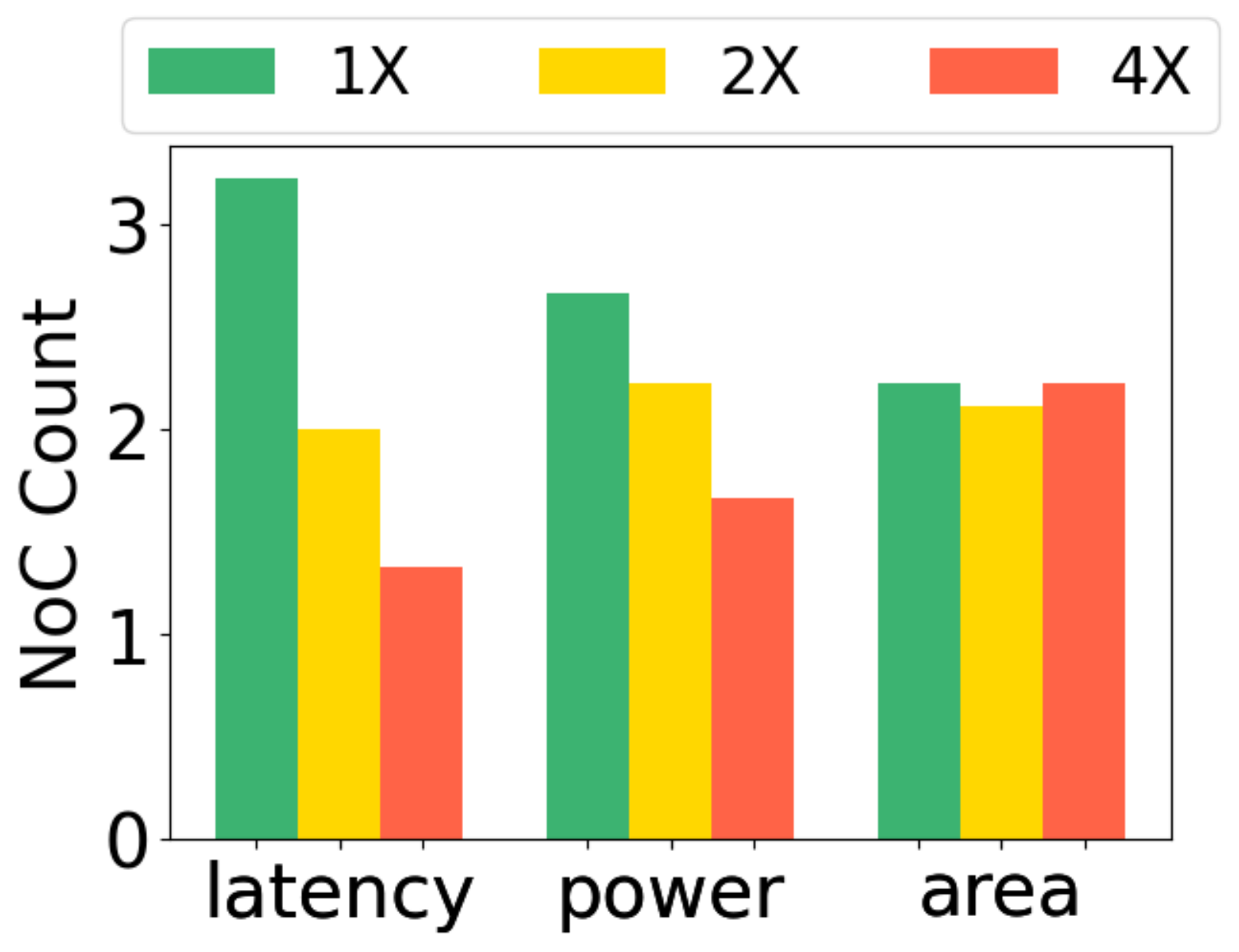}
        \label{fig:case_study:noc_cnt}
    }
    
    \subfloat[Scratchpad memory size.]{
        \includegraphics[trim=6.5 6.5 6.5 6.5, clip, width=.34\linewidth]{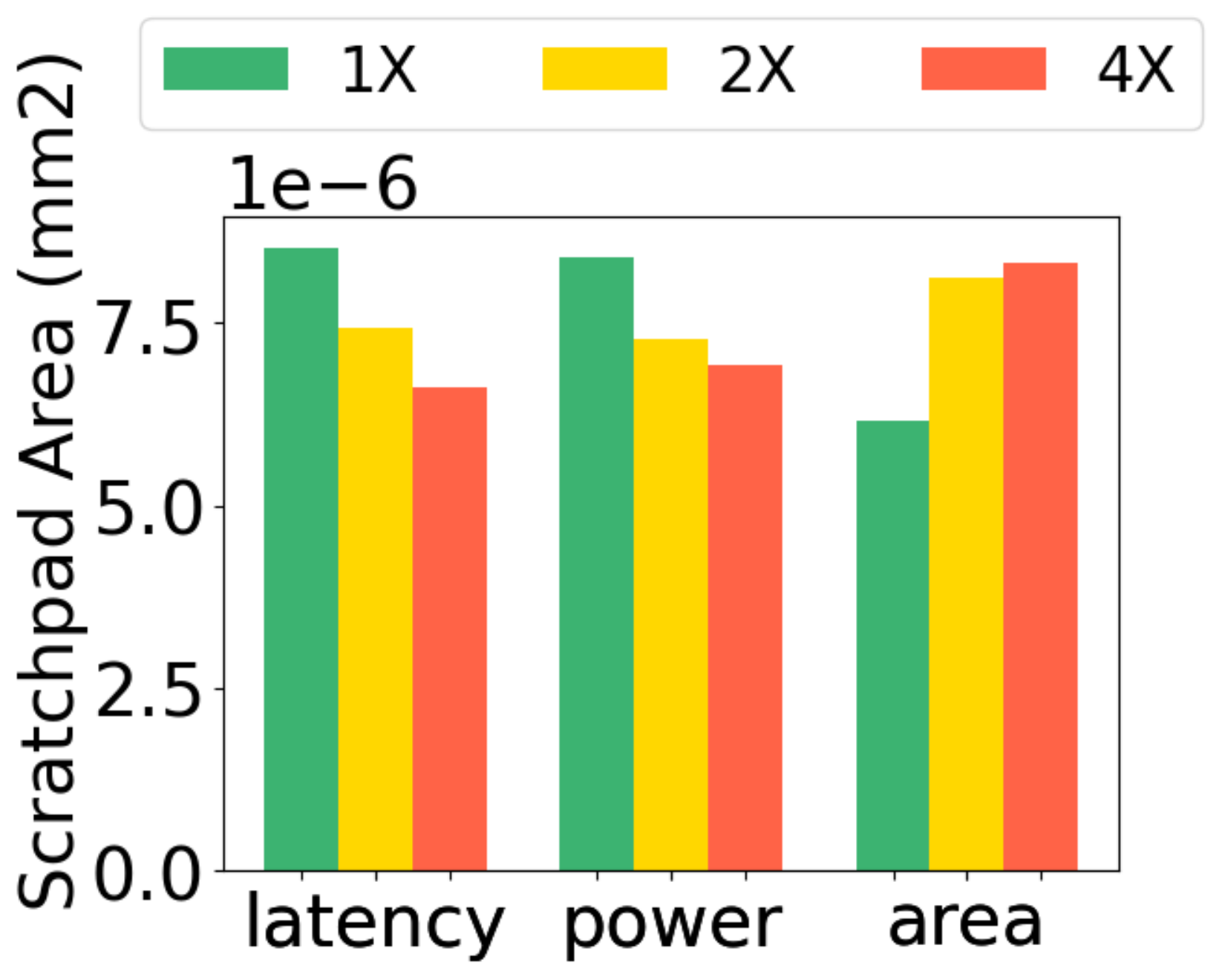}
        \label{fig:case_study:memory_size}
    }
    \subfloat[NoC average frequency.]{
        \includegraphics[trim=6.5 6.5 6.5 6.5, clip, width=.34\linewidth]{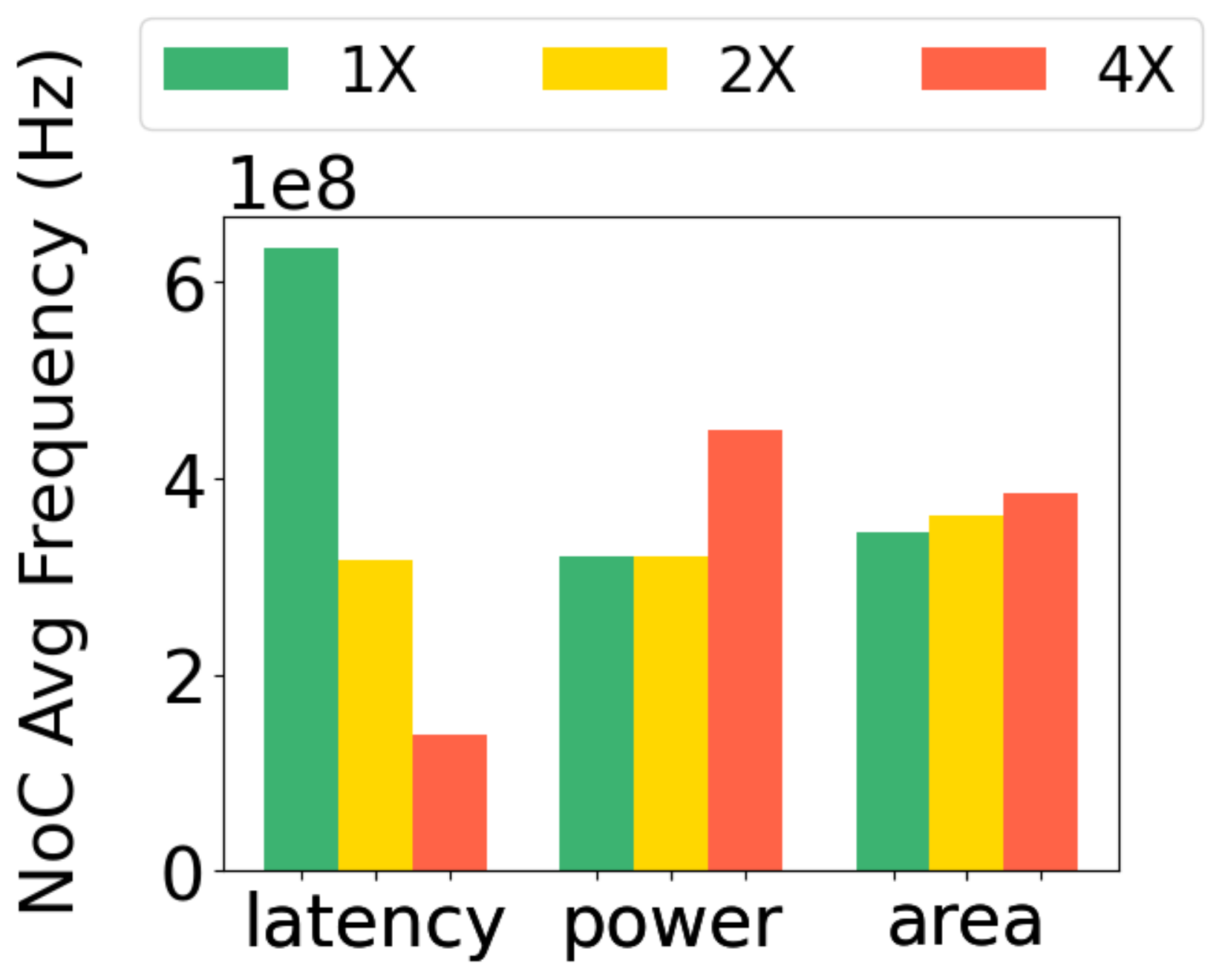}
        \label{fig:case_study:bus_freq}
    }
\caption{System implications for different budgets. FARSI lowers block count and thus lowers design complexity when the latency and power budgets are relaxed by 2$\times$ and 4$\times$. FARSI is also careful about keep hardware knobs low such as memory size/frequency when possible.}
    \label{fig:case_study:SI_count_all}
\end{figure}

\textbf{Component Implications:} Figures~\ref{fig:case_study:ip_cnt} and~\ref{fig:case_study:noc_cnt} show the impact of different budgets on IP and Network on a chip (NoC) subsystems. As shown, increasing the latency budget from 1X to 4X reduces the IP and NoC counts by 17\% and 150\% respectively.
This shows that FARSI lowers the block count when system budgets are relaxed.
In addition, this indicates that in delivering performance, NoC design/integration should be prioritized as the performance has a higher sensitivity to NoC than IP count. We also observe that a power budget increase of 1X to 4X results in a 36\% and 59\% reduction in IP and NoC count. This means that IP's impact on power reduction is higher than its impact on performance. However, the reverse is true for NoC's. We also do not see a meaningful relationship with the area.

\begin{figure}[t!]
\centering
\subfloat[NoC's link count heterogeneity.]{
        \includegraphics[trim=6.5 6.5 6.5 6.5, clip, width=.33\linewidth]{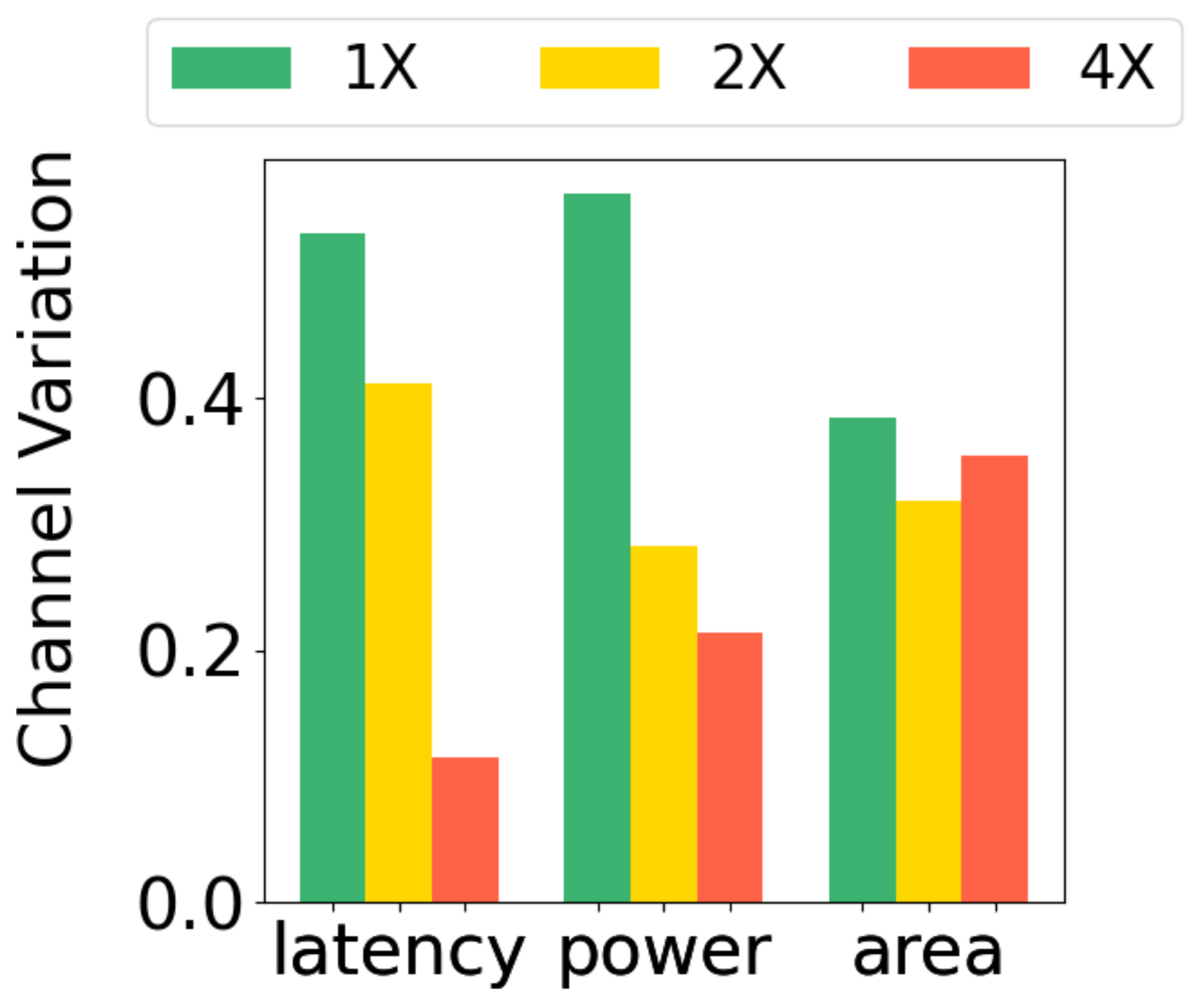}
        \label{fig:case_study:channel_cnt_htero}
    }
     \subfloat[Scratchpad size heterogeneity.]{
        \includegraphics[trim=6.5 6.5 6.5 6.5, clip, width=.33\linewidth]{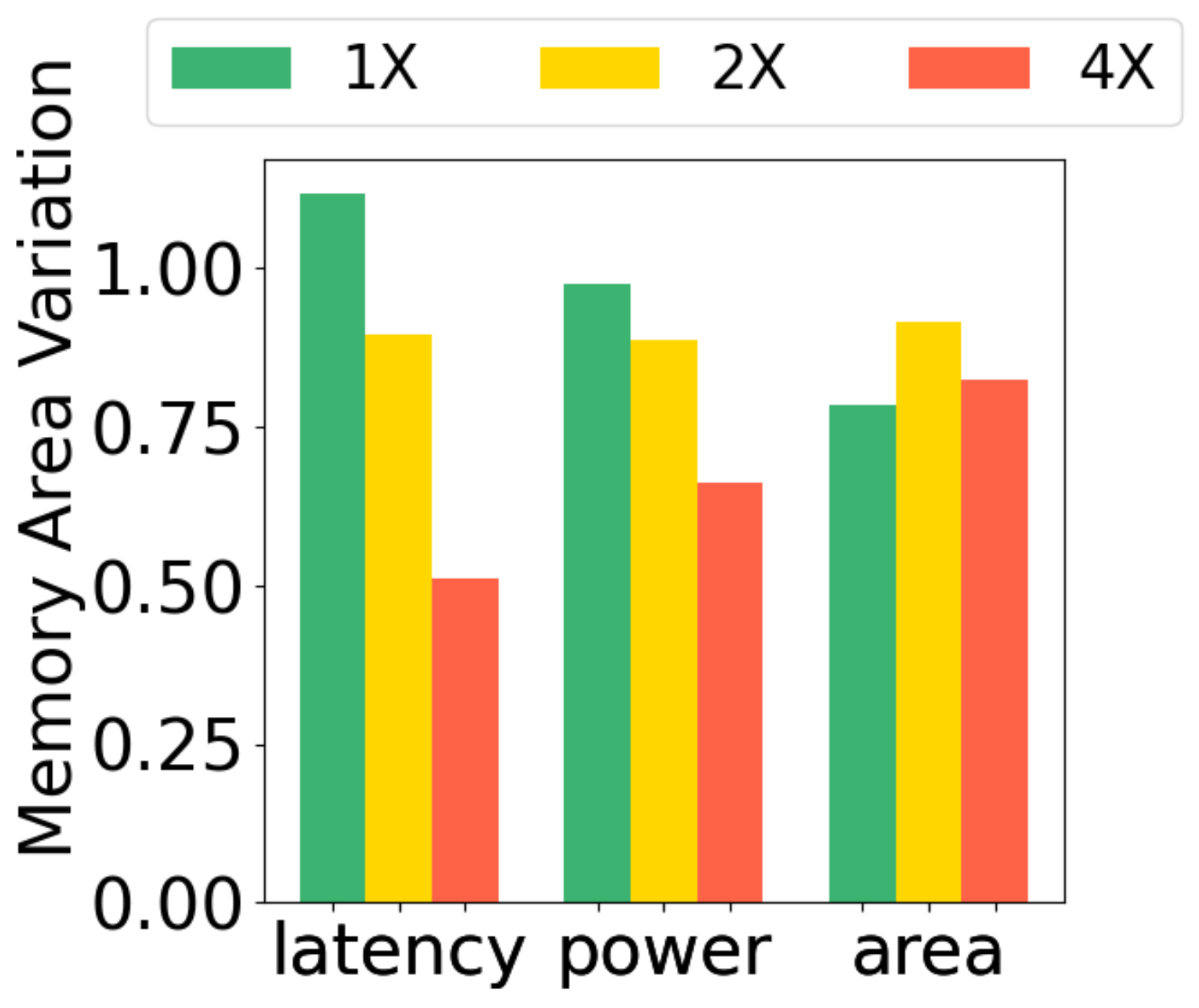}
        \label{figs:case_study:memory_size_hetero}
    }
  \subfloat[NoC frequency heterogeneity.]{
        \includegraphics[trim=6.5 6.5 6.5 6.5, clip, width=.33\linewidth]{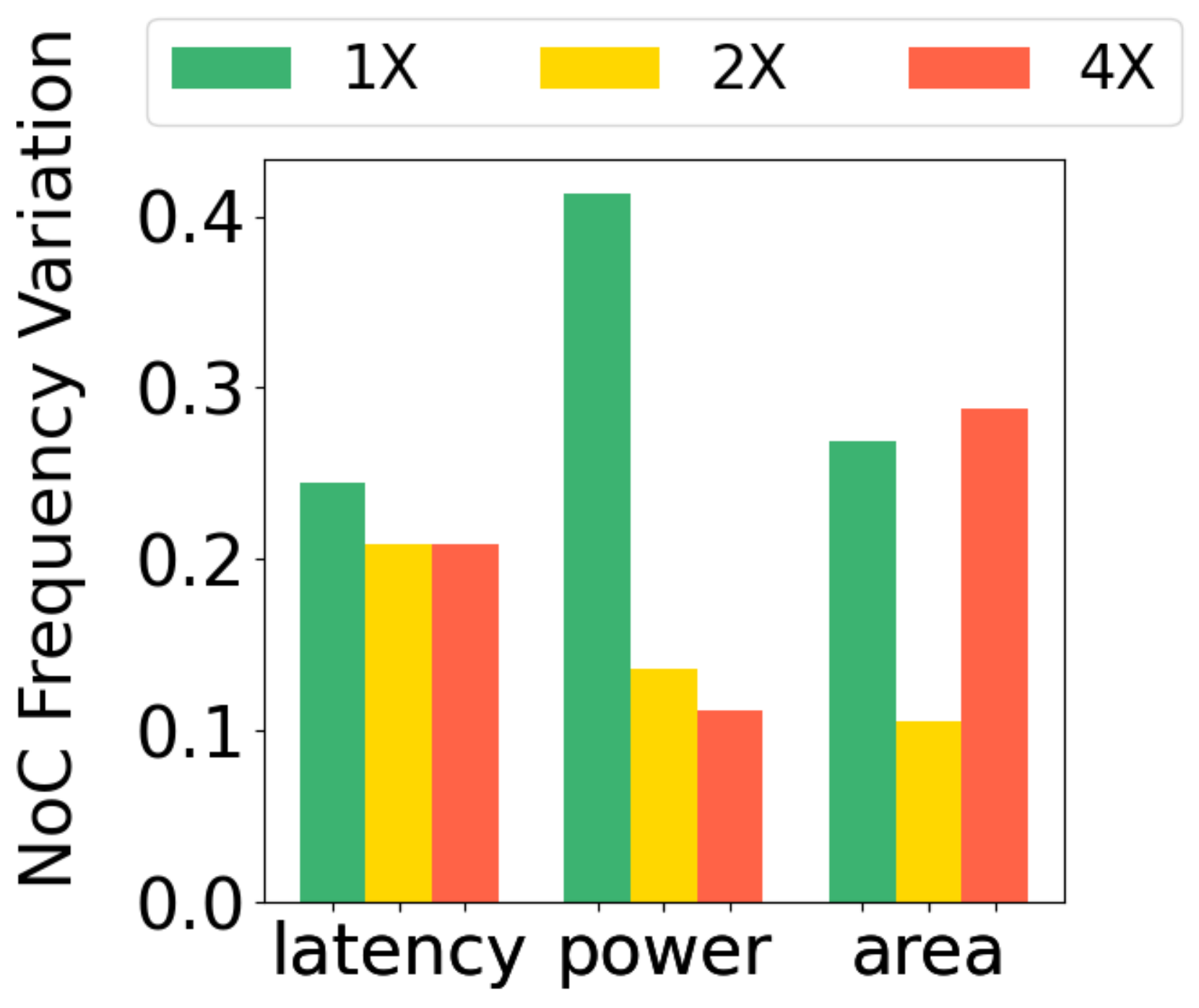}
        \label{fig:case_study:bus_freq_hetero}
    }
\caption{System heterogeneity. FARSI lowers the block heterogeneity and thus design complexity when the latency and power budgets are relaxed by 2$\times$ and 4$\times$.}
    \label{fig:case_study:SI_hetero_all}
\end{figure}

Figure~\ref{fig:case_study:memory_size} shows memory size sensitivity to the budgets.
Increasing the latency and power budget from 1X to 4X results in a total on-chip scratchpad memory size reduction of 28\% and 21\%, respectively. This is because the relaxed budget can tolerate higher global data movement latency and energy of DRAM. However, for the same budget increase in area, the (low-density) SRAM area can be freely increased by 25\% to move the data back locally.
System designers can use FARSI and conduct such studies to find optimal memory allocations across SRAM and DRAM.

System designers can also use FARSI to investigate the impact of budget on NoC Frequency. 
Figure ~\ref{fig:case_study:bus_freq} details this sensitivity.
Concretely, we observe that as the latency budget is relaxed to 4X, NoC frequency is scaled down by 360\%. This is because lower performance budgets do not require high-frequency operating NoCs. In contrast, relaxing the power budget by 4X instead scales up the frequency by 70\%. This is because, for higher power budgets, the system can tolerate higher frequency NoCs. Overall, we show that FARSI can exploit system-budget trade-offs and tune the design according to the budget needs.

\textbf{System Heterogeneity:} System heterogeneity (block variations) is also a determinant of the design complexity/development effort. 
To quantify a system variable's heterogeneity (e.g., scratch pad size variation), we use the \textit{coefficient of variation} ($CV = \frac{standard\  deviation}{mean}$)~\cite{cv} which captures the variable's deviation from its mean. Higher CV equates to a higher variation; as an example, in Figure~\ref{figs:case_study:memory_size_hetero}, the height of leftmost bar, i.e., 1.1, indicates a 110\% mean variation of scratchpad memory sizes within the system. 

As shown in Figure~\ref{fig:case_study:channel_cnt_htero}, increasing the performance or power budgets by 4X allow for a 450\% and 150\% reduction in the link (channel) count heterogeneity among NoCs. Simply put, this means lower variation in the number of links across NoCs. This lowers the design effort as variation in NoC design increases the design complexity.
We also observe that higher NoC customization is necessary for power budget comparing to performance budget variations. This is because the lowest and the highest power budgets (1X and 4X) tend to cause more heterogeneity than their corresponding latency scaling values.
 

A similar trend can be observed in both memory size (Figure~\ref{figs:case_study:memory_size_hetero}) and NoC frequency heterogeneity (Figure~\ref{fig:case_study:bus_freq_hetero}). The former experiences a 118\% and
47\% variation reduction as a result of relaxing the system performance and power budget. The latter experiences a 17\% and 267\% reduction for relaxation of the same metrics. This renders memory variation more critical for latency delivery while bus frequency variation more critical for power.   
Note that memory size and NoC frequency variations impact complexity as the former increases the optimization or customization effort, and the latter increases the clock tree and PLL complexity.

Overall, we see that FARSI lowers system heterogeneity in response to budget relaxation. In addition, system designers can use FARSI to investigate the product quality impact on system heterogeneity for different components. This then can guide various product decisions while keeping the development effort in mind.

\begin{figure}[t]
    \centering
    \subfloat[Accelerator-level parallelism (ALP).]{
        \includegraphics[trim=6.5 6.5 6.5 6.5, clip, width=.33\linewidth]{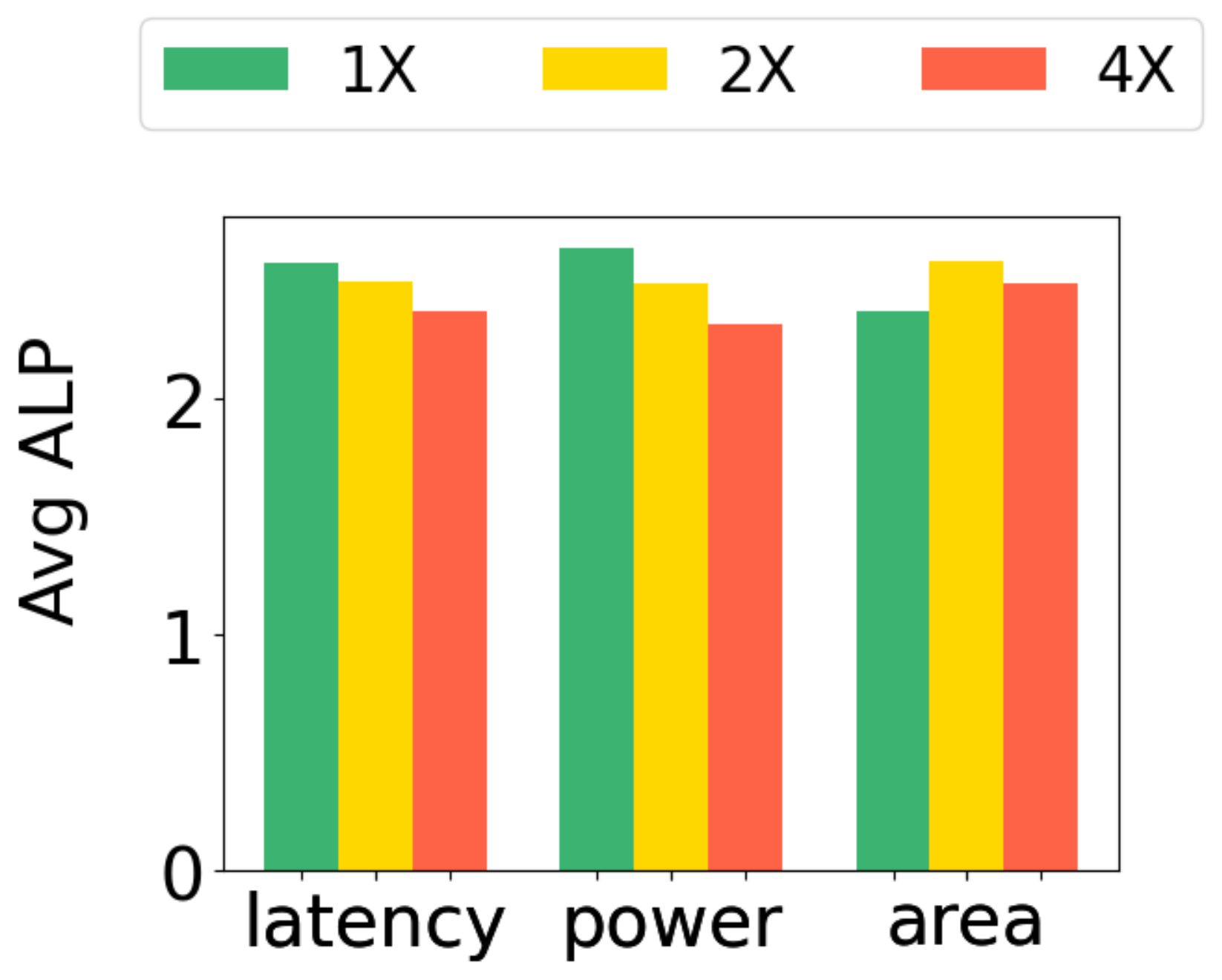}
        \label{figs:case_study:accel_paralellism}
    }
    \subfloat[Scratchpad memory traffic.]{
        \includegraphics[trim=6.5 6.5 6.5 6.5, clip, width=.33\linewidth]{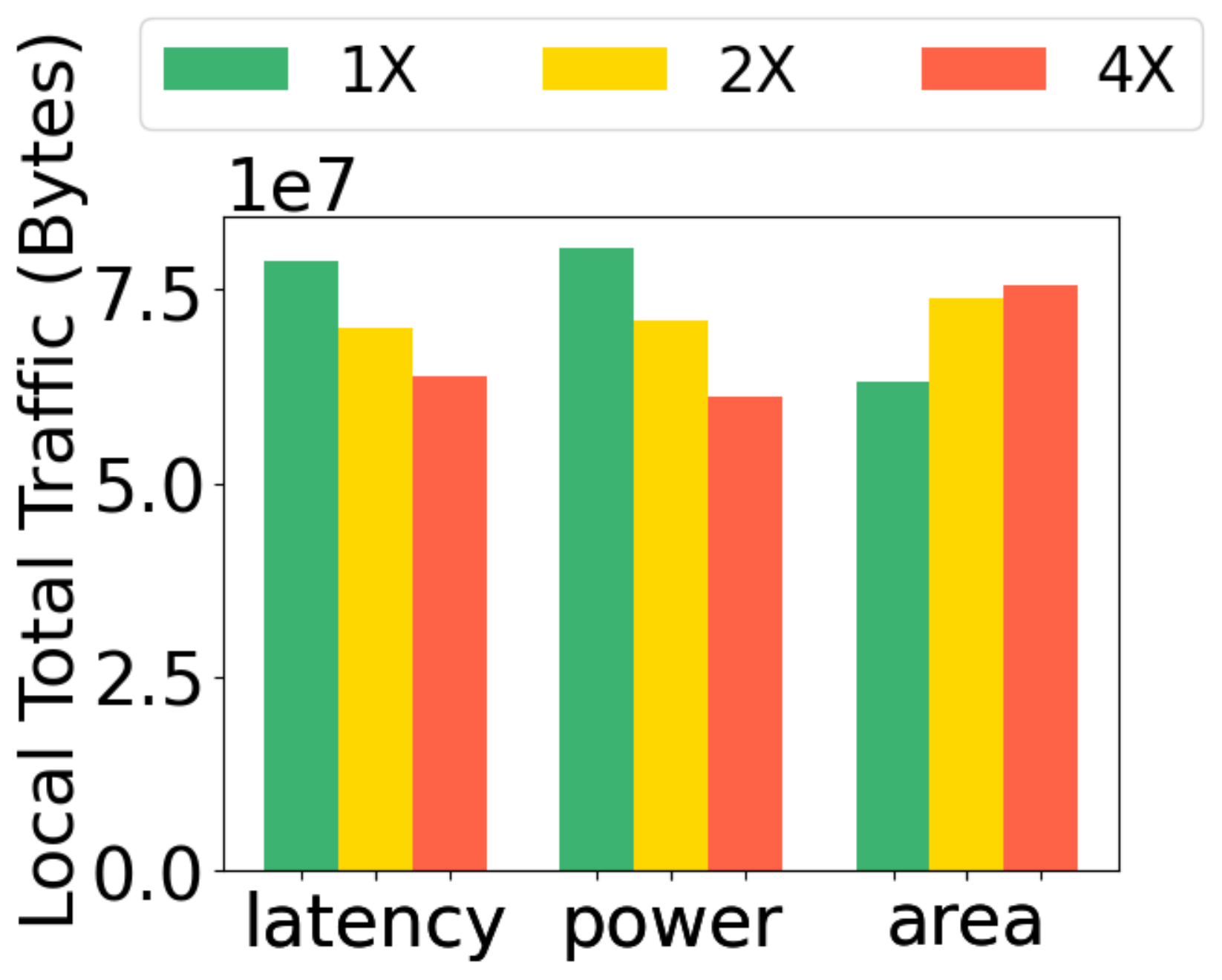}
        \label{fig:case_study:memory_traffic}
    }
    \subfloat[Scratchpad memory reuse.]{
        \includegraphics[trim=6.5 6.5 6.5 6.5, clip, width=.33\linewidth]{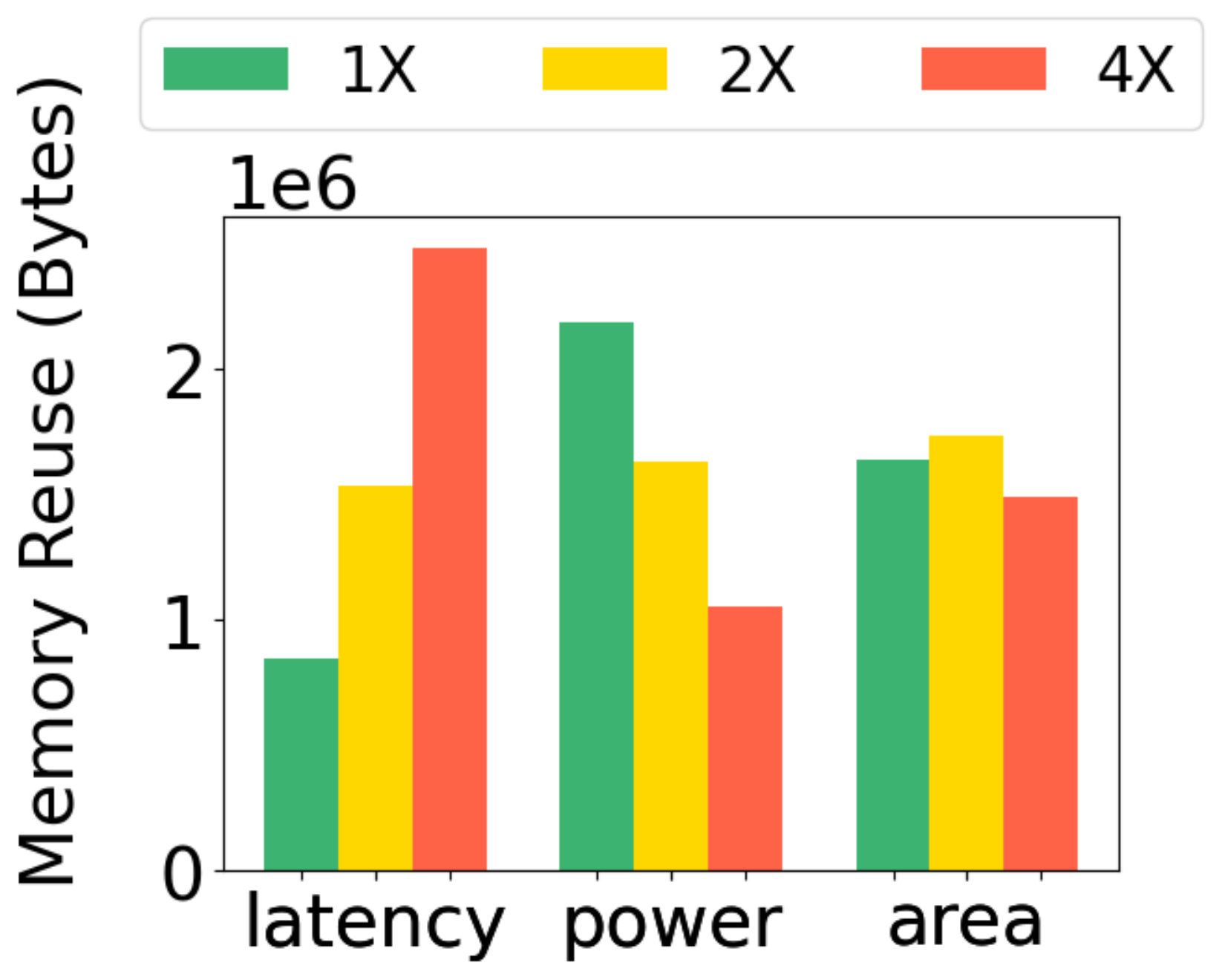}
        \label{fig:case_study:memory_reuse}
    }
    \caption{System dynamics. FARSI's capability to capture various system dynamics can help system designers to make design decisions.}
    \label{fig:case_study:local_mem_traffic}
\end{figure}

\textbf{System Dynamics:}
We provide designers with system dynamics' analysis to guide the design planning. 
Here we detail some of these dynamics. 
Figure~\ref{figs:case_study:accel_paralellism} quantifies the accelerator level parallelism, i.e., the average number of accelerators that run in parallel~\cite{HillReddiALP}. As shown, an increase of 8\% and 13\% are needed to meet the tighter 1X performance and power budgets. Such parallelism increases the local traffic by 23\% and 31\% (Figure~\ref{fig:case_study:memory_traffic}) and demands
a 107\% increase in reuse (how many times a memory is re-accessed, measured in bytes) to keep the memory size small, specifically when power efficiency is required (Figure~\ref{fig:case_study:memory_reuse}). However, to deliver for performance, FARSI requires lower memory reuse of 65\% to prevent memory contention. Note that the reuse's opposite direction between performance and power indicates the importance of memory mapping and its delicate balancing act. Such optimizations can only be achieved using an agile (rather than manual) methodology with a holistic system-level lens to explore sufficient scenarios.

\subsection{Divide and Conquer Suboptimality}
Lack of access to holistic design methodologies forces designers to tame the complexity by taking a divide and conquer approach. This means splitting the system into subsystems (one for each workload), imposing ad hoc (estimated) power and area budgets on each, and finally using best efforts to reach them in isolation. This approach is specifically common in domains that consists of a diverse set of sub-domains such as AR (video, audio, graphics, ...), however as we will show, it can lead to sub-optimal designs due to myopic budget estimations (Problem 1) and myopic optimizations (Problem 2). 

\textbf{Problem 1, Myopic Budget Estimation:}
In the absence of automated DSEs, each workload's power/area budget is decided manually using architects' insights and back-of-the-envelope estimates. If the estimates for a workload are too tight, the entire chip budget needs to be expanded to incorporate the workload's best design. However, if a workload's budgets are too loose, optimization opportunities targeting a tighter budget are left unexploited. (Note that the extra budget can then be redistributed to the other workloads in need).
We emulate this problem by setting the power budgets according to isolated power estimates provided in~\cite{huzaifa2021illixr}, while for the area, we use power as a proxy and budget each workload according to its power ratio relative to the entire system~\cite{compArchTech2008}.
Note that from~\cite{huzaifa2021illixr}, data are scaled for 5nm according to~\cite{TSMC40nm}, \cite{TSMC5nm} and \cite{cheetah}. Latency values are set similar to previous sections to ensure a high-quality user experience. We use FARSI to find a design that met this pre-determined budget for each workload.

\textbf{Problem 2, Myopic Optimizations in Isolation:}
Focusing on each workload in isolation misses out on the cross-workload optimization opportunities such as memory sharing. 
To isolate this issue and relieve the said budgeting problem (problem 1), for each workload, we sweep all budget values from 0 to the SoC budget with increments of 5\%.
Concretely, for each workload in isolation, we run FARSI with the mentioned budget sweep and generate a power/area Pareto front. Then, final SoCs (that run all the workloads) are put together by combining all the permutations of the designs on the workload's Pareto Fronts.
 
\begin{figure}[!t]
\centering
\includegraphics[trim=2 2 2 2, clip, width=0.64\linewidth]{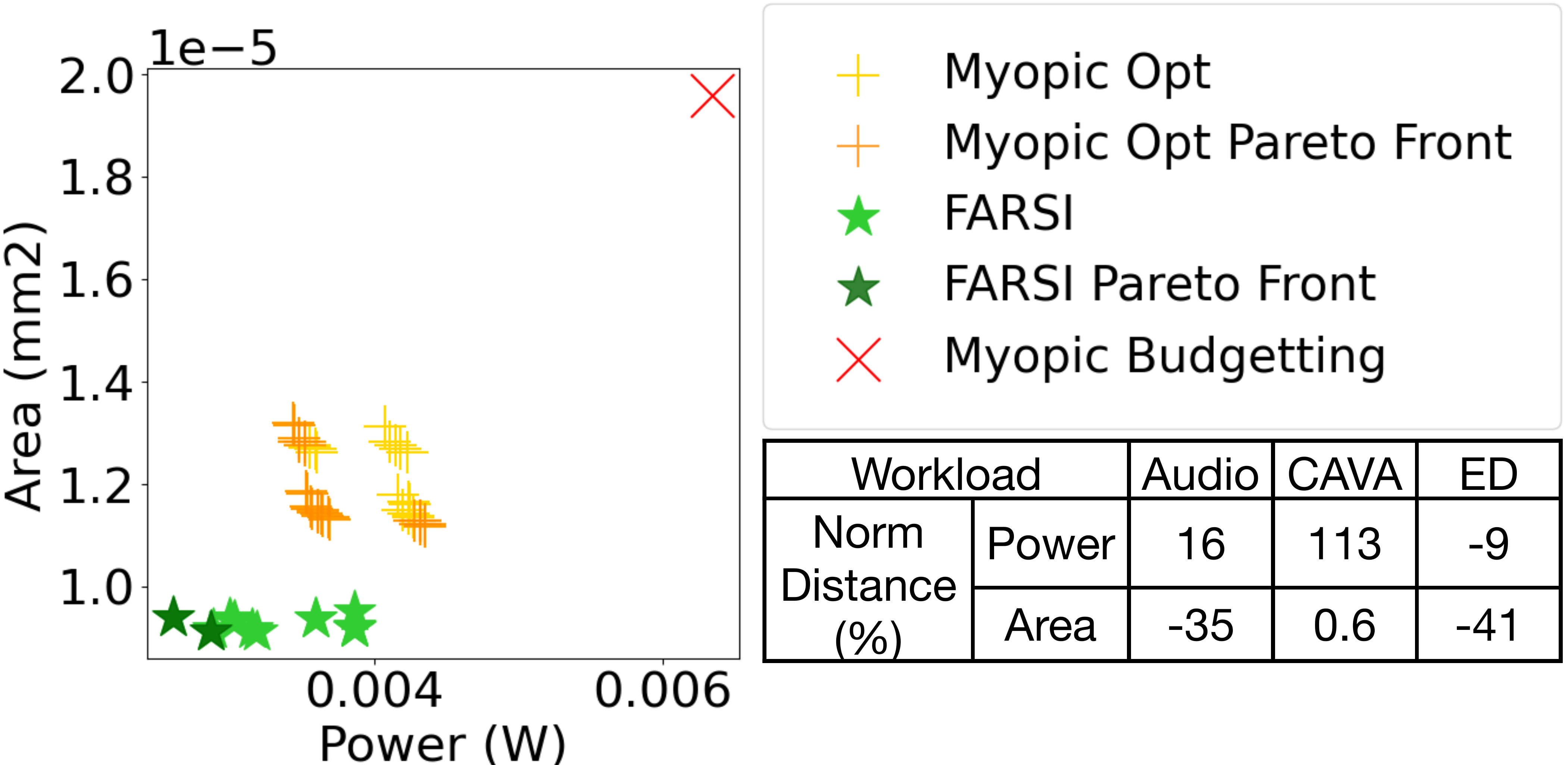}
\caption{Divide and conquer sub-optimality. Myopic optimization/budgeting (left). Myopic budgeting details (right). }
\label{fig:divide_and_con}
\end{figure} 
 
\textbf{System Degradation Associated with Problem 1 and 2:} 
Here we quantify the degradation by comparing the methodology used in Problem 1 and 2 with full-fledged FARSI. Note that at the high level, full-fledged FARSI automatically solves problem 1 as it does not require individual workload budgets. Instead, it finds the optimal sub-budgeting as it explores the space. Furthermore, FARSI circumvents Problem 2 by conducting cross-workload optimizations.  

Figure~\ref{fig:divide_and_con} illustrates the suboptimality of both approaches with the x-axis and y-axis denoting the power and area associated with the designs generated for each methodology.
Myopic Budgeting predictably does the worst with its point (far top right) experiencing a 56\% and 52\% power and area degradation comparing to FARSI's Pareto Front points (on average). The table shown sheds light on the cause behind this issue by presenting the distance between Myopic Budgetting's best design and its pre-determined budget (normalized and multiplied by 100). Negative values mean the design never met the budget, and thus, the budget was too tight and vice versa. As shown, the power/area budgets for Edge Detection and the area for Audio were too tight. However, CAVA's power budget was set too loose and should have been distributed across the other two workloads. The area budget for CAVA was optimal.

Myopic Optimization (Myopic Opt) provides suboptimal solutions as well. As shown in Figure~\ref{fig:divide_and_con}, both its Pareto Fronts and almost all of its generated designs are less optimal comparing to full-fledged FARSI. Concretely, points on the Pareto Front of this methodology, on average, experience a 27\% and 21\% power and area degradation comparing to FARSI. Myopic optimization cannot use cross-workload mapping solutions to share memory and save area or keep the congestion low and improve performance without improving frequency.

%% file: tex/related.tex
\section{Related Work}
\label{sec:related}

\textbf{Simulators:}
Many pre-RTL simulators ranging from accurate yet slow cycle-accurate simulators \cite{gem5, GPGPUSIM} to fast yet low fidelity analytical models \cite{gables,logCA, roofline, ETP, GPGPUAn} have been proposed. The formers' low agility and the latters' inaccuracy render them suboptimal for DSSoC DSEs. 
Trace-driven frameworks address some of the said problems although they focus on individual components, e.g., \cite{aladdin, gem5-salam, accelsim} target accelerators and \cite{cachesim, memorysim} target memory. In contrast, our work is agile, accurate, and captures system-level dynamics. Although works such as \cite{aladdin, gem5-salam} are later augmented to enable system-level modeling, they revert to cycle-accurate non-agile simulation.

\textbf{Design Space Exploration (DSE) Frameworks:}
Works such as \cite{RankArch, RpStacks-MT, EOA, MemDSE} target DSE for individual components such as the processor, accelerator, or memory. Our work improves on them by providing a holistic system view.
Others only target one design stage. For example,  \cite{tangram, schedtask, AutoScale, Gloss} target scheduling/runtime resource management, \cite{MemMap, CommMapping, CGRAMapping} target mapping, and \cite{CSS, MVBD, NASA} combine allocation and mapping. Our work improves on them by providing multi-stage optimization and three other co-design vectors.

SoC DSE has used various heuristics, such simulated annealing \cite{eles1997annealing}, integer linear programming \cite{MVBD},  particle swarm \cite{PSO}, genetic algorithms \cite{dick1998generic-mogac, palesi2002generic}, and reinforcement learning 
\cite{AutoScale}; Our work improves on them by incorporating architectural reasoning. Note we are not limited to simulated annealing as our architectural reasoning can be integrated into other heuristics. 

%% file: tex/conclusion.tex
\section{Conclusion}
\label{sec:conclusion}
This work presents FARSI, a DSSoC DSE equipped with an agile system simulator, and an automated heuristic with built-in and expandable architectural reasoning. We identify critical ingredients of an optimal DSSoC DSE and quantify their impact on the convergence time, and further show FARSI's heavy use of them.
We achieve 8,400X, and 98.5\% simulation speed up and accuracy compared to Platform Architect. We also achieve 16X convergence speed up exploiting architectural reasoning and co-design comparing to simulated annealing. 
We further present two case studies show-casing FARSI's importance in the design challenges of future complex systems.
FARSI is an open-source project and it has been evaluated at a major industry organization for industry-relevant use cases that require DSSoC solutions in a highly resource-constrained environment such as AR/VR/MR glasses.

%% file: main.bbl
\begin{thebibliography}{10}

\bibitem{Apple_AR}
Apple in google.
\newblock \url{https://www.apple.com/augmented-reality/}, 2021.

\bibitem{AR_education}
Ar for education.
\newblock
  \url{https://www.forbes.com/sites/bernardmarr/2021/07/23/10-best-examples-of-vr-and-ar-in-education/?sh=668f90021f48},
  2021.

\bibitem{AR_repair}
Ar for repair.
\newblock
  \url{https://www.wikitude.com/blog-augmented-reality-maintenance-and-remote-assistance/},
  2021.

\bibitem{AR_tourism}
Ar for toursim.
\newblock
  \url{https://www.forbes.com/sites/bernardmarr/2021/04/12/the-amazing-ways-vr-and-ar-are-transforming-the-travel-industry/?sh=5120735636e0},
  2021.

\bibitem{Facebook_AR}
Ar in facebook.
\newblock
  \url{https://research.fb.com/category/augmented-reality-virtual-reality/},
  2021.

\bibitem{Smart_glasses}
Ar in glasses.
\newblock \url{https://en.wikipedia.org/wiki/Smartglasses}, 2021.

\bibitem{Google_AR}
Ar in google.
\newblock \url{https://www.google.com/glass/start/}, 2021.

\bibitem{Nikon}
Nikon d7000.
\newblock \url{https://en.wikipedia.org/wiki/Nikon_D7000}, 2021.

\bibitem{logCA}
Muhammad Shoaib~Bin Altaf and David~A. Wood.
\newblock Logca: A high-level performance model for hardware accelerators.
\newblock {\em SIGARCH Comput. Archit. News}, 45(2):375–388, June 2017.

\bibitem{cachesim}
Mario Badr and Natalie~Enright Jerger.
\newblock Synfull: Synthetic traffic models capturing cache coherent behaviour.
\newblock In {\em 2014 ACM/IEEE 41st International Symposium on Computer
  Architecture (ISCA)}, pages 109--120, 2014.

\bibitem{GPGPUSIM}
Ali Bakhoda, George~L. Yuan, Wilson W.~L. Fung, Henry Wong, and Tor~M. Aamodt.
\newblock Analyzing cuda workloads using a detailed gpu simulator.
\newblock In {\em 2009 IEEE International Symposium on Performance Analysis of
  Systems and Software}, PPoPP '12, page 11–22, New York, NY, USA, 2009.
  Association for Computing Machinery.

\bibitem{gem5}
Nathan Binkert, Bradford Beckmann, Gabriel Black, Steven~K. Reinhardt, Ali
  Saidi, Arkaprava Basu, Joel Hestness, Derek~R. Hower, Tushar Krishna, Somayeh
  Sardashti, Rathijit Sen, Korey Sewell, Muhammad Shoaib, Nilay Vaish, Mark~D.
  Hill, and David~A. Wood.
\newblock The gem5 simulator.
\newblock {\em SIGARCH Comput. Archit. News}, 39(2):1–7, August 2011.

\bibitem{ETP}
T.~{Chen}, A.~{Rucker}, and G.~E. {Suh}.
\newblock Execution time prediction for energy-efficient hardware accelerators.
\newblock In {\em 2015 48th Annual IEEE/ACM International Symposium on
  Microarchitecture (MICRO)}, pages 457--469, 2015.

\bibitem{MemDSE}
Mohammad Dashti, Alexandra Fedorova, Justin Funston, Fabien Gaud, Renaud
  Lachaize, Baptiste Lepers, Vivien Quema, and Mark Roth.
\newblock Traffic management: A holistic approach to memory placement on numa
  systems.
\newblock {\em SIGARCH Comput. Archit. News}, 41(1):381–394, March 2013.

\bibitem{dick1998generic-mogac}
R.P. Dick and N.K. Jha.
\newblock Mogac: a multiobjective genetic algorithm for hardware-software
  cosynthesis of distributed embedded systems.
\newblock {\em IEEE Transactions on Computer-Aided Design of Integrated
  Circuits and Systems}, (10), 1998.

\bibitem{eles1997annealing}
Petru Eles, Zebo Peng, Krzysztof Kuchcinski, and Alex Doboli.
\newblock System level hardware/software partitioning based on simulated
  annealing and tabu search.
\newblock {\em Design Automation for Embedded Systems}, 2, 07 1997.

\bibitem{EOA}
S.~{Eyerman}, L.~{Eeckhout}, and K.~{De Bosschere}.
\newblock Efficient design space exploration of high performance embedded
  out-of-order processors.
\newblock In {\em Proceedings of the Design Automation Test in Europe
  Conference}, volume~1, 2006.

\bibitem{material}
Godot.
\newblock Material testers.
\newblock
  \url{https://github.com/godotengine/godot-demo-projects/tree/master/3d/material_testers},
  2020.

\bibitem{platformer}
Godot.
\newblock Platformer {3D}.
\newblock
  \url{https://github.com/godotengine/godot-demo-projects/tree/master/3d/platformer},
  2020.

\bibitem{MVBD}
S.~{Graf}, M.~{Glaß}, J.~{Teich}, and C.~{Lauer}.
\newblock Multi-variant-based design space exploration for automotive embedded
  systems.
\newblock In {\em 2014 Design, Automation Test in Europe Conference Exhibition
  (DATE)}, pages 1--6, 2014.

\bibitem{gables}
Mark Hill and Vijay Janapa~Reddi.
\newblock Gables: A roofline model for mobile socs.
\newblock In {\em 2019 IEEE International Symposium on High Performance
  Computer Architecture (HPCA)}, pages 317--330, 2019.

\bibitem{HillReddiALP}
Mark~D. Hill and Vijay~Janapa Reddi.
\newblock Accelerator-level parallelism.
\newblock {\em CoRR}, abs/1907.02064, 2019.

\bibitem{CGRAMapping}
{Hongsik Lee}, {Dong Nguyen}, and J.~{Lee}.
\newblock Optimizing stream program performance on cgra-based systems?
\newblock In {\em 2015 52nd ACM/EDAC/IEEE Design Automation Conference (DAC)},
  2015.

\bibitem{huzaifa2021illixr}
Muhammad Huzaifa, Rishi Desai, Samuel Grayson, Xutao Jiang, Ying Jing, Jae Lee,
  Fang Lu, Yihan Pang, Joseph Ravichandran, Finn Sinclair, Boyuan Tian, Hengzhi
  Yuan, Jeffrey Zhang, and Sarita~V. Adve.
\newblock {Exploring Extended Reality with ILLIXR: A New Playground for
  Architecture Research}.
\newblock {\em CoRR}, abs/2004.04643, 2021.

\bibitem{RpStacks-MT}
Hanhwi Jang, Jae-Eon Jo, Jaewon Lee, and Jangwoo Kim.
\newblock Rpstacks-mt: A high-throughput design evaluation methodology for
  multi-core processors.
\newblock In {\em Proceedings of the 51st Annual IEEE/ACM International
  Symposium on Microarchitecture}, MICRO-51, page 586–599. IEEE Press, 2018.

\bibitem{NASA}
Z.~J. {Jia}, A.~D. {Pimentel}, M.~{Thompson}, T.~{Bautista}, and A.~{Núñez}.
\newblock Nasa: A generic infrastructure for system-level mp-soc design space
  exploration.
\newblock In {\em 2010 8th IEEE Workshop on Embedded Systems for Real-Time
  Multimedia}, pages 41--50, 2010.

\bibitem{schedtask}
Prathmesh Kallurkar and Smruti Sarangi.
\newblock Schedtask: a hardware-assisted task scheduler.
\newblock pages 612--624, 10 2017.

\bibitem{AR_med}
Carolien Kamphuis, Esther Barsom, Marlies Schijven, and L.H.(Noor) Christoph.
\newblock Augmented reality in medical education?
\newblock {\em Perspectives on medical education}, 3, 01 2014.

\bibitem{compArchTech2008}
Stefanos Kaxiras and Margaret Martonosi.
\newblock Computer architecture techniques for power-efficiency, 2008.

\bibitem{accelsim}
Mahmoud Khairy, Zhesheng Shen, Tor~M. Aamodt, and Timothy~G. Rogers.
\newblock Accel-sim: An extensible simulation framework for validated gpu
  modeling.
\newblock In {\em 2020 ACM/IEEE 47th Annual International Symposium on Computer
  Architecture (ISCA)}, pages 473--486, 2020.

\bibitem{AutoScale}
Y.~G. {Kim} and C.~J. {Wu}.
\newblock Autoscale: Energy efficiency optimization for stochastic edge
  inference using reinforcement learning.
\newblock In {\em 2020 53rd Annual IEEE/ACM International Symposium on
  Microarchitecture (MICRO)}, 2020.

\bibitem{hpvm}
Maria Kotsifakou, Prakalp Srivastava, Matthew~D Sinclair, Rakesh Komuravelli,
  Vikram Adve, and Sarita Adve.
\newblock {HPVM}: Heterogeneous parallel virtual machine.
\newblock In {\em Proceedings of the 23rd ACM SIGPLAN Symposium on Principles
  and Practice of Parallel Programming}, pages 68--80, 2018.

\bibitem{kotsifakou2018hpvm}
Maria Kotsifakou, Prakalp Srivastava, Matthew~D Sinclair, Rakesh Komuravelli,
  Vikram Adve, and Sarita Adve.
\newblock {HPVM}: Heterogeneous parallel virtual machine.
\newblock In {\em Proceedings of the 23rd ACM SIGPLAN Symposium on Principles
  and Practice of Parallel Programming}, pages 68--80, 2018.

\bibitem{RankArch}
J.~{Lee}, H.~{Jang}, and J.~{Kim}.
\newblock Rpstacks: Fast and accurate processor design space exploration using
  representative stall-event stacks.
\newblock In {\em 2014 47th Annual IEEE/ACM International Symposium on
  Microarchitecture}, pages 255--267, 2014.

\bibitem{CommMapping}
{Liang-Yu Lin}, {Cheng-Yeh Wang}, {Pao-Jui Huang}, {Chih-Chieh Chou}, and
  {Jing-Yang Jou}.
\newblock Communication-driven task binding for multiprocessor with latency
  insensitive network-on-chip.
\newblock In {\em Proceedings of the ASP-DAC 2005. Asia and South Pacific
  Design Automation Conference, 2005.}, volume~1, pages 39--44 Vol. 1, 2005.

\bibitem{online_learning}
Barbara~B Lockee.
\newblock Online education in the post-covid era.
\newblock {\em Nature Electronics}, 4(1):5--6, 2021.

\bibitem{CSS}
M.~{Lukasiewycz}, M.~{Streubuhr}, M.~{Glass}, C.~{Haubelt}, and J.~{Teich}.
\newblock Combined system synthesis and communication architecture exploration
  for mpsocs.
\newblock In {\em 2009 Design, Automation Test in Europe Conference
  Exhibition}, 2009.

\bibitem{sponza}
Monado.
\newblock {Sponza} scene in {Godot} with {OpenXR} addon.
\newblock
  \url{https://gitlab.freedesktop.org/monado/demos/godot-sponza-openxr}, 2019.

\bibitem{perf}
N/A.
\newblock perf.
\newblock \url{https://perf.wiki.kernel.org/index.php/Main_Page}, 2020.

\bibitem{cacti}
N/A.
\newblock Hp labs: Cacti.
\newblock \url{https://www.hpl.hp.com/research/cacti/}, 2021.

\bibitem{cv}
N/A.
\newblock Variation coefficient wikipedia.
\newblock \url{https://en.wikipedia.org/wiki/Coefficient_of_variation}, 2021.

\bibitem{MemMap}
Heikki Orsila, Tero Kangas, Erno Salminen, Timo~D. H\"{a}m\"{a}l\"{a}inen, and
  Marko H\"{a}nnik\"{a}inen.
\newblock Automated memory-aware application distribution for multi-processor
  system-on-chips.
\newblock {\em J. Syst. Archit.}, 53(11):795–815, November 2007.

\bibitem{PSO}
G.~{Palermo}, C.~{Silvano}, and V.~{Zaccaria}.
\newblock Discrete particle swarm optimization for multi-objective design space
  exploration.
\newblock In {\em 2008 11th EUROMICRO Conference on Digital System Design
  Architectures, Methods and Tools}, pages 641--644, 2008.

\bibitem{palesi2002generic}
M.~Palesi and T.~Givargis.
\newblock Multi-objective design space exploration using genetic algorithms.
\newblock In {\em Proceedings of the Tenth International Symposium on
  Hardware/Software Codesign. CODES 2002 (IEEE Cat. No.02TH8627)}, pages
  67--72, 2002.

\bibitem{PA}
{Platform Architect}.
\newblock
  \url{https://www.synopsys.com/verification/virtual-prototyping/platform-architect.html},
  2021.

\bibitem{memorysim}
R.~{Pellizzoni}, A.~{Schranzhofer}, {Jian-Jia Chen}, M.~{Caccamo}, and
  L.~{Thiele}.
\newblock Worst case delay analysis for memory interference in multicore
  systems.
\newblock In {\em 2010 Design, Automation Test in Europe Conference Exhibition
  (DATE 2010)}, pages 741--746, 2010.

\bibitem{tangram}
Raghavendra~Pradyumna Pothukuchi, Joseph~L. Greathouse, Karthik Rao,
  Christopher Erb, Leonardo Piga, Petros~G. Voulgaris, and Josep Torrellas.
\newblock Tangram: Integrated control of heterogeneous computers.
\newblock In {\em Proceedings of the 52nd Annual IEEE/ACM International
  Symposium on Microarchitecture}, MICRO '52, page 384–398, New York, NY,
  USA, 2019. Association for Computing Machinery.

\bibitem{Gloss}
Sumanaruban Rajadurai, Jeffrey Bosboom, Weng-Fai Wong, and Saman Amarasinghe.
\newblock Gloss: Seamless live reconfiguration and reoptimization of stream
  programs.
\newblock ASPLOS '18, page 98–112, New York, NY, USA, 2018. Association for
  Computing Machinery.

\bibitem{cheetah}
Brandon Reagen, Woo-Seok Choi, Yeongil Ko, Vincent~T. Lee, Hsien-Hsin~S. Lee,
  Gu-Yeon Wei, and David Brooks.
\newblock Cheetah: Optimizing and accelerating homomorphic encryption for
  private inference.
\newblock In {\em 2021 IEEE International Symposium on High-Performance
  Computer Architecture (HPCA)}, pages 26--39, 2021.

\bibitem{gem5-salam}
S.~{Rogers}, J.~{Slycord}, M.~{Baharani}, and H.~{Tabkhi}.
\newblock gem5-salam: A system architecture for llvm-based accelerator
  modeling.
\newblock In {\em 2020 53rd Annual IEEE/ACM International Symposium on
  Microarchitecture (MICRO)}, pages 471--482, 2020.

\bibitem{TSMC5nm}
David Schor.
\newblock {TSMC Starts 5-Nanometer Risk Production}.
\newblock
  \url{https://fuse.wikichip.org/news/2207/tsmc-starts-5-nanometer-risk-production/},
  2019.

\bibitem{aladdin}
Yakun~Sophia Shao, Brandon Reagen, Gu-Yeon Wei, and David Brooks.
\newblock Aladdin: A pre-rtl, power-performance accelerator simulator enabling
  large design space exploration of customized architectures.
\newblock In {\em 2014 ACM/IEEE 41st International Symposium on Computer
  Architecture (ISCA)}, pages 97--108, 2014.

\bibitem{diephoto}
Yakun~Sophia Shao and Yu~Emma Wang.
\newblock Die photo analysis.
\newblock
  \url{http://vlsiarch.eecs.harvard.edu/research/accelerators/die-photo-analysis/},
  2015.

\bibitem{GPGPUAn}
Jaewoong Sim, Aniruddha Dasgupta, Hyesoon Kim, and Richard Vuduc.
\newblock A performance analysis framework for identifying potential benefits
  in gpgpu applications.
\newblock {\em SIGPLAN Not.}, 47(8):11–22, February 2012.

\bibitem{zedmini}
Stereolabs.
\newblock {ZED Mini - Mixed-Reality Camera}.
\newblock \url{https://www.stereolabs.com/zed-mini/}.

\bibitem{takahashi2018}
Dean Takahashi.
\newblock Oculus chief scientist mike abrash still sees the rosy future through
  ar/vr glasses.
\newblock
  \url{https://venturebeat.com/2018/09/26/oculus-chief-scientist-mike-abrash-still-sees-the-rosy-future-through-ar-vr-glasses/},
  September 2018.

\bibitem{TSMC40nm}
TSMC.
\newblock {TSMC 40 nm Technology}.
\newblock
  \url{https://www.tsmc.com/english/dedicatedFoundry/technology/logic/l_40nm}.

\bibitem{roofline}
Samuel Williams, Andrew Waterman, and David Patterson.
\newblock Roofline: An insightful visual performance model for multicore
  architectures.
\newblock {\em Commun. ACM}, 52(4):65–76, April 2009.

\bibitem{cava}
Yuan Yao and Saketh Rama.
\newblock yaoyuannnn/cava.
\newblock \url{https://github.com/yaoyuannnn/cava}.

\bibitem{accelSeeker}
Georgios Zacharopoulos, Lorenzo Ferretti, Giovanni Ansaloni, Giuseppe
  Di~Guglielmo, Luca Carloni, and Laura Pozzi.
\newblock Compiler-assisted selection of hardware acceleration candidates from
  application source code.
\newblock pages 1--9, 2019.

\end{thebibliography}
